\documentclass[12pt
]{article}%
\usepackage[font=footnotesize]{caption}
\usepackage{graphicx}
\usepackage[update]{epstopdf}
\usepackage[mathscr]{euscript}
\usepackage{enumerate}
\usepackage{dcolumn}
    \newcolumntype{d}[1]{D{.}{.}{#1}}
\usepackage{bbm}
\usepackage{soul}
\usepackage{dsfont}
\usepackage{lscape}
\usepackage{placeins}
 \usepackage{longtable}
\usepackage{threeparttable}
\usepackage[centertags,sumlimits, intlimits, namelimits,]{amsmath}
\usepackage{amsfonts}
\usepackage[english]{babel}
\usepackage{amssymb}
\usepackage{appendix}
\usepackage{fancyhdr}
\usepackage{booktabs}
\usepackage[normalem]{ulem}
\usepackage{natbib}
\usepackage[utf8]{inputenc}
\usepackage{url}
\usepackage{longtable}
\usepackage[small]{titlesec}
\usepackage{geometry}
\usepackage[gen]{eurosym}
\usepackage{comment}
\usepackage{tabularx}
\usepackage{array}
\usepackage{multirow}
\usepackage{makecell}
\usepackage{pbox}
\usepackage{hhline}
\usepackage{rotating} 
\usepackage{pifont} 
\usepackage{float} 
\usepackage{outlines} 
\usepackage{subcaption} 
\usepackage{datatool} 
\DTLsetseparator{,} 
\usepackage{xltabular} 

\usepackage[usenames,dvipsnames]{xcolor}
\usepackage{hyperref}

\definecolor{mycolor}{rgb}{0.55, 0.0, 0.0}

\hypersetup{
		pdftoolbar=true,        
    pdfmenubar=true,        
    pdffitwindow=false,     
    pdfstartview={FitH},    
		pdftitle={Debt Aversion: Theory and Experiment},    
		pdfauthor={Thomas Meissner, David Albrecht},     
    pdfsubject={Debt Aversion: Theory and Experiment},   
    pdfkeywords={Debt Aversion}, 
    pdfnewwindow=true,      
    colorlinks=true,       
    linkcolor=mycolor,          
    citecolor=mycolor,        
    filecolor=mycolor,      
    urlcolor=mycolor,           
		pdfpagemode = UseNone,
}
\usepackage[hyperpageref]{backref}
   \renewcommand*{\backref}[1]{}
   \renewcommand*{\backrefalt}[4]{
      \ifcase #1
         Not cited
      \or
         Cited on page~#2.
      \else
         Cited on pages~#2.
      \fi}

\def\sym#1{\ifmmode^{#1}\else\(^{#1}\)\fi}
\renewcommand\abstract{{\noindent\bfseries Abstract\\[1.5ex]}}
\def\keywordname{{\bfseries Keywords}}%
\def\keywords#1{\par\addvspace\medskipamount{\rightskip=0pt plus1cm
\def\and{\ifhmode\unskip\nobreak\fi\ $\cdot$
}\noindent\keywordname\enspace\ignorespaces#1\par}}
\def\JELname{{\bfseries JEL Classification}\enspace}
\def\JEL#1{\par\addvspace\medskipamount{\rightskip=0pt plus1cm
\def\and{\ifhmode\unskip\nobreak\fi\ $\cdot$
}\noindent\JELname\ignorespaces#1\par}}
\def\ackname{Acknowledgments}%
\def\acknowledgement{\par\addvspace{17pt}\small\rmfamily
\trivlist\if!\ackname!\item[]\else
\item[\hskip\labelsep
{\bfseries\ackname}]\fi}

\renewcommand{\tablename}{TABLE}
\renewcommand\thetable{\arabic{table}}
\def\fnum@table{\tablename\nobreakspace\thetable}
\usepackage{tikz} 
\usepackage{pgfplots} 
\usepackage{pgfplotstable} 
\usepackage{verbatim} 
\usetikzlibrary{trees} 
\usetikzlibrary{patterns}
\pgfplotsset{width=\textwidth,compat=1.9} 

\newcolumntype{C}[1]{>{\centering\arraybackslash}m{#1}}
\newcolumntype{L}[1]{>{\raggedleft\arraybackslash}m{#1}}
\newcolumntype{J}[1]{>{\arraybackslash}m{#1}}

\include{data}

\begin{document}

\title{Debt Aversion: Theory and Measurement%
\footnote{Corresponding author: Thomas Meissner (\href{mailto: meissnet@gmail.com}{meissnet@gmail.com}). We are thankful for helpful comments from Arno Riedl, Chris Woolnough and participants of several conferences and seminars. Thomas Meissner acknowledges funding from the European Union's Horizon 2020 research and innovation programme under grant agreement No. 795958. Replication files can be found via OSF: \url{https://osf.io/7sm2f/?view_only=8466d18f89d1492b81baf7ec84ef12ea}}}
\author{Thomas Meissner\footnote{Maastricht University} \\ David Albrecht\footnotemark[2] }
\maketitle

\bigskip
\bigskip

\renewcommand{\baselinestretch}{1} \normalsize

\begin{abstract}
\noindent 
 
Debt aversion can have severe adverse effects on financial decision-making. We propose a model of debt aversion, and design an experiment involving real debt and saving contracts, to elicit and jointly estimate debt aversion with preferences over time, risk and losses. Structural estimations reveal that the vast majority of participants (89\%) are debt averse, and that this has a strong impact on choice. We estimate the ``borrowing premium'' -- the compensation a debt averse person would require to accept getting into debt -- to be around 16\% of the principal for our average participant.

\end{abstract}
\vspace*{5mm}


\renewcommand{\baselinestretch}{1.45} \normalsize
\setlength{\footnotesep}{0.75\baselineskip}

\newpage
\section{Introduction}

\rightline{\textit{Rather go to bed without dinner than rise in debt.}}
\rightline{--- Benjamin Franklin}
\hspace{12pt}

Borrowing and saving decisions are among the most important and economically significant choices people face in their lives. An unwillingness to save may have severe implications such as insufficient retirement savings. In the same way, borrowing too much or too little can have negative economic consequences. Debt aversion, defined as an intrinsic unwillingness to take on debt, has received increased attention by researchers lately, for its adverse effects on financial decision-making such as failure to invest in tertiary education \citep{field2009,caetano2019} and energy-efficient technologies \citep{schleich2021} or credit self-rationing of entrepreneurs \citep{nguyen2020}.

However, in absence of a theory of debt aversion, it is difficult to measure debt aversion. To the best of our knowledge, no satisfactory theory or measurement exist so far. The difficulty to measure debt aversion stems from the fact that many other preferences and constraints may influence borrowing (and also saving) behavior. To illustrate this point, consider a prospective student who deliberates about taking on a student loan to finance their studies. Clearly, time preferences, such as discounting and the elasticity of intertemporal substitution will influence this decision.\footnote{In the standard discounted utility model, the elasticity of intertemporal substitution is determined by the parameter of risk aversion. For convenience, we will make use of that assumption too, but test it's implications.} Moreover, if the repayment of the loan is counted as a future loss, loss aversion may affect this choice as well. To understand whether a person is truly debt averse, these other preferences need to be taken into consideration. Therefore, to cleanly identify debt aversion, a model is necessary that allows to disentangle and identify debt aversion separately from these other preferences.  

The goal of this paper is to understand whether debt aversion is a preference in it's own right, or whether it is merely an emergent behavioral property of other preferences, biases, beliefs and constraints. To this end, we propose a formal model of debt aversion, and design and conduct an experiment to elicit and to jointly estimate debt aversion with preferences over time, risk and losses. Debt aversion is difficult to identify with field data, because many factors that influence borrowing and saving decisions are typically unobservable. Coming back to the student who considers accepting a student loan, this decision may be influenced by their belief about their individual future return of college education, their access to credit, peer effects, and many other potential factors that are not debt aversion. Lab experiments are an excellent tool in this case, as they allow to control for confounding factors, such as beliefs about potential returns or access to credit.

In the experiment participants can accept or reject a series of different debt and saving contracts involving real money, that is, participants can actually save and borrow with the experimenter. Using participants' choices, we can identify whether they systematically prefer saving contracts over debt contracts, controlling for time preferences, risk aversion and loss aversion. To this end, we employ maximum likelihood estimations to structurally estimate the preference parameters of our model of debt aversion.

We find that participants are on average debt averse, thus establishing debt aversion as a dimension of individual preference in its own right, that is distinct from other relevant preferences. Comparing the choice of our average participant to a counterfactual debt neutral participant, reveals that debt aversion has a quantitatively meaningful impact on choice: Our participants require a ``borrowing premium'' of around 16\% of the principal in order to accept getting into debt. Further, testing the relation of debt aversion and individual characteristics, we find a weak negative association between debt aversion and cognitive ability: People who score higher on our tests of cognitive ability have lower levels of debt aversion. Other individual characteristics, such as age, gender, financial literacy and personality appear unrelated to debt aversion. To test for potential interdependence of the different preference domains, we estimate the joint distribution of our preference parameters using simulated maximum likelihood estimations. We find that that debt aversion is positively correlated with loss aversion, but not related to risk or time preferences. Further, according to our estimated distribution of the debt aversion parameter, we find that around 89\% of individuals exhibit debt aversion. In an extension of the main experiment, using a subset of participants that received additional saving and borrowing choices, we try to explore potential mechanisms behind debt aversion. Results indicate that debt aversion increases in the time that people spend indebted. Finally, we demonstrate robustness of debt aversion to a wide array of alternative modeling specifications.

In the following, we first provide an overview of the related literature in Section~\ref{sec:lit}. In Section~\ref{sec:theory} we introduce the theoretical framework for modeling debt aversion and in Section~\ref{sec:experiment} we describe the experiment. The results are reported in Section~\ref{sec:results}, and Section~\ref{sec:conclusion} provides a discussion of the findings and concluding remarks.

\subsection{Related Literature}
\label{sec:lit}

Our study connects to a growing literature on debt aversion. Existing theoretical work on debt aversion has produced models of intertemporal choice that feature debt aversion as an emergent behavioral property of other preferences. \cite{loewenstein1992} present a model of intertemporal choice incorporating variable utility curvature as well as discounting for positive and negative money streams. Decision makers in their model require much more favorable rates to borrow than to save. \cite{prelec1998} introduce a framework that differentiates mental accounts for consumption and associated (loan) payments. The model allows utility of consumption and disutility of payments to vary depending on the relative timing of consumption and payments. This so called prospective accounting predicts aversion to debt where debt might either be seen as consuming before paying, or receiving payment for future, yet undone work. Both frameworks explain debt averse behavior through variations in utility curvature, time discounting and loss aversion. Advancing on the existing theoretical work, we aim to model debt aversion as a preference in it's own right, that cannot be explained by preferences  over time, risk and losses. To this end, we model debt aversion explicitly, while also accounting for other relevant preferences.

A large part of the empirical work on debt aversion focuses on its influence on investment in higher education, with somewhat mixed results. In field experiments offering differently labeled loan contracts to students, \cite{field2009} and \cite{caetano2019} find that debt aversion might indeed deter investment in education and influence career choices. Using a representative survey on UK final year high-school students, \citep{callender2005} find that more debt averse individuals, who often-times also have low socioeconomic status, are far less likely to actually go to university. Results on the existence of debt aversion among (prospective) students have later been supported by large-scale surveys for the US \citep{boatman2017} and the Netherlands \citep{oosterbeek2009}. Furthermore, \cite{gopalan2021} find that a positive income shocks lead students to substantially decrease their debt, while non-students do not change their borrowing behavior. In contrast, \cite{eckel2016} find little evidence that debt-aversion poses a barrier to investing in higher studies among a sample of Canadian adults. Besides investment in education, debt aversion has been associated with investment decisions by small and medium size business owners \citep{nguyen2020}, with low uptake of debt-based public support programs related to COVID-19 \citep{paaso2020} and with hesitancy to invest in retrofit measures to increase energy efficiency of private buildings \citep{schleich2021}. \cite{helka2021} find that openness to being indebted is a far more important predictor of borrowing for hedonistic purposes than of borrowing for investments and necessities. Lastly, \cite{ikeda2015} find more debt averse people in a sample of Japanese adults to engage less in activities they classify as overborrowing, such as taking unsecured consumer loans, engaging in debt restructuring or declaring personal bankruptcy. Note that most of these studies either use measures of debt aversion that could potentially be confounded by other preferences, and/or qualitative measures that ask participants for their stated debt aversion. Such qualitative measures are convenient to include in studies where time is critical, but it is not clear whether they actually measure debt aversion, as no validated survey module exists.\footnote{In a companion paper \citep{albrecht2022the}, we develop such a module, based on choices in this experiment.} Also, in most field and survey settings, it is difficult to identify whether taking on debt would actually be optimal or not - making it hard to identify biases in borrowing behavior and thus debt aversion.

An advantage of lab experiments is that optimal saving and borrowing can be controlled by the experimenter, which allows to identify debt aversion. \cite{Meissner2016} conducts an experiment in which participants play consumers in a life-cycle consumption problem. He creates two treatments, in which participants have to either save or borrow within an experimental session in order to consume optimally. He finds that people are on average less willing to borrow than to save in order to smooth consumption over the experimental life-cycle. \cite{ahrens2022intertemporal} replicate \cite{Meissner2016} using a student sample from the US and find similar levels of debt aversion. \cite{duffy2020} attribute sub-optimal borrowing on the extensive and intensive margin in an intertemporal consumption and saving experiment to debt aversion.\footnote{See \cite{Duffy2016} for an overview on dynamic intertemporal consumption and saving experiments.} Focusing on debt repayment, rather than borrowing, \cite{martinezmarquina2021} report that participants forgo profitable investments and substantial monetary gains in order to repay debt as soon as possible. Relatedly, \cite{besharat2015} and \cite{amar2019} find people to exhibit debt account aversion, i.e. when holding debt on multiple accounts, people tend to repay the account with the lowest outstanding debt first, to reduce the overall amount of debt accounts, despite forgoing monetary gains. These studies have in common, that debt and indebtedness is either entirely hypothetical or restricted to an experimental account. The latter arises in the context of consumption/saving experiments where monetary gains and losses, such as debt repayments, accumulate over the course of the experiment but real payments are never effected until participants receive their final payment at the end of the  overall session. In contrast, we implement the first experiment, that actually encompasses real indebtedness: Participants may first receive a loan payout, take the money out of the lab, be indebted with the experimenter over a period of multiple weeks, and face the obligation to repay their debt afterwards. Moreover, existing approaches only identify debt aversion on the aggregate level. Our study is the first attempt to identify and estimate a parameter of debt aversion on the individual level.\footnote{ \cite{ahrens2022intertemporal} introduce an individual index of debt aversion. However, this index does not measure debt aversion itself, as is constructed based on deviations from optimal consumption.}

Summing up, we are first to propose a theory of debt aversion, in which debt aversion is a preference in its own right rather than an emergent behavioral property of other preferences. We are also first to implement actual indebtedness in a laboratory experiment, which improves external validity compared to other experimental approaches in which indebtedness is only implemented hypothetically or within one experimental session. Finally, we are first to identify and to structurally estimate debt aversion on the individual level.

\section{A Theory of Debt Aversion}
\label{sec:theory}

We consider agents who choose between intertemporal prospects that are defined over streams of monetary gains or losses in up to two periods.\footnote{The model can be extended to $n$ periods. However, this makes the model considerably less tractable. As our experiment only involves trade-offs between payments in up to two periods, we favor the two-period approach.} $\boldsymbol{x}=(x_t,x_T)$ denotes a stream of payments that offers $x_t$ at time $t$, and $x_T$ at time $T$, where $0\leq t < T$. $X=(\boldsymbol{x}_1,p_1;\boldsymbol{x}_2,p_2;...;\boldsymbol{x}_N,p_N)$ denotes an intertemporal prospect, that gives the payment stream $\boldsymbol{x}_n$ with probability $p_n$. Intertemporal utility is written as:

\begin{equation*}
\label{eq:U}
    U(X)=\mathbb{E}\left[\phi(t)v(x_t)+\phi \left(T\right)v(x_T) - \mathbbm{1}_{debt} c\left(\boldsymbol{x}\right)\right]
\end{equation*}

where $v(x_t)$ denotes atemporal utility of monetary gains and losses at time $t$. Agents discount future gains and losses with the discount function $\phi$. 

\emph{Saving contracts} are payment streams characterized by $x_t<0$ and $x_T>0$.  Inversely, \emph{debt contracts} are payment streams characterized by $x_t>0$ and $x_T<0$. We allow agents to evaluate debt contracts differently than other contracts. To this end we introduce debt aversion as a cost of being in debt $c(\boldsymbol{x})$, which is only incurred for debt contracts:

\begin{equation*}
    \mathbbm{1}_{debt}=\begin{cases}
    1 & \mbox{if } x_t > 0 \mbox{ and } x_T <0\\
    0 & \mbox{ otherwise.}
    \end{cases}
\end{equation*}

To estimate preferences structurally, we require specific forms of the functions defined above. Below is a summary of our preferred specification. In section~\ref{sec:robustness} we conduct a series of robustness checks to test whether our finding of debt aversion is robust to different specifications of these functional forms. Debt aversion remains robust.

Following Prospect Theory \citep{RN384}, we allow gains and losses of money to be evaluated differently, relative to a reference point $(x=0)$:

\begin{align}
v(x)=
\begin{cases}
u(x) & \mbox{if } x\geq 0\\
-\lambda u(-x) & \mbox{if } x<0,\\
\end{cases}
\end{align}

We assume normalized power utility to model curvature of utility in gains and losses:\footnote{Note that we assume the same curvature of utility in the gain and loss domain. While this is not uncommon in the literature, allowing curvature to vary would be a more flexible approach. In our experimental setting this assumption is benign, as all potential losses can only take one value: \euro{-15}. In Appendix \ref{sec:robust_appendix} we show that debt aversion is robust to this assumption.} 

\begin{align}
    u(x)=\frac{(x+\varepsilon)^{1-\alpha}-\varepsilon^{1-\alpha}}{1-\alpha}
\end{align}

For $\varepsilon=0$, this function is characterized by constant relative risk aversion (CRRA). However, for values of $\alpha$ above unity and $\varepsilon=0$, the derivatives of the utility function diverge around the reference point. We set $\varepsilon=0.0001$, to maintain a close approximation of CRRA utility, while ensuring that preferences are well behaved around the reference point.\footnote{See \citet{Wakker2008} for an illustration, and \cite{Meissner2022} for a recent application.} Agents are assumed to discount the future exponentially:\footnote{The experiment allows to identify other forms of discounting, such as quasi-hyperbolic discounting. In Section \ref{sec:extension_hdisc}, we test a quasi-hyperbolic discount function \citep{RN386}, but find no empirical support for present bias.}
\begin{align}
    \phi(\tau) = \frac{1}{(1+\delta)^{\tau}}   
\end{align}

Finally, the utility cost of borrowing is modelled as cost incurred at the time of debt repayment, depending on the amount owed $x_T$:
\begin{align}
   c(\boldsymbol{x})=(1-\gamma)\phi(T)v(x_T)
\end{align}

Here, $\gamma$ is the parameter of debt aversion. A parameter of $\gamma=1$ implies debt neutrality, $\gamma>1$ implies debt aversion and $\gamma<1$ implies debt affinity. We chose this specification mostly for tractability: Using this functional form ensures that the debt aversion parameter scales the disutility associated with the loss of having to repay the owed amount in a similar way as the parameter of loss aversion in a standard prospect theory model. To illustrate this point, note that the intertemporal utility of a deterministic saving contract simplifies to: 
\begin{align}
    U(X)=-\lambda\phi(t)u(-x_t)+\phi(T)u(x_T),
\end{align}
while the utility of a deterministic debt contract collapses to:
\begin{align}
    U(X)=\phi(t)u(x_t)-\lambda\gamma\phi(T)u(-x_T)
\end{align}
Further, note that the cost of being in debt is assumed to be discounted with $\phi(T)$. As a consequence, the effect of debt aversion lessens the further a repayment is shifted to the future. As an extension we also consider an adjusted specification of debt aversion in Section \ref{sec:extension}, where the cost of being in debt is allowed to depend on the time an agent spends in debt, $T-t$.

\section{Experiment}
\label{sec:experiment}

We utilized multiple price lists (MPLs) to elicit preferences over saving and borrowing, as well as time discounting, risk aversion and loss aversion. As depicted in Figure~\ref{fig:timeline} we introduce a real time dimension, by requiring participants to come to the laboratory on a total of three dates, equally spaced four weeks apart. This ensured that we could offer and enforce real saving and debt contracts. All payment relevant choices were made in the first session, lasting around 90 minutes. Four and eight weeks after the first session, Session~2 and~3 took place. In all three sessions participants were asked to answer questionnaires, and all payments based on participants' choices were paid to the participants or collected from them.

The main experiment consisted of a total of 90 binary choices, defined over payments at different points in time, lotteries, as well as saving and debt contracts. Although all choices were presented sequentially, we  will  refer  to the underlying MPLs for expository purposes throughout the paper. The 90 choices can be grouped into seven MPLs (see Table~\ref{table:pl_overview}).\footnote{ Appendix~\ref{sec:mpls} contains details on all 90 choices.} We adapted this common elicitation method for risk and time preferences, going back to \cite{coller1999, harrison2002} and \cite{holt2002}, to also elicit preferences over losses, as well as debt and savings. 
While the order in which participants completed the MPLs was fixed, within each MPL the order of choices was randomized. 

\usetikzlibrary{arrows,decorations.pathmorphing,backgrounds,fit,positioning,shapes.symbols,chains}
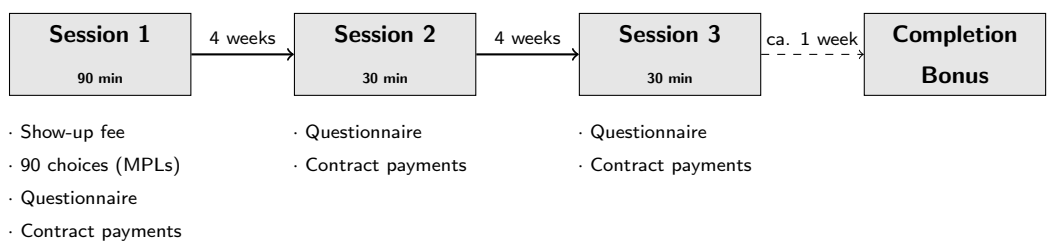
\begin{figure}[t!]
\caption{Timeline of the experiment}
\label{fig:timeline}

\centering
\begin{tikzpicture}
[node distance = 5.5cm, auto,font=footnotesize,
every node/.style={node distance=4cm, font=\scriptsize\sffamily},
comment/.style={rectangle, inner sep= 5pt, text width=2.7cm, node distance=0.25cm, font=\scriptsize\sffamily},
force/.style={rectangle, draw, fill=black!10, inner sep=5pt, text width=2.3cm, text badly centered, minimum height=1.2cm, font=\bfseries\footnotesize\sffamily}, scale=.9,transform shape] 


\node [force] (date2) {Session 2\\ {\tiny 30 min}};
\node [force, left=1.5cm of date2] (date1) {Session 1\\ {\tiny 90 min}};
\node [force, right=1.5cm of date2] (date3) {Session 3\\ {\tiny 30 min}};
\node [force, right=1.5cm of date3] (payment) {Completion\\ Bonus};


\node [comment, below=0.25 of date2] (comment-date2) {
$\cdot$ Questionnaire\\
$\cdot$ Contract payments\\
};

\node [comment, below=0.25cm of date1] {
$\cdot$ Show-up fee\\
$\cdot$ 90 choices (MPLs)\\
$\cdot$ Questionnaire\\
$\cdot$ Contract payments};

\node [comment, below=0.25 of date3] {
$\cdot$ Questionnaire\\
$\cdot$ Contract payments\\
};



\path[->,thick] 
(date1) edge node {4 weeks} (date2)
(date2) edge node {4 weeks} (date3);

\path[->,dashed] 
(date3) edge node {ca. 1 week} (payment);

\end{tikzpicture} 

\label{timeline}
\end{figure}


\begin{table}[ht]
\caption{Stylized overview of choices \label{table:pl_overview}}
\centering
\begin{tabularx}{\textwidth}{ 
    c
    @{\hspace{24pt}}
    c
    @{\hspace{24pt}}
  >{\centering\arraybackslash}X 
  >{\centering\arraybackslash}X 
  >{\centering\arraybackslash}X }
\toprule

    \textbf{MPL} 
    & \textbf{Choice} 
    & \multicolumn{3}{c}{\textbf{Payment}}\\
    \cline{3-5}

    &
    &\small{Session 1} 
    & \small{Session 2} 
    & \small{Session 3} \\
    
\midrule

    \small{1} 
    & \makecell{\footnotesize{receive money} \\[-0.3cm] \footnotesize{in Session 1 or 2}} 
    & \makecell{\footnotesize{varying per choice} \\[-0.3cm] \footnotesize{\euro{8-18.2}}} 
    & \footnotesize{\euro{18}} 
    & \footnotesize{--}\\
    
\hline

    \small{2} 
    & \makecell{\footnotesize{safe amount} \\[-0.3cm] \footnotesize{or lottery} \\[-0.3cm] \footnotesize{in Session 1}} 
    & \makecell{\footnotesize{varying safe amount} \\[-0.3cm] \footnotesize{(\euro{1-30}) or }\\[-0.3cm] \footnotesize{coin flip (\euro{1 or 30})}}  
    & \footnotesize{--} 
    & \footnotesize{--}\\

    \small{3} 
    & \makecell{\footnotesize{safer lottery or} \\[-0.3cm] \footnotesize{or riskier lottery} \\[-0.3cm] \footnotesize{in Session 1}} 
    & \makecell{\footnotesize{coin flip (\euro{14 or 17})} \\[-0.3cm] \footnotesize{or riskier coin}\\[-0.3cm] \footnotesize{flips (EV: \euro{9}-23)}}
    & \footnotesize{--}  
    & \footnotesize{--}\\


\hline

    \small{4} 
    & \makecell{\footnotesize{(not) accept} \\[-0.3cm] \footnotesize{savings contract}} 
    & \footnotesize{pay \euro{15}} 
    & \makecell{\footnotesize{receive} \\[-0.3cm] \footnotesize{\euro{12-45}}} 
    & \footnotesize{--}\\

    \small{5} 
    & \makecell{\footnotesize{(not) accept} \\[-0.3cm] \footnotesize{savings contract}} 
    & \footnotesize{--} 
    & \footnotesize{pay \euro{15}} 
    & \makecell{\footnotesize{receive} \\[-0.3cm] \footnotesize{\euro{7-40}}}\\

\hline

    \small{6} 
    & \makecell{\footnotesize{(not) accept} \\[-0.3cm] \footnotesize{debt contract}} 
    & \makecell{\footnotesize{receive} \\[-0.3cm] \footnotesize{\euro{3-31}}} 
    & \footnotesize{pay \euro{15}}  
    & \footnotesize{--}\\

    \small{7} 
    & \makecell{\footnotesize{(not) accept} \\[-0.3cm] \footnotesize{debt contract}} 
    & \footnotesize{--} 
    & \makecell{\footnotesize{receive} \\[-0.3cm] \footnotesize{\euro{3-33}}} 
    & \footnotesize{pay \euro{15}}\\

\bottomrule

\end{tabularx}

\end{table}

The first three MPLs elicit risk and time preferences. MPL1 offers the choice between a varying amount at Session~1 and fixed amount at Session~2. MPL2 and MPL3 offer the choice between safer and riskier lotteries. 

In order to identify preferences over saving and borrowing, we introduce MPL4 - MPL7. These MPLs consist of real saving and debt contracts. MPL4 consists of a list of saving contracts, involving the payment of \euro{15} by the participant to the experimenter in Session~1, followed by a varying repayment from the experimenter to the participant at Session~2. MPL5 consists of similar saving contracts, shifted by four weeks. MPL6 consists of real debt contracts offering a payment of \euro{15} to the participant at Session~1, followed by a varying repayment by the participant to the experimenter at Session~2. As with the saving contracts, MPL7 contains similar debt contracts as MPL6, but all payments are shifted into the future by four weeks. Note that in all choices involving debt and saving contracts, participants could either accept or reject each contract. Rejection implied that no further payments take place. 
The questionnaires asked participants about a number of individual characteristics, such as age, gender, cognitive ability and financial literacy (see Appendix~\ref{sec:indiv_char_app}). We also elicited preferences using non-incentivized survey items, and asked participants about their actual saving and borrowing behavior outside of the experiment. These data are used in \cite{albrecht2022the} to identify a survey module that best predicts debt aversion as measured with this experiment.

\subsection{Procedures}

All participants were required to come to all three sessions, irrespective of their financial choices. The scheduling of sessions took into account that participants knew potential conflicts with their university schedule at the time of sign-up. Moreover, the importance of attending all sessions was emphasized in the invitation mails as well as in person before the start of the experiment. It was announced that participants who fail to participate in all sessions for reasons other than force majeure will be exempted from payment of the completion bonus and will be counted as no-shows for the entire experiment, which leads to a removal from the experimental participants pool. Informed consent with the rules and procedures of the experiment was indicated via electronic acceptance of the invitation and confirmed verbally at the lab facilities.

Data collection took place at the Behavioral and Experimental Economics Laboratory at Maastricht University during winter 2019/20, autumn 2020 and autumn 2021. A total of 148 participants (62 in winter 2019/20, 53 in autumn 2020 and 33 in autumn 2021) 
attended the opening session and hence made all choices relevant for the estimation of preferences. Over the course of the experiment attrition amounted to 21, such that 127  participants (54 in winter 2019/20, 46 in autumn 2020 and 27 in autumn 2021) 
completed all three sessions.\footnote{Compared to other experiments conducted with the same participants pool these numbers seem normal, if not below average.} All participants were recruited through ORSEE \citep{greiner2015}. Around 74\% followed an undergrad program and  25\%  pursued a masters degree. 
Their backgrounds ranged from music to law, with a clear mode in the field of business and economics. 
The experiment was programmed and conducted using z-Tree \citep{fischbacher2007}.\footnote{The complete instructions can be found in Appendix~\ref{sec:inst}.}

\subsection{Payment}
At the end of the opening session and after completion of all payment relevant choices, one decision was randomly drawn as the `decision that counts'. The random draw was conducted with help of a bingo cage containing 90 numbered balls. Decision-based payouts for the whole experiment were determined by this decision. If the decision involved payments not only in in the opening session, a physical, individualized contract  delineating all payments to be made and received was drawn up and signed by the experimenter as well as the participant (see Appendix~\ref{sec:contract}). All payments due on the opening session were directly executed. Later payments were executed at the end of the respective session. Payments to the participants were always effected in cash. Participants were allowed to make payments to the experimenters in cash or via PayPal, to minimize the potential transaction burden of payment.  All participants with due payments in later sessions received respective email-reminders the day before due date.\footnote{To minimize risk of confounding preference elicitation, the overall setup was directed at increasing perceived payment reliability, i.e. trust that promised payments by the experimenters will be made, and confidence that promised payments by the participants actually have to be made in the future. Undertaken measures included issuing of physical contracts with contact details of the principal experimenter as well as the emphasize in the experimental instructions that the principal experimenter guarantees for all payments and multiple ways for contacting any of the experimenters and the associated economics department at Maastricht University were given in case any issues regarding payment should arise.}

In addition to the decision-based payments, all participants received a show-up fee of \euro{15} for all three sessions at the beginning of the opening session. This money was handed to participants in cash before any decisions took place. As payments in MPL4 contained saving contracts that required participants to pay \euro{15} to the experimenter in the first session, we allowed participants to pay this out of the show-up fee.

After completion of all three sessions and settlement of all due payments participants received a completion bonus of \euro{20}. This payment was implemented via bank transfer and with a delay of one week, to prevent participants to settle any outstanding debt with the completion bonus. Had we paid the completion bonus in cash during the last session, participants may not have thought that they are really in debt, which could have impeded the identification of debt aversion.\footnote{Some experimental sessions in spring 2020 were affected by the closure of Dutch universities including the BEElab facilities at Maastricht University due to the COVID-19 pandemic. For 44  participants it was not possible to conduct Session~3 in the lab as planned. We transformed the respective experimental protocol to an online survey keeping all content identical and visual appearance as similar as possible. As cash payments were no longer possible, all payments were made via bank transfers. This only affected the collection of non-incentivized questionnaires in Session~3; all payment relevant choices had been made in January 2020 before the COVID-19 pandemic hit Europe and the Netherlands. The same option to conduct Session~3 online was also offered to participants in autumn 2020 and 2021 who needed to quarantine.}

\section{Results}
\label{sec:results}

In this section, we will estimate the preference parameters of our debt aversion model. We will start in subsection \ref{sec:main} with our main specification, which estimates the model as specified in \ref{sec:theory}. We will then consider extensions and employ various robustness checks in subsection \ref{sec:robustness}.  
 
\subsection{Main specification}
\label{sec:main}

We will start out by estimating preference parameters on the aggregate, that is for our average participant. We will then account for observed heterogeneity, by allowing preference parameters to vary with oberserved individual characteristics. Finally we will account for unobserved individual heterogeneity by estimating the joint population distribution of all preference parameters. 

\subsubsection{Aggregate Structural Estimation}
\label{sec:agg_estimation}

We jointly estimate preference parameters for risk aversion, time discounting, loss aversion and debt aversion, according to the model specified in Section~\ref{sec:theory}, and broadly following the estimation strategies described in e.g. \cite{andersen2008,harrison2008} and \cite{abdeallaoui2019}.


As basis for all estimations we consider a random utility model incorporating errors in the decision making process. Decision makers may make errors when evaluating the expected utility of different options captured by noise parameter $\mu$. In particular, choices between option A and B are evaluated at their expected intertemporal utility, as specified in Equation~\ref{eq:U} plus a stochastic error term $\varepsilon$. A decision maker with preference parameters $\omega=(\alpha,\beta,\gamma,\lambda)$ chooses option B if $U(X^B,\omega)+\varepsilon^{B}\geq U(X^A,\omega) +\varepsilon^{A}$. The probability of observing choice B can then be written as:

\begin{align}
\label{eq:cond_prob}
P^B(\theta)& = F\left(\frac{U(X^B,\omega)-U(X^A,\omega)}{\mu}\right) = F(\Delta U(\theta)),
\end{align}

where $F$ is the cumulative distribution function of $(\varepsilon^A-\varepsilon^B)$ and $\theta=(\alpha,\delta,\gamma,\lambda,\mu)$ denotes the vector of preference parameters and the error parameter. We assume $(\varepsilon^A-\varepsilon^B)$ to follow a standard logistic distribution with  $F(\xi)=(1+\mathrm{e}^{-\xi})^{-1}$ in our main specification. This specification is often termed Luce model \citep{luce1965preference, holt2002} or Fechner error with logit link \citep{drichoutis2014}. 
Overall, we estimate four preference parameters and one error parameter: risk aversion $\alpha$, time discounting $\delta$, debt aversion $\gamma$, loss aversion $\lambda$ and the Fechner error $\mu$, respectively. Intuitively, the error parameter can be interpreted as follows: for $\mu \rightarrow 0$ choice becomes deterministic, and for $\mu \rightarrow \infty$, choice approaches uniform randomization.
Aggregating over all choices and individuals the log-likelihood function writes as:

\begin{align}
\label{eq:logL1}
ln \left( L(\theta) \right) = \sum_i\sum_j\left[ln\left(F(\Delta U(\theta))\right)c_{ij}+ln(1-F(\Delta U(\theta)))(1-c_{ij})\right],
\end{align}

where $c_{ij}=0$ if individual~$i$ chooses~A in choice~$j$ and $c_{ij}=1$ if individual~$i$ chooses~B in choice~$j$. 

By maximizing the log-likelihood function over $\theta$ we derive point estimates for all preference parameters and the error parameter. These estimates describe preferences of the average decision maker. 
To account for dependency of choices made by the same person we cluster standard errors at the individual level. Estimates are calculated using STATA's modified Newton-Raphson algorithm.

\begin{filecontents*}{RESULTS/agg_main.csv}
"","result","alpha","delta","gamma","lambda","mu"
"1","parameter",0.643,0.036,1.0535,1.1074,0.4484
"2","robust se",0.0345,0.006,0.0112,0.0118,0.0402
"3","p-value",18.6477646579102,6.00372726057775,93.9827413288365,93.7367030251837,11.1470064535478
"4","z",0.0000000000000000000000000000000000000000000000000000000000000000000000000000131668317229303,0.00000000192838529577414,0,0,0.0000000000000000000000000000740560277524288
"5","lowerlim",0.58,0.02,1.03,1.08,0.37
"6","upperlim",0.71,0.05,1.08,1.13,0.53
"7",NA,NA,NA,NA,NA,NA
"8",NA,1.95996398454005,1.95996398454005,1.95996398454005,1.95996398454005,1.95996398454005
"9",NA,0,0,0,0,0
"10","log-likelihood",-4107.91,NA,NA,NA,NA
"11","n",12240,NA,NA,NA,NA
"12","AIC",8225.81,NA,NA,NA,NA
"13","BIC",8262.88,NA,NA,NA,NA
"14","cluster",127,NA,NA,NA,NA
\end{filecontents*}
\DTLloaddb{agg_main}{RESULTS/agg_main.csv}

\begin{table}[t!]
\caption{Aggregate parameter estimates}
\begin{tabularx}{\textwidth}{ 
    c
    @{\hspace{36pt}}
  >{\centering\arraybackslash}X 
  >{\centering\arraybackslash}X 
  >{\centering\arraybackslash}X
  >{\centering\arraybackslash}X
  >{\centering\arraybackslash}X
  }
\toprule

& $\boldsymbol{\alpha}$ 
& $\boldsymbol{\delta}$ 
&  $\boldsymbol{\gamma}$ 
& $\boldsymbol{\lambda}$ 
& $\boldsymbol{\mu}$\\[-6pt]

& \scriptsize{risk aversion} 
& \scriptsize{discounting} 
& \scriptsize{debt aversion} 
& \scriptsize{loss aversion} 
& \scriptsize{Fechner error} \\

\midrule
\scriptsize{\textbf{point estimate}} 
& \DTLfetch{agg_main}{result}{parameter}{alpha}
& \DTLfetch{agg_main}{result}{parameter}{delta}
& \DTLfetch{agg_main}{result}{parameter}{gamma}
& \DTLfetch{agg_main}{result}{parameter}{lambda}
& \DTLfetch{agg_main}{result}{parameter}{mu}\\[-6pt]

\scriptsize{95\% confidence interval} 
& \scriptsize{\DTLfetch{agg_main}{result}{lowerlim}{alpha}\,/\,\DTLfetch{agg_main}{result}{upperlim}{alpha}}
& \scriptsize{\DTLfetch{agg_main}{result}{lowerlim}{delta}\,/\,\DTLfetch{agg_main}{result}{upperlim}{delta}}
& \scriptsize{\DTLfetch{agg_main}{result}{lowerlim}{gamma}\,/\,\DTLfetch{agg_main}{result}{upperlim}{gamma}}
& \scriptsize{\DTLfetch{agg_main}{result}{lowerlim}{lambda}\,/\,\DTLfetch{agg_main}{result}{upperlim}{lambda}}
&\scriptsize{\DTLfetch{agg_main}{result}{lowerlim}{mu}\,/\,\DTLfetch{agg_main}{result}{upperlim}{mu}}\\[-9pt]

\scriptsize{robust standard error} 
& \scriptsize{\DTLfetch{agg_main}{result}{robust se}{alpha}}
& \scriptsize{\DTLfetch{agg_main}{result}{robust se}{delta}}
& \scriptsize{\DTLfetch{agg_main}{result}{robust se}{gamma}}
& \scriptsize{\DTLfetch{agg_main}{result}{robust se}{lambda}}
& \scriptsize{\DTLfetch{agg_main}{result}{robust se}{mu}}\\
\midrule
 
\multicolumn{6}{l}{\scriptsize{estimation details: n = \DTLfetch{agg_main}{result}{n}{alpha},  log-likelihood = \DTLfetch{agg_main}{result}{log-likelihood}{alpha},
AIC = \DTLfetch{agg_main}{result}{AIC}{alpha}, 
BIC = \DTLfetch{agg_main}{result}{BIC}{alpha}, logit Fechner error}}\\

\bottomrule
\multicolumn{6}{l}{\scriptsize{Robust standard errors (SE) clustered at the individual level, \DTLfetch{agg_main}{result}{cluster}{alpha} clusters}}
\end{tabularx}
\label{table:agg_main}
\end{table}



Estimation results are presented in Table~\ref{table:agg_main}. Most importantly for this study, the estimate of the parameter indicating debt aversion $\gamma=1.0535$ is significantly larger than one, suggesting that participants are on average debt averse. To put this estimate in perspective, a decision-maker with the preference parameters as in Table~\ref{table:agg_main} would be indifferent between accepting or rejecting a debt contract that involves a loan amount of \euro{20.93} today, with an associated repayment of \euro{15} in four weeks. That is, our average participant would require a negative interest rate to accept a debt contract. This in itself, however, is not yet evidence of debt aversion, as other preferences such as loss aversion could potentially partly explain this. To understand the impact of debt aversion, we calculate what a counterfactual debt-neutral decision-maker with the same preferences, except $\gamma=1$, would do. Such a decision-maker would accept a loan already as soon as it pays at least \euro{18.08} today, everything else equal. Our average debt averse participants thus requires \euro{2.85} more upfront, in order to be indifferent between accepting or rejecting a debt contract compared to the counterfactual debt-neutral decision-maker. We define the ``borrowing premium'' as the relative increase in the upfront payment (i.e. the principal) a debt averse person would require compared to a debt neutral person in order to accept a debt contract. For the average participant this would be $2.85/18.08=15.76\%$. In other words, the average, debt averse decision maker requires a borrowing premium of 15.76\%  larger loan sizes, while keeping repayment constant, to be willing to accept a debt contract compared to their debt-neutral counterpart. 

Regarding preferences other than debt aversion, participants are on average risk averse with a parameter of relative risk aversion $\alpha=0.643$. This estimate is in range of previous studies with large scale samples using a similar utility specifications. \cite{andersen2008} find a parameter of relative risk aversion in the adult Danish population of $0.741$, while \cite{ Meissner2022} report $0.456$ based on a representative sample from eight European countries. The four-week discount rate $\delta$ is estimated at around $0.036$ which is in the range of other lab studies on time preferences as summarized by recent meta-study results \citep{matousek2022}. Further, participants are on average loss averse, with $\lambda=1.1074$. This is lower than estimates typically observed in the literature, where $\lambda$ is usually found to be around 2 \citep{brown2021}. However, in most studies loss aversion is elicited with risky prospects, i.e. gains and losses are separated by state at one point in time. \cite{abdellaoui2013} show that when gains and losses are separated by time and do not involve risk, such as in our savings and debt contracts, the loss aversion parameter is substantially lower with an estimate at around $\lambda=1.15$.

Note that in our main specification we are assuming that the elasticity of intertemporal substitution is the reciprocal of the parameter of risk aversion, and thus, that the elasticity of intertemporal substitution can be identified with the help of a-temporal lottery choices. While this is a common assumption in the literature, there is considerable debate whether it is warranted \citep{andreoni2012risk}. \cite{andersen2008} were first to use utility curvature elicited with lotteries to correct the estimates of discount rates.
Since then, most studies find that while utility over time is also concave, it exhibits less curvature than utility over risk \citep{abdellaoui2013,cheung2020eliciting}. Further, while utility curvature over time appears to be different from utility curvature over risk, the two appear to be correlated \citep{meissner2022measuring}. For our study, correcting for utility curvature appears to be the most conservative approach with respect to the estimate of the debt aversion parameter. In Appendix \ref{sec:robust_appendix} and Table \ref{table:rob_riskneutral} we relax this assumption, and show that not accounting for utility curvature leads to considerably larger estimates of the debt aversion parameter.

\subsubsection{Debt Aversion and Observable Individual Characteristics}

In this section, we expand the previous aggregate estimation, by allowing preference and error parameters to vary with observable individual characteristics. Preference and error parameters are estimated as linear functions of individual characteristics including age, gender, cognitive ability, financial literacy and personality traits as described in Table~\ref{tab:covariate_description}. The estimation results are presented in Table~\ref{table:covariates}.

\begin{table}[t!]
    \caption{Description of variables of observable characteristics}
    \centering
    \begin{tabularx}{\textwidth}{
        C{3.5cm}    
        >{\centering\arraybackslash}X  
        >{\setlength{\baselineskip}{-1.5\baselineskip}}J{11.15cm}
        }
        \toprule
        \textbf{\footnotesize{Variable label}} && \textbf{\footnotesize{Variable description}}\\
        \midrule
        
        \footnotesize{Age} && \footnotesize{Participant age in years} \\[6pt]
        
        \footnotesize{Cognitive ability} && \footnotesize{Number of correct answers in cognitive reflection, numeracy and raven tests weighted according to number of items per category (z-score, see Appendix~Table~\ref{tab:covariate_description} for more details)} \\[12pt]
        
        \footnotesize{Female} && \footnotesize{Dummy coded $=1$ if female} \\[6pt]
        
        \footnotesize{Financial Literacy} && \footnotesize{Number of correct answers in financial literacy quiz (z-score)}\\[12pt]
        
        \footnotesize{Agreeableness} && \footnotesize{Big-5 personality trait agreeableness (z-score)} \\[6pt]
        
        \footnotesize{Conscientiousness} && \footnotesize{Big-5 personality trait conscientiousness (z-score)} \\[6pt]
                
        \footnotesize{Extraversion} && \footnotesize{Big-5 personality trait extraversion (z-score)} \\[6pt]
        
        \footnotesize{Negative emotionality} && \footnotesize{Big-5 personality trait negative emotionality (z-score)} \\[6pt]
        
        \footnotesize{Openmindedness} && \footnotesize{Big-5 personality trait open mindedness (z-score)} \\[6pt]
        
        
        
        
        
        \bottomrule
    \end{tabularx}
    \label{tab:covariate_description}
\end{table}

First, focusing on individual characteristics associated with debt aversion we can identify a weak negative association between debt aversion and cognitive ability: People who score higher on our measure of cognitive ability, which includes tests on cognitive reflection, numeracy and fluid intelligence, appear to have lower levels of debt aversion. This finding is interesting, as it has the opposite sign of what is reported in \cite{ahrens2022intertemporal}, who report a weak positive association. However, the two findings are difficult to compare, as different measures of individual debt aversion as well as cognitive ability are used. 

Other individual characteristics, such as age, gender, financial literacy and personality appear unrelated to debt aversion.

Further findings include a positive correlation between age and risk aversion and strong evidence for a negative correlation between age and loss aversion. These findings are in line with the thrust of the literature \citep{Meissner2022}. Finally, we find weak evidence suggesting that females are more risk but less loss averse, and that people with higher agreeableness scores tend to be more loss averse.

\begin{table}[t!]
    \centering
    \caption{Individual characteristics associated with preference parameters}
    \label{table:covariates}
    \footnotesize
{
\def\sym#1{\ifmmode^{#1}\else\(^{#1}\)\fi}
\begin{tabular*}{\hsize}{@{\hskip\tabcolsep\extracolsep\fill}l*{5}{c}}
\toprule
                &\makecell{Risk aversion \\ $\alpha$}         &\makecell{Discounting \\ $\delta$}         &\makecell{Debt Aversion \\ $\gamma$}         &\makecell{Loss Aversion \\ $\lambda$}         &\makecell{Fechner error \\ $\mu$}         \\
\midrule
Age             &    0.035\sym{**} &   -0.003         &   -0.006         &   -0.012\sym{***}&   -0.038\sym{***}\\
                &  (0.015)         &  (0.002)         &  (0.005)         &  (0.004)         &  (0.013)         \\
Cognitive ability&   -0.007         &   -0.012         &   -0.022\sym{*}  &   -0.015         &   -0.034         \\
                &  (0.037)         &  (0.008)         &  (0.012)         &  (0.011)         &  (0.057)         \\
Female          &    0.161\sym{*}  &   -0.008         &    0.010         &   -0.063\sym{*}  &   -0.283\sym{*}  \\
                &  (0.097)         &  (0.015)         &  (0.034)         &  (0.038)         &  (0.158)         \\
Financial literacy&   -0.033         &    0.003         &   -0.003         &   -0.006         &    0.009         \\
                &  (0.025)         &  (0.007)         &  (0.013)         &  (0.006)         &  (0.027)         \\
Agreeableness   &   -0.027         &    0.005         &    0.004         &    0.013\sym{*}  &    0.010         \\
                &  (0.027)         &  (0.005)         &  (0.010)         &  (0.007)         &  (0.026)         \\
Conscientiousness&   -0.040         &   -0.005         &   -0.016         &    0.005         &    0.055         \\
                &  (0.037)         &  (0.007)         &  (0.014)         &  (0.014)         &  (0.050)         \\
Extraversion    &   -0.005         &   -0.003         &    0.001         &   -0.005         &    0.003         \\
                &  (0.051)         &  (0.009)         &  (0.014)         &  (0.010)         &  (0.059)         \\
Negative emotionality&    0.043         &   -0.002         &   -0.007         &   -0.015         &   -0.037         \\
                &  (0.076)         &  (0.011)         &  (0.017)         &  (0.017)         &  (0.102)         \\
Openmindedness  &    0.021         &    0.001         &    0.004         &   -0.014         &   -0.008         \\
                &  (0.030)         &  (0.007)         &  (0.014)         &  (0.009)         &  (0.032)         \\
Constant        &   -0.199         &    0.107\sym{**} &    1.176\sym{***}&    1.414\sym{***}&    1.424\sym{***}\\
                &  (0.289)         &  (0.053)         &  (0.110)         &  (0.080)         &  (0.278)         \\
\midrule
N               &    12240         &                  &                  &                  &                  \\
Log. Likelihood &    -3695         &                  &                  &                  &                  \\
BIC             &     7860         &                  &                  &                  &                  \\
\bottomrule
\multicolumn{6}{l}{\footnotesize Standard errors (clustered at the subject level) in parentheses}\\
\multicolumn{6}{l}{\footnotesize \sym{*} \(p<0.1\), \sym{**} \(p<0.05\), \sym{***} \(p<0.01\)}\\
\end{tabular*}
}

\end{table}

\subsubsection{Population Distributions of Parameters}

As a further generalization, we account for unobserved heterogeneity of preferences between individuals in our sample by estimating a structural model of the joint distribution of preference parameters in the population 
as in \cite{conte2011} and \cite{gaudecker2011}. To this end, we extend our stochastic specification to be in line with the non-linear-mixed-logit routine introduced by \citep{andersen2012non}.

In particular, we assume that the vector of preference parameters and the error parameter $\theta = (\alpha, \delta, \gamma, \lambda, \mu)$ follows a joint normal distribution $f$ with distribution hyper-parameter vector $\Theta$. Given the joint normal form, $\Theta$ comprises mean as well as standard deviation for each parameter in $\theta$ and the covariances between all possible pairings of these parameters. 

Let $\theta_i$ denote a realization of $\theta$ for a particular individual $i$. Analogously to Equation \ref{eq:cond_prob}, in a particular decision, individual $i$ will choose Option~B, conditional on $\theta_i$, with the following probability:

\begin{align}
    P_i^B(\theta_i)=F(\Delta U(\theta_i))
\end{align}


Aggregating over all choices $j$, the probability of all observed choices by individual $i$ is:

\begin{align}
P_i(\theta_i)= \prod_j (P_{ij}^B(\theta_i)c_{ij}+(1-P_{ij}^B(\theta_i))(1-c_{ij})),
\end{align}

where, analogously to the aggregate specification, the index $c_{ij}=0$ if individual~$i$ chooses Option~A in choice~$j$ and $c_{ij}=1$ if individual~$i$ chooses Option~B in choice~$j$. Deriving the probability of observed choices conditional on the population distribution hyper-parameters $\Theta$ rather than an individual realization $\theta_i$ involves integration over the distribution of $\theta$: 

\begin{align}
P_i(\Theta)=\int P_i(\theta_i) \ f(\theta|\Theta) \ \mathrm{d}\theta
\end{align}

In particular, $P_i(\Theta)$ for any individual $i$ is given by integrating over the weighted average of conditional probabilities of observed choices $P_i(\theta_i)$ aggregated over all choices~$j$ evaluated at different values of~$\theta$ and weights given by the density of model parameters~$f$. The log-likelihood function over all individuals then writes as:

\begin{align}
\label{eq:logL2}
ln \ L(\Theta)= \sum_iln(P_i(\Theta))
\end{align}

\begin{figure}[t!]
\caption{Probability density functions of preference parameters} 
\begin{tikzpicture}
	\begin{axis}[
		width  = \textwidth,
		height = 10cm,
		ymin=0,
		ymax=20,
		axis lines = left,
		ylabel = \footnotesize{densities},
		ymajorgrids = true,
		xmin = -0.2,
		xmax = 1.6,
		xlabel= \footnotesize{preference parameter estimates ($\gamma\text{, }\alpha\text{, }\delta\text{ and }\lambda$)}, 
		tick label style={font=\footnotesize},
		extra x ticks={0},
	    extra x tick style={xticklabel=0},
	    xticklabels={,,,0.2,0.4,0.6,0.8,1.0,1.2,1.4,},
	    yticklabels={,,,,,,,,},
		area legend,
		legend cell align={left}, 
        legend style={
            font=\footnotesize,
            at={(0.95,0.75)},
            anchor=east},
		]
		\addplot	[
					red,ultra thick,
					domain=0.8:1.3,
					samples=100,
					forget plot,
					]
				 	{(1/(0.0519*sqrt(2*pi)))*exp((-1/2)*((x-1.0639)/0.0519)^2)};
	    \addplot+	[
	                mark=none,
					fill,
					red,
					opacity=0.3,
					domain=0.8:1.3,
					samples=100
					]
				 	{(1/(0.0519*sqrt(2*pi)))*exp((-1/2)*((x-1.0639)/0.0519)^2)};
				 	\addlegendentry{\textbf{debt aversion ($\gamma$)}}
		\node[
				 	label={\scriptsize{$\gamma\sim\mathcal{N}(\DTLfetch{means}{result}{parameter_rounded}{gamma},
				    \DTLfetch{varcov}{result}{parameter_rounded}{v33}$)}} =2pt] at (axis cs:0.8,4) {};
				 	    
		\addplot	[
					blue, ultra thick,
					domain=-0.2:1.6,
					samples=100,
					forget plot,
					]
				 	{(1/(0.178*sqrt(2*pi)))*exp((-1/2)*((x-0.5319)/0.178)^2)};
	    \addplot+	[   
	                mark=none,
					fill,
					blue,
					opacity=0.3,
					samples=100
					]
				 	{(1/(0.178*sqrt(2*pi)))*exp((-1/2)*((x-0.5319)/0.178)^2)};
				 	\addlegendentry{risk aversion ($\alpha$)}
		\node[
				 	label={\scriptsize{$\alpha\sim\mathcal{N}( \DTLfetch{means}{result}{parameter_rounded}{alpha}, \DTLfetch{varcov}{result}{parameter_rounded}{v11})$}} =2pt] at (axis cs:0.5,2) {};
				 	
				 	
		\addplot	[
					violet, ultra thick,
					domain=-0.2:0.2,
					samples=100,
					forget plot,
					]
				 	{(1/(0.0361*sqrt(2*pi)))*exp((-1/2)*((x-0.0391)/0.0361)^2)};
	    \addplot+	[
	                mark=none,
					fill,
					violet,
					opacity=0.1,
					domain=-0.2:0.2,
					samples=100
					]
				 	{(1/(0.0361*sqrt(2*pi)))*exp((-1/2)*((x-0.0391)/0.0361)^2)};
				 	\addlegendentry{time discounting ($\delta$)}
		\node[
				 	label={\scriptsize{$\delta\sim\mathcal{N}( \DTLfetch{means}{result}{parameter_rounded}{delta},
				 	 \DTLfetch{varcov}{result}{parameter_rounded}{v22})$}} =2pt] at (axis cs:0.25,10) {};
				 	
		\addplot	[
					yellow, ultra thick,
					domain=0.6:1.6,
					samples=100,
					forget plot,
					]
				 	{(1/(0.1578*sqrt(2*pi)))*exp((-1/2)*((x-1.1444)/0.1578)^2)};
	    \addplot+	[
	                mark=none,
					fill,
					yellow,
					opacity=0.25,
					domain=0.6:1.6,
					samples=100
					]
				 	{(1/(0.1578*sqrt(2*pi)))*exp((-1/2)*((x-1.1444)/0.1578)^2)};
				 	\addlegendentry{loss aversion ($\lambda$)}
		\node[
				 	label={\scriptsize{$\lambda\sim\mathcal{N}( \DTLfetch{means}{result}{parameter_rounded}{lambda},
				 	 \DTLfetch{varcov}{result}{parameter_rounded}{v44})$}} =2pt] at (axis cs:1.425,2) {};
	\end{axis}
\end{tikzpicture}

\label{fig:pop_alldistrib_all}
\end{figure}
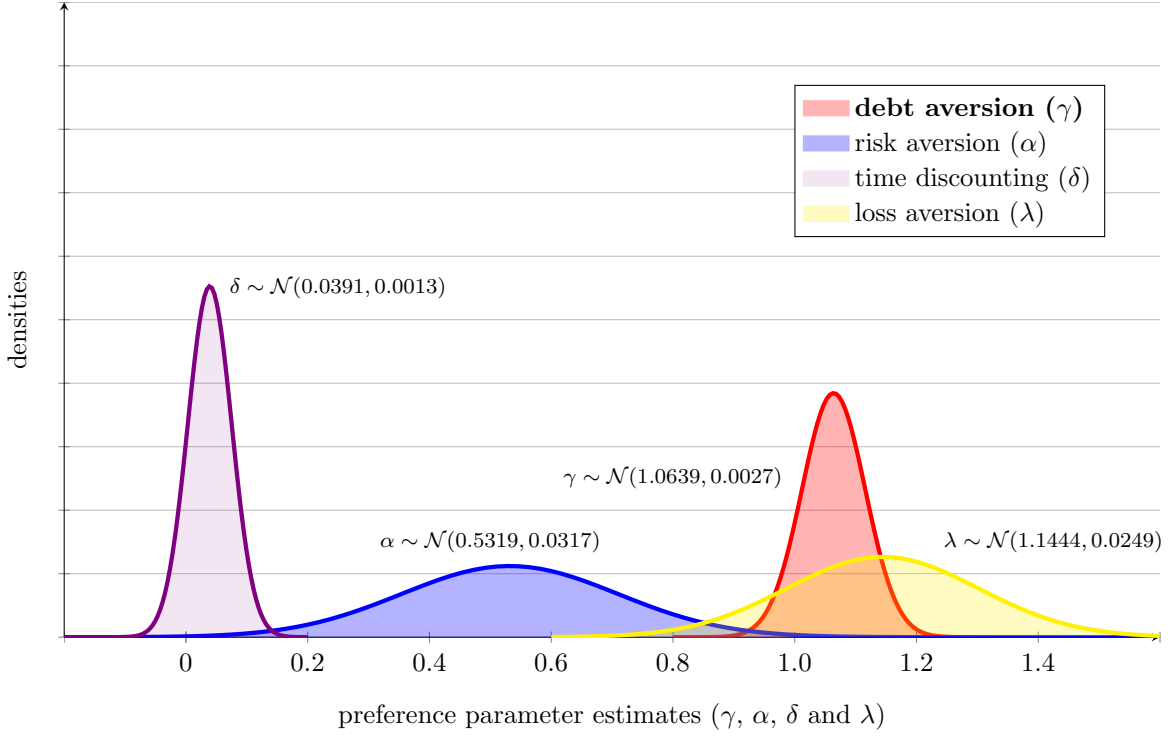

We maximize the log likelihood numerically, using simulated maximum likelihood, as suggested by \cite{andersen2012non} and reviewed earlier in \cite{cameron2005} and \cite{ train2009}. In particular, we employ STATA's modified Newton-Raphson algorithm to maximize the likelihood function in Equation~\ref{eq:logL2}. Resulting estimates of the distributional parameters for preferences over risk, time, losses and debt are displayed in Table~\ref{table:normal_dist_means} (means) and Table~\ref{table:normal_dist_varcov} (variance-covariance matrix). The two-dimensional cross sections of the probability density function for all parameters are illustrated in  Figure~\ref{fig:pop_alldistrib_all}.

\begin{filecontents*}{RESULTS/distribution_means_completed.csv}
"","result","alpha","delta","gamma","lambda","sigma","NA.","NA..1","NA..2","NA..3","NA..4","NA..5","NA..6","NA..7","NA..8","NA..9","NA..10","NA..11","NA..12","NA..13","NA..14"
"1","parameter",0.5319,0.0391,1.0639,1.1444,0.3116,0.1782,-0.0075,0.0024,-0.0891,-0.167,0.0347,0.0163,0.1029,0.0824,0.0493,0.0493,0.0882,0.063,-0.0212,-0.0254
"2","se",0.0127,0.003,0.0074,0.0072,0.0141,0.0067,0.0026,0.0066,0.0049,0.0101,0.0035,0.0088,0.006,0.0128,0.0059,0.0049,0.0141,0.0035,0.0098,0.0118
"3","parameter_rounded",0.5319,0.0391,1.0639,1.1444,0.3116,NA,NA,NA,NA,NA,NA,NA,NA,NA,NA,NA,NA,NA,NA,NA
"4","se_rounded",0.0127,0.003,0.0074,0.0072,0.0141,NA,NA,NA,NA,NA,NA,NA,NA,NA,NA,NA,NA,NA,NA,NA
"5","lowerlim",0.507,0.0332,1.0494,1.1304,0.2839,NA,NA,NA,NA,NA,NA,NA,NA,NA,NA,NA,NA,NA,NA,NA
"6","upperlim",0.5567,0.0449,1.0783,1.1585,0.3393,NA,NA,NA,NA,NA,NA,NA,NA,NA,NA,NA,NA,NA,NA,NA
"7","lowerlim_rounded",0.507,0.0332,1.0494,1.1304,0.2839,NA,NA,NA,NA,NA,NA,NA,NA,NA,NA,NA,NA,NA,NA,NA
"8","upperlim_rounded",0.5567,0.0449,1.0783,1.1585,0.3393,NA,NA,NA,NA,NA,NA,NA,NA,NA,NA,NA,NA,NA,NA,NA
"9","log-likelihood",-2531.17,NA,NA,NA,NA,NA,NA,NA,NA,NA,NA,NA,NA,NA,NA,NA,NA,NA,NA,NA
"10","n",24480,NA,NA,NA,NA,NA,NA,NA,NA,NA,NA,NA,NA,NA,NA,NA,NA,NA,NA,NA
"11","AIC",5102.33,NA,NA,NA,NA,NA,NA,NA,NA,NA,NA,NA,NA,NA,NA,NA,NA,NA,NA,NA
"12","BIC",5264.44,NA,NA,NA,NA,NA,NA,NA,NA,NA,NA,NA,NA,NA,NA,NA,NA,NA,NA,NA
"13","cluster",127,NA,NA,NA,NA,NA,NA,NA,NA,NA,NA,NA,NA,NA,NA,NA,NA,NA,NA,NA
\end{filecontents*}
\DTLloaddb{means}{RESULTS/distribution_means_completed.csv} 
\begin{filecontents*}{RESULTS/distribution_sd_completed.csv}
"","result","alpha","delta","gamma","lambda","sigma"
"1","parameter",0.1782,0.0355,0.0519,0.1579,0.2087
"2","se",0.0067,0.0036,0.007,0.0073,0.015
"3","parameter_rounded",0.1782,0.0355,0.0519,0.1579,0.2087
"4","se_rounded",0.0067,0.0036,0.007,0.0073,0.015
"5","lowerlim",0.165,0.0285,0.0383,0.1435,0.1793
"6","upperlim",0.1913,0.0426,0.0656,0.1723,0.2381
"7","lowerlim_rounded",0.165,0.0285,0.0383,0.1435,0.1793
"8","upperlim_rounded",0.1913,0.0426,0.0656,0.1723,0.2381
\end{filecontents*}
\DTLloaddb{sds}{RESULTS/distribution_sd_completed.csv} 

\begin{table}[t!]
\caption{Maximum simulated likelihood estimates}
\begin{tabularx}{\textwidth}{ 
    c
    @{\hspace{36pt}}
  >{\centering\arraybackslash}X 
  >{\centering\arraybackslash}X 
  >{\centering\arraybackslash}X
  >{\centering\arraybackslash}X
  >{\centering\arraybackslash}X
  }
\toprule

& $\boldsymbol{\alpha}$ 
& $\boldsymbol{\delta}$ 
&  $\boldsymbol{\gamma}$ 
& $\boldsymbol{\lambda}$ 
& $\boldsymbol{\mu}$\\[-6pt]

& \scriptsize{risk aversion} 
& \scriptsize{discounting} 
& \scriptsize{debt aversion} 
& \scriptsize{loss aversion} 
& \scriptsize{Fechner error} \\
 \midrule
 
\scriptsize{\textbf{mean}} 
& \DTLfetch{means}{result}{parameter_rounded}{alpha}
& \DTLfetch{means}{result}{parameter_rounded}{delta}
& \DTLfetch{means}{result}{parameter_rounded}{gamma}
& \DTLfetch{means}{result}{parameter_rounded}{lambda}
& \DTLfetch{means}{result}{parameter_rounded}{sigma} \\[-6pt]

\scriptsize{95\% CI} 
& \scriptsize{\DTLfetch{means}{result}{lowerlim_rounded}{alpha}\,/\,\DTLfetch{means}{result}{upperlim_rounded}{alpha}}
& \scriptsize{\DTLfetch{means}{result}{lowerlim_rounded}{delta}\,/\,\DTLfetch{means}{result}{upperlim_rounded}{delta}}
& \scriptsize{\DTLfetch{means}{result}{lowerlim_rounded}{gamma}\,/\,\DTLfetch{means}{result}{upperlim_rounded}{gamma}}
& \scriptsize{\DTLfetch{means}{result}{lowerlim_rounded}{lambda}\,/\,\DTLfetch{means}{result}{upperlim_rounded}{lambda}}&
\scriptsize{\DTLfetch{means}{result}{lowerlim_rounded}{sigma}\,/\,\DTLfetch{means}{result}{upperlim_rounded}{sigma}}\\[-9pt]

\scriptsize{SE} 
& \scriptsize{\DTLfetch{means}{result}{se_rounded}{alpha}}
& \scriptsize{\DTLfetch{means}{result}{se_rounded}{delta}}
& \scriptsize{\DTLfetch{means}{result}{se_rounded}{gamma}}
& \scriptsize{\DTLfetch{means}{result}{se_rounded}{lambda}}
& \scriptsize{\DTLfetch{means}{result}{se_rounded}{sigma}}\\
\midrule
 


 
\multicolumn{6}{l}{\scriptsize{estimation details: n = \DTLfetch{means}{result}{n}{alpha},  log-likelihood = \DTLfetch{means}{result}{log-likelihood}{alpha},
AIC = \DTLfetch{means}{result}{AIC}{alpha}, 
BIC = \DTLfetch{means}{result}{BIC}{alpha}, logit Fechner error}}\\
\bottomrule
\multicolumn{6}{l}{\scriptsize{Standard errors (SE) clustered at the individual level, \DTLfetch{means}{result}{cluster}{alpha} clusters}}
\end{tabularx}

\label{table:normal_dist_means}
\end{table}
\begin{filecontents*}{RESULTS/distribution_variances_completed.csv}
"","result","v11","v21","v31","v41","v51","v22","v32","v42","v52","v33","v43","v53","v44","v54","v55"
"1","parameter",0.0317,-0.0013,0.0004,-0.0159,-0.0298,0.0013,0.0005,0.0042,0.0041,0.0027,0.0039,0.0053,0.0249,0.0264,0.0435
"2","se",0.0024,0.0005,0.0012,0.0012,0.0024,0.0003,0.0004,0.0005,0.0008,0.0007,0.0013,0.002,0.0023,0.0029,0.0063
"3","parameter_rounded",0.0317,-0.0013,0.0004,-0.0159,-0.0298,0.0013,0.0005,0.0042,0.0041,0.0027,0.0039,0.0053,0.0249,0.0264,0.0435
"4","se_rounded",0.0024,0.0005,0.0012,0.0012,0.0024,0.0003,0.0004,0.0005,0.0008,0.0007,0.0013,0.002,0.0023,0.0029,0.0063
"5","lowerlim",0.0271,-0.0023,-0.0019,-0.0183,-0.0345,0.0008,-0.0002,0.0034,0.0026,0.0013,0.0014,0.0013,0.0204,0.0206,0.0313
"6","upperlim",0.0364,-0.0004,0.0027,-0.0135,-0.025,0.0018,0.0013,0.0051,0.0056,0.0041,0.0064,0.0093,0.0295,0.0321,0.0558
"7","lowerlim_rounded",0.0271,-0.0023,-0.0019,-0.0183,-0.0345,0.0008,-0.0002,0.0034,0.0026,0.0013,0.0014,0.0013,0.0204,0.0206,0.0313
"8","upperlim_rounded",0.0364,-0.0004,0.0027,-0.0135,-0.025,0.0018,0.0013,0.0051,0.0056,0.0041,0.0064,0.0093,0.0295,0.0321,0.0558
\end{filecontents*}
\DTLloaddb{varcov}{RESULTS/distribution_variances_completed.csv} 

\begin{table}[t!]
\caption{Variance-covariance matrix}
\begin{tabularx}{\textwidth}{ 
    c
    c
    @{\hspace{36pt}}
  >{\centering\arraybackslash}X 
  >{\centering\arraybackslash}X 
  >{\centering\arraybackslash}X
  >{\centering\arraybackslash}X
  >{\centering\arraybackslash}X
  }
\toprule
& 
& $\boldsymbol{\alpha}$ 
& $\boldsymbol{\delta}$ 
&  $\boldsymbol{\gamma}$ 
& $\boldsymbol{\lambda}$ 
& $\boldsymbol{\mu}$\\[-6pt]

& 
& \scriptsize{risk aversion} 
& \scriptsize{discounting} 
& \scriptsize{debt aversion} 
& \scriptsize{loss aversion} 
& \scriptsize{Fechner error} \\
\midrule
 
& \scriptsize{\textbf{var/cov}} 
& \DTLfetch{varcov}{result}{parameter}{v11} 
& 
& 
&
& \\[-6pt]

$\boldsymbol{\alpha}$ 
& \scriptsize{95\% CI} 
&  \scriptsize{\DTLfetch{varcov}{result}{lowerlim}{v11}\,/\,\DTLfetch{varcov}{result}{upperlim}{v11}} 
& 
& 
& \\[-9pt]

&\scriptsize{SE} 
& \scriptsize{\DTLfetch{varcov}{result}{se}{v11}} 
& 
& 
& 
& \\

& 
& \DTLfetch{varcov}{result}{parameter}{v21}
& \DTLfetch{varcov}{result}{parameter}{v22} 
& 
& 
& \\[-6pt]

$\boldsymbol{\delta}$ 
& 
& \scriptsize{\DTLfetch{varcov}{result}{lowerlim}{v21}\,/\,\DTLfetch{varcov}{result}{upperlim}{v21}}
& \scriptsize{\DTLfetch{varcov}{result}{lowerlim}{v22}\,/\,\DTLfetch{varcov}{result}{upperlim}{v22}}
& 
& 
&\\[-9pt]

& 
& \scriptsize{\DTLfetch{varcov}{result}{se}{v21}}
& \scriptsize{\DTLfetch{varcov}{result}{se}{v22}}
& 
&
& \\

& 
& \DTLfetch{varcov}{result}{parameter}{v31}
& \DTLfetch{varcov}{result}{parameter}{v32}
& \DTLfetch{varcov}{result}{parameter}{v33}
& 
& \\[-6pt]

$\boldsymbol{\gamma}$  
& 
& \scriptsize{\DTLfetch{varcov}{result}{lowerlim}{v31}\,/\,\DTLfetch{varcov}{result}{upperlim}{v31}}
& \scriptsize{\DTLfetch{varcov}{result}{lowerlim}{v32}\,/\,\DTLfetch{varcov}{result}{upperlim}{v32}}
& \scriptsize{\DTLfetch{varcov}{result}{lowerlim}{v33}\,/\,\DTLfetch{varcov}{result}{upperlim}{v33}}
& 
& \\[-9pt]

& 
& \scriptsize{\DTLfetch{varcov}{result}{se}{v31}}
& \scriptsize{\DTLfetch{varcov}{result}{se}{v32}}
& \scriptsize{\DTLfetch{varcov}{result}{se}{v33}}
&
& \\

& 
& \DTLfetch{varcov}{result}{parameter}{v41}
& \DTLfetch{varcov}{result}{parameter}{v42}
& \DTLfetch{varcov}{result}{parameter}{v43}
& \DTLfetch{varcov}{result}{parameter}{v44}
& \\[-6pt]

$\boldsymbol{\lambda}$ 
& 
& \scriptsize{\DTLfetch{varcov}{result}{lowerlim}{v41}\,/\,\DTLfetch{varcov}{result}{upperlim}{v41}}
& \scriptsize{\DTLfetch{varcov}{result}{lowerlim}{v42}\,/\,\DTLfetch{varcov}{result}{upperlim}{v42}}
& \scriptsize{\DTLfetch{varcov}{result}{lowerlim}{v43}\,/\,\DTLfetch{varcov}{result}{upperlim}{v43}}
& \scriptsize{\DTLfetch{varcov}{result}{lowerlim}{v44}\,/\,\DTLfetch{varcov}{result}{upperlim}{v44}}
& \\[-9pt]

& 
& \scriptsize{\DTLfetch{varcov}{result}{se}{v41}}
& \scriptsize{\DTLfetch{varcov}{result}{se}{v42}}
& \scriptsize{\DTLfetch{varcov}{result}{se}{v43}}
& \scriptsize{\DTLfetch{varcov}{result}{se}{v44}}
& \\

& 
& \DTLfetch{varcov}{result}{parameter}{v51}
& \DTLfetch{varcov}{result}{parameter}{v52}
& \DTLfetch{varcov}{result}{parameter}{v53}
& \DTLfetch{varcov}{result}{parameter}{v54}
& \DTLfetch{varcov}{result}{parameter}{v55} \\[-6pt]

$\boldsymbol{\mu}$ 
& 
& \scriptsize{\DTLfetch{varcov}{result}{lowerlim}{v51}\,/\,\DTLfetch{varcov}{result}{upperlim}{v51}}
& \scriptsize{\DTLfetch{varcov}{result}{lowerlim}{v52}\,/\,\DTLfetch{varcov}{result}{upperlim}{v52}}
& \scriptsize{\DTLfetch{varcov}{result}{lowerlim}{v53}\,/\,\DTLfetch{varcov}{result}{upperlim}{v53}}
& \scriptsize{\DTLfetch{varcov}{result}{lowerlim}{v54}\,/\,\DTLfetch{varcov}{result}{upperlim}{v54}}
& \scriptsize{\DTLfetch{varcov}{result}{lowerlim}{v55}\,/\,\DTLfetch{varcov}{result}{upperlim}{v55}} \\[-9pt]

& 
& \scriptsize{\DTLfetch{varcov}{result}{se}{v51}}
& \scriptsize{\DTLfetch{varcov}{result}{se}{v52}}
& \scriptsize{\DTLfetch{varcov}{result}{se}{v53}}
& \scriptsize{\DTLfetch{varcov}{result}{se}{v54}}
& \scriptsize{\DTLfetch{varcov}{result}{se}{v55}} \\
\midrule
 
\multicolumn{7}{l}{\scriptsize{estimation details: n = \DTLfetch{means}{result}{n}{alpha},  log-likelihood = \DTLfetch{means}{result}{log-likelihood}{alpha},
AIC = \DTLfetch{means}{result}{AIC}{alpha}, 
BIC = \DTLfetch{means}{result}{BIC}{alpha}, logit Fechner error}}\\
\bottomrule

\multicolumn{7}{l}{\scriptsize{Standard errors (SE) clustered at the individual level, \DTLfetch{means}{result}{cluster}{alpha} clusters}}
\end{tabularx}

\label{table:normal_dist_varcov}
\end{table}

Distribution estimation results support the finding of debt aversion in the aggregate estimations: around 89\% of the population is estimated to have a debt aversion parameter above one, i.e. exhibits debt aversion.


A key advantage of estimating the joint distribution of all preference parameters, including the variance-covariance matrix, is that we can identify correlations of the structural preference parameters based on covariances of the estimated population distributions. In this regard, we find that debt aversion is positively correlated with loss aversion, with a correlation coefficient of $\rho=0.4756$.\footnote{Pearson's correlation coefficient is calculated as $\rho_{x,y}=\frac{Cov_{x,y}}{\sigma_x\sigma_y}$ where $Cov$ is the covariance as reported in Table~\ref{table:normal_dist_varcov}, and $\sigma$ denotes standard deviations, which can be retrieved as $\sqrt{var}$ using variances reported in Table~\ref{table:normal_dist_varcov}.} Notably, no other preference parameter appears to be correlated with debt aversion. 
Regarding preferences other than debt aversion, risk aversion appears to be negatively correlated with time discounting ($\rho=-0.2025$) and loss aversion ($\rho=-0.5659$), and time discounting is positively correlated with loss aversion ($\rho=0.7382$). These results are in line with \cite{schleich2019large}, who test for correlation of preference parameters in a large-scale multi-country representative survey.


\subsection{Extensions and Robustness Checks}
\label{sec:robustness}

\subsubsection{Duration of Indebtedness}
\label{sec:extension}

In this extension we aim to investigate potential mechanisms of how debt aversion takes effect. Specifically, we will test whether the cost of being in debt could depend on the time a person spends indebted. We find support that debt aversion does not only depend on amount and timing of repayment but also substantially on the time of indebtedness.

To test this, we extended the main experiment in the final wave of data collection. This extended experiment contained 30 additional choices on savings and debt contracts spanning not four, but a longer period of eight weeks. The remaining 90 choices and all general procedures are the same as in the main experiment without extension. Just as in the 90 original choices, all payments to the experimenter in the additional choices are held constant at \euro{15}, i.e. all loans require the same amount of repayment.

We consider an extended model specification of debt aversion that additionally depends on the time of being indebted $T-t$:

\begin{align}
   c(\boldsymbol{x})=(1-\gamma\zeta^{(T-t-1)})\phi(T)v(x_T)
\end{align}

Maintaining the components and interpretation of the previous specification of debt aversion, in this extension $\zeta$ scales debt aversion based on the time of being indebted.  In particular, $\zeta>1$ implies increasing cost of being indebted if time of being indebted increases, $\zeta=1$ implies invariance of debt aversion with respect to time of indebtedness and $\zeta<1$ describes decreasing cost of being indebted if time of indebtedness increases.

In this setting the utility of a short debt contract, i.e. $T-t=1$, is the same as in the specification without debt duration dependent scaling of debt aversion ($\zeta$):

\begin{align}
    U(x)=\phi(t)u(x_t)-\lambda\gamma\phi(T)u(-x_T),
\end{align}

for long debt contracts, i.e. $T-t=2$ in the setting of our experiment, the utility of debt contracts simplifies to

\begin{align}
   U(x)=\phi(t)u(x_t)-\lambda\gamma\zeta\phi(T)u(-x_T)
\end{align}

Using aggregate maximum likelihood estimations and pooled choice data from the standard experiment with 90 choices and the extended experiment with 120 choices yields results as summarized in Table \ref{table:agg_extended}. Parameter estimates, including debt aversion as established in the main specification remain largely unchanged. Interestingly however, the debt duration aversion parameter $\zeta$ is significantly larger than one. This implies that the cost of being indebted increases in the the time of indebtedness.

To illustrate what this implies in terms of behavior, the average decision maker characterized by parameter estimates derived in the frame of the extended model is indifferent between accepting and rejecting a debt contract that offers a loan of \euro{20.67} today and requires repayment of \euro{15} in four weeks. However, the same decision maker requires a larger loan of \euro{21.11} today if repayment of \euro{15} is not due in four but eight weeks. In contrast, the hypothetical debt and debt duration neutral counterpart (i.e. $\gamma=1$ and $\zeta=1$), requires a four-week loan of size \euro{17.43} and and eight-week loan of size \euro{15.51} to be indifferent. Based on the differences, we can calculate borrowing premia of 
18.59\% for short, four week debt contracts and
36.10\% for long, eight week debt contracts. In other words, the borrowing premium increases by around $93\%$ if duration of indebtedness doubles. Note, however, that the identification of the debt duration aversion parameter relies on only 30 participants who completed the extended list of MPLs, and should therefore be interpreted with caution. 


\begin{filecontents*}{RESULTS/agg_extended.csv}
"","result","alpha","delta","gamma","zeta","lambda","mu"
"1","parameter",0.6398,0.0429,1.0635,1.851,1.1007,0.448
"2","robust se",0.034,0.0075,0.0134,0.2919,0.0121,0.0402
"3","p-value",18.8462857424598,5.73162969066009,79.343780216163,6.34081236287896,91.2145273700457,11.1441304861789
"4","z",0.00000000000000000000000000000000000000000000000000000000000000000000000000000031521226782442,0.00000000994702498084236,0,0.000000000228556768383659,0,0.0000000000000000000000000000764877249780894
"5","lowerlim",0.57,0.03,1.04,1.28,1.08,0.37
"6","upperlim",0.71,0.06,1.09,2.42,1.12,0.53
"7",NA,NA,NA,NA,NA,NA,NA
"8",NA,1.95996398454005,1.95996398454005,1.95996398454005,1.95996398454005,1.95996398454005,1.95996398454005
"9",NA,0,0,0,0,0,0
"10","log-likelihood",-4095.55,NA,NA,NA,NA,NA
"11","n",12240,NA,NA,NA,NA,NA
"12","AIC",8203.1,NA,NA,NA,NA,NA
"13","BIC",8247.57,NA,NA,NA,NA,NA
"14","cluster",127,NA,NA,NA,NA,NA
\end{filecontents*}
\DTLloaddb{agg_extended}{RESULTS/agg_extended.csv} 

\begin{table}[t!]
\caption{Aggregate parameter estimates including debt duration aversion}
\begin{tabularx}{\textwidth}{ 
    c
    @{\hspace{36pt}}
  >{\centering\arraybackslash}X 
  >{\centering\arraybackslash}X 
  >{\centering\arraybackslash}X
  >{\centering\arraybackslash}X
  >{\centering\arraybackslash}X  >{\centering\arraybackslash}X
  }
\toprule

& $\boldsymbol{\alpha}$ 
& $\boldsymbol{\delta}$ 
&  $\boldsymbol{\gamma}$ 
& $\boldsymbol{\zeta}$ 
& $\boldsymbol{\lambda}$ & $\boldsymbol{\mu}$\\[-6pt]

& \scriptsize{risk aversion} 
& \scriptsize{discounting} 
& \scriptsize{debt} 
& \scriptsize{debt duration} 
& \scriptsize{loss aversion} 
& \scriptsize{Fechner error} \\[-12pt]

& 
& 
& \scriptsize{aversion} 
& \scriptsize{aversion} 
&
& \\
 \midrule
 
\makecell{\scriptsize{\textbf{point}} \\[-12pt] \scriptsize{\textbf{estimate}}}
& \DTLfetch{agg_extended}{result}{parameter}{alpha}
& \DTLfetch{agg_extended}{result}{parameter}{delta}
& \DTLfetch{agg_extended}{result}{parameter}{gamma}
& \DTLfetch{agg_extended}{result}{parameter}{zeta}
& \DTLfetch{agg_extended}{result}{parameter}{lambda}
& \DTLfetch{agg_extended}{result}{parameter}{mu}\\[-6pt]

\scriptsize{95\% CI} 
& \scriptsize{\DTLfetch{agg_extended}{result}{lowerlim}{alpha}\,/\,\DTLfetch{agg_extended}{result}{upperlim}{alpha}}
& \scriptsize{\DTLfetch{agg_extended}{result}{lowerlim}{delta}\,/\,\DTLfetch{agg_extended}{result}{upperlim}{delta}}
& \scriptsize{\DTLfetch{agg_extended}{result}{lowerlim}{gamma}\,/\,\DTLfetch{agg_extended}{result}{upperlim}{gamma}}
& \scriptsize{\DTLfetch{agg_extended}{result}{lowerlim}{zeta}\,/\,\DTLfetch{agg_extended}{result}{upperlim}{zeta}}
& \scriptsize{\DTLfetch{agg_extended}{result}{lowerlim}{lambda}\,/\,\DTLfetch{agg_extended}{result}{upperlim}{lambda}}
&\scriptsize{\DTLfetch{agg_extended}{result}{lowerlim}{mu}\,/\,\DTLfetch{agg_extended}{result}{upperlim}{mu}}\\[-9pt]

\scriptsize{robust SE} 
& \scriptsize{\DTLfetch{agg_extended}{result}{robust se}{alpha}}
& \scriptsize{\DTLfetch{agg_extended}{result}{robust se}{delta}}
& \scriptsize{\DTLfetch{agg_extended}{result}{robust se}{gamma}}
& \scriptsize{\DTLfetch{agg_extended}{result}{robust se}{zeta}}
& \scriptsize{\DTLfetch{agg_extended}{result}{robust se}{lambda}}
& \scriptsize{\DTLfetch{agg_extended}{result}{robust se}{mu}}\\
\midrule
 
\multicolumn{7}{l}{\scriptsize{estimation details: n = \DTLfetch{agg_extended}{result}{n}{alpha},  log-likelihood = \DTLfetch{agg_extended}{result}{log-likelihood}{alpha},
AIC = \DTLfetch{agg_extended}{result}{AIC}{alpha}, 
BIC = \DTLfetch{agg_extended}{result}{BIC}{alpha}, logit Fechner error}}\\
\bottomrule
\multicolumn{7}{l}{\scriptsize{Robust standard errors (SE) clustered at the individual level, \DTLfetch{agg_extended}{result}{cluster}{alpha} clusters}}
\end{tabularx}
\label{table:agg_extended}
\end{table}

\subsubsection{Present Bias}
\label{sec:extension_hdisc}

In principle our experimental setup allows to identify present bias, as we include debt and saving contract that are shifted into the future, while maintaining the same temporal distance between involved dates. 
To test whether present bias is existent in our sample, we consider an alternative discount function, that incorporates quasi-hyperbolic discounting \citep{RN392, RN386}:

\begin{align}
    \phi'(\tau) = 
\begin{cases}
1  & \mbox{if } \tau=0 \\
\frac{1}{(1+\beta)} \frac{1}{(1+\delta)^{\tau}}   & \mbox{if }\tau\neq0.
\end{cases}
\end{align}

In this specification, $\delta$ is the exponential discount rate, and $\beta$ is the parameter that determines present bias. A parameter of $\beta>0$ indicates present bias, $\beta=0$ indicates no present bias and $\beta<0$ indicates future bias. The results of the aggregate maximum likelihood estimation using this alternative specification are presented in Table~\ref{table:agg_beta}. Present bias is estimated precisely at, and statistically indistinguishable from 0. The evidence against present bias also extends to a setting considering different parameters of present bias for gains and losses (see also Appendix~\ref{sec:robust_appendix}). These findings, appear in line with recent meta-study results on present bias elicited in experiments \citep{imai2020}.

\begin{filecontents*}{RESULTS/agg_beta.csv}
"","result","alpha","beta","delta","gamma","lambda","mu"
"1","parameter",0.6431,"-0.0012",0.037,1.0545,1.1069,0.4484
"2","robust se",0.0345,"0.0029",0.0065,0.0114,0.012,0.0402
"3","p-value",18.6399409420025,"-0.425562840672493",5.67674897139591,92.2512474793961,92.2340108006171,11.1497559465896
"4","z",0.0000000000000000000000000000000000000000000000000000000000000000000000000000152408966951382,"0.670426422133573",0.0000000137278696841417,0,0,0.0000000000000000000000000000718030340937313
"5","lowerlim",0.58,"-0.01",0.02,1.03,1.08,0.37
"6","upperlim",0.71,"\ensuremath{{\scriptscriptstyle <}.01}",0.05,1.08,1.13,0.53
"7",NA,NA,NA,NA,NA,NA,NA
"8",NA,1.95996398454005,"1.95996398454005",1.95996398454005,1.95996398454005,1.95996398454005,1.95996398454005
"9",NA,0,"0",0,0,0,0
"10","log-likelihood",-4107.86,NA,NA,NA,NA,NA
"11","n",12240,NA,NA,NA,NA,NA
"12","AIC",8227.71,NA,NA,NA,NA,NA
"13","BIC",8272.19,NA,NA,NA,NA,NA
"14","cluster",127,NA,NA,NA,NA,NA
\end{filecontents*}
\DTLloaddb{agg_beta}{RESULTS/agg_beta.csv} 

\begin{table}[t!]
\caption{Aggregate parameter estimates including present bias}
\begin{tabularx}{\textwidth}{ 
    c
    @{\hspace{36pt}}
  >{\centering\arraybackslash}X 
  >{\centering\arraybackslash}X 
  >{\centering\arraybackslash}X
  >{\centering\arraybackslash}X
  >{\centering\arraybackslash}X  >{\centering\arraybackslash}X
  }
\toprule
& $\boldsymbol{\alpha}$ 
& $\boldsymbol{\beta}$ 
& $\boldsymbol{\delta}$ 
&  $\boldsymbol{\gamma}$ 
& $\boldsymbol{\lambda}$ 
& $\boldsymbol{\mu}$\\[-6pt]

& \scriptsize{risk aversion} 
& \scriptsize{present bias}
& \scriptsize{discounting} 
& \scriptsize{debt aversion} 
& \scriptsize{loss aversion} 
& \scriptsize{Fechner error} \\
 \midrule
 
\makecell{\scriptsize{\textbf{point}} \\[-12pt] \scriptsize{\textbf{estimate}}}
& \DTLfetch{agg_beta}{result}{parameter}{alpha}
& \DTLfetch{agg_beta}{result}{parameter}{beta}
& \DTLfetch{agg_beta}{result}{parameter}{delta}
& \DTLfetch{agg_beta}{result}{parameter}{gamma}
& \DTLfetch{agg_beta}{result}{parameter}{lambda}
& \DTLfetch{agg_beta}{result}{parameter}{mu}\\[-6pt]

\scriptsize{95\% CI} 
& \scriptsize{\DTLfetch{agg_beta}{result}{lowerlim}{alpha}\,/\,\DTLfetch{agg_beta}{result}{upperlim}{alpha}}
& \scriptsize{\DTLfetch{agg_beta}{result}{lowerlim}{beta}\,/\,\DTLfetch{agg_beta}{result}{upperlim}{beta}}
& \scriptsize{\DTLfetch{agg_beta}{result}{lowerlim}{delta}\,/\,\DTLfetch{agg_beta}{result}{upperlim}{delta}}
& \scriptsize{\DTLfetch{agg_beta}{result}{lowerlim}{gamma}\,/\,\DTLfetch{agg_beta}{result}{upperlim}{gamma}}
& \scriptsize{\DTLfetch{agg_beta}{result}{lowerlim}{lambda}\,/\,\DTLfetch{agg_beta}{result}{upperlim}{lambda}}
&\scriptsize{\DTLfetch{agg_beta}{result}{lowerlim}{mu}\,/\,\DTLfetch{agg_beta}{result}{upperlim}{mu}}\\[-9pt]

\scriptsize{robust SE} 
& \scriptsize{\DTLfetch{agg_beta}{result}{robust se}{alpha}}
& \scriptsize{\DTLfetch{agg_beta}{result}{robust se}{beta}}
& \scriptsize{\DTLfetch{agg_beta}{result}{robust se}{delta}}
& \scriptsize{\DTLfetch{agg_beta}{result}{robust se}{gamma}}
& \scriptsize{\DTLfetch{agg_beta}{result}{robust se}{lambda}}
& \scriptsize{\DTLfetch{agg_beta}{result}{robust se}{mu}}\\
\midrule
 
\multicolumn{7}{l}{\scriptsize{estimation details: n = \DTLfetch{agg_beta}{result}{n}{alpha},  log-likelihood = \DTLfetch{agg_beta}{result}{log-likelihood}{alpha},
AIC = \DTLfetch{agg_beta}{result}{AIC}{alpha}, 
BIC = \DTLfetch{agg_beta}{result}{BIC}{alpha}, logit Fechner error}}\\
\bottomrule
\multicolumn{7}{l}{\scriptsize{Robust standard errors (SE) clustered at the individual level, \DTLfetch{agg_beta}{result}{cluster}{alpha} clusters}}
\end{tabularx}
\label{table:agg_beta}
\end{table}

\subsubsection{Robustness Checks}

Our finding of debt aversion may be sensitive to the assumptions underlying our estimations. 
We therefore employ a wide array of robustness checks. 
In particular, we first consider alternative forms of the cost of borrowing by modelling debt aversion as fixed cost of being indebted as well as scaling of utility from borrowed money. Second, we alter various characteristics of our utility specification in general. These comprise the distinction of risk aversion and time discounting in the gain and loss domain, abstracting from risk aversion as well as considering alternative forms of the utility function such as CARA utility and CRRA utility without $\varepsilon$-transformation. Third, we scrutinize different error structures: we introduce an additional tremble error, exchange the logit for a probit Fechner error, and allow distinct probit Fechner errors per choice domain. Moreover, we also test the effect of excluding participants who did not complete the entire experimental sequence or expressed some doubt in the trustworthiness of the experimental environment. Summing up the results, debt aversion remains robust regardless of the utilized functional forms and sample selection criteria. Detailed descriptions and results on all robustness checks can be found in Appendix~\ref{sec:robust_appendix}.

\section{Discussion and conclusion}
\label{sec:conclusion}

In this paper we introduce a novel theoretical framework and experiment that allows to model and measure debt aversion. We are able to separately identify debt aversion from other relevant preferences, such as risk aversion, loss aversion, and time preferences. In this way, we aim to establish debt aversion as a preference in it's own right, as opposed to an emergent behavioral property of other preferences, beliefs and constraints. 

Using a structural maximum likelihood estimation, we find that our participants are on average debt averse. They would be willing to forgo a substantial amount of money in order to avoid getting into debt. We estimate the ``borrowing premium'', that is the increase in the upfront payment our average participant would require compared to a counterfactual debt neutral participant to accept a debt contract, to be around 16\%. Testing how observed individual characteristics correlate with debt aversion, we find weak evidence supporting a negative correlation between debt aversion and cognitive ability. Other individual characteristics, such as age, gender, financial literacy and personality appear unrelated to debt aversion. Further, we estimate the joint population distribution of all preference parameters, using simulated maximum likelihood. We find that a substantial share of 89\% of our participants exhibit debt aversion. Furthermore, debt aversion appears to be correlated positively with loss aversion, but not with other preference parameters. Finally, we find evidence that debt aversion depends positively on the duration a person spends indebted. Notably, present bias does not appear to be existent in our data, and debt aversion remains robust after a series of robustness checks. Summing up, we find robust evidence supporting debt aversion as a preference in it's own right. Most participants are debt averse, and debt aversion appears to have a meaningful impact on choice. 

The existence of debt aversion could have far-reaching implication for individual financial decision-making. Debt averse individuals could invest less in otherwise profitable investment projects, such as education or energy efficient technologies, and make consumption and saving decisions that deviate from the standard model of intertemporal choice. While we have made a first step in cleanly identifying debt aversion, many open questions on the mechanisms that underlie debt aversion, and it's implications for financial decision-making, remain. We test a multitude of ways of modelling debt aversion, and while all specifications lend clear support to it's existence, our setup is not well suited to discriminate between different models and different mechanisms of how debt aversion works. In an extension of the base experiment, we show that debt aversion appears to increase in the duration participants spend indebted, but many other interesting questions remain. We hope that our theoretical and experimental framework can pave the way for future research that could improve the knowledge on the exact mechanisms at play.  

Debt aversion could also have implications on the macroeconomic level. \cite{hundtofte2019credit} show that, consistent with debt aversion, individuals in the US and Iceland are reluctant to use credit to smooth negative transitory income shocks. They argue that while standard theory predicts countercyclical credit demand, credit demand appears to be pro-cyclical which could deepen business cycle fluctuations.

Further, our findings also have implications for policy design. Many policies rely on offering favourable loans to subsidize particular behaviors, such as investment in tertiary education or energy-efficient technologies. However, in face of a largely debt averse population these loans might not be very effective. Moreover, if debt aversion correlates with individual characteristics, such as income or socio-economic status, such policies could have unintended effects. For instance, loan-based policies to facilitate tertiary education for students from weak financial backgrounds might be particularly unattractive to these students if they are also more debt averse. For these reasons, we believe that more research on how debt aversion relates to individual characteristics of representative populations is required. To facilitate this, we have constructed a short and easy-to-use survey module for measuring individual debt aversion in a companion paper \citep{albrecht2022the}. Using the data from this experiment, we identify a set of survey items that best predicts the debt aversion parameter as elicited with our experiment. The survey module contains three short items, and predicts debt aversion reasonably well. We hope that this survey module will prove useful for future research on debt aversion on a larger scale, where complicated and incentivized experiments are often not feasible. 

Finally, we believe that our setup provides a valuable methodological contribution. To our knowledge, we are first to put participants into actual debt in a laboratory experiment. As the vast majority of participants did not default on their obligations, we believe that such experiments could prove useful to analyze debt related behavior in the future.

\newpage

\bibliographystyle{ecta}
\bibliography{Debt}


\appendix
\clearpage
\section{Multiple Price Lists} \label{sec:mpls}
\FloatBarrier
{\def\arraystretch{0.75}
\begin{table}[htp]
\begin{tabularx}{\textwidth}{ 
    c
    @{\hspace{36pt}}
  >{\centering\arraybackslash}X 
  >{\centering\arraybackslash}X 
  }
\toprule

\textbf{Choice} 
& \textbf{Option A}
& \textbf{Option B}\\
\hline

\footnotesize{1}
& \footnotesize{Receive an amount of \euro{18.2} today}
& \footnotesize{Receive an amount of  \euro{18.0} in 4 weeks}\\

\footnotesize{2}
& \footnotesize{Receive an amount of \euro{18.0} today}
& \footnotesize{Receive an amount of  \euro{18.0} in 4 weeks}\\

\footnotesize{3} 
& \footnotesize{Receive an amount of  \euro{17.8} today} 
& \footnotesize{Receive an amount of  \euro{18.0} in 4 weeks}\\

\footnotesize{4}
& \footnotesize{Receive an amount of  \euro{17.3} today}
& \footnotesize{Receive an amount of  \euro{18.0} in 4 weeks}\\

\footnotesize{5}
& \footnotesize{Receive an amount of  \euro{16.8} today}
& \footnotesize{Receive an amount of  \euro{18.0} in 4 weeks}\\

\footnotesize{6}
& \footnotesize{Receive an amount of  \euro{16.0} today}
& \footnotesize{Receive an amount of  \euro{18.0} in 4 weeks}\\

\footnotesize{7}
& \footnotesize{Receive an amount of  \euro{14.0} today}
& \footnotesize{Receive an amount of  \euro{18.0} in 4 weeks}\\

\footnotesize{8}
& \footnotesize{Receive an amount of  \euro{12.0} today}
& \footnotesize{Receive an amount of  \euro{18.0} in 4 weeks}\\

\footnotesize{9}
& \footnotesize{Receive an amount of  \euro{10.0} today}
& \footnotesize{Receive an amount of  \euro{18.0} in 4 weeks}\\

\footnotesize{10}
& \footnotesize{Receive an amount of  \euro{8.0} today}
& \footnotesize{Receive an amount of  \euro{18.0} in 4 weeks}\\

\bottomrule
\end{tabularx}
\caption{Multiple price list of intertemporal choices (MPL1)}
\label{tab:MPLI}
\end{table}

\begin{table}[htp]
\begin{tabularx}{\textwidth}{ 
c
@{\hspace{36pt}}
C{0.1cm} 
>{\centering\arraybackslash}X 
>{\centering\arraybackslash}X 
C{0.1cm} 
>{\centering\arraybackslash}X 
>{\centering\arraybackslash}X  
C{0.1cm}
}
\toprule
& & \multicolumn{2}{c}{\textbf{Option A}} & & \multicolumn{2}{c}{\textbf{Option B}} & \\
\cline{3-4} \cline{6-7}
\textbf{Choice}  & &  \makecell{\small{ Coin shows  } \\[-0.2cm] \small{ Heads }} & \makecell*{\small{ Coin shows  } \\[-0.2cm] \small{ Tails }} & & \makecell*{\small{ Coin shows  } \\[-0.2cm] \small{ Heads }} & \makecell*{\small{ Coin shows  } \\[-0.2cm] \small{ Tails }} & \\
\hline
\footnotesize{1} & & \footnotesize{\euro{30}  today} & \footnotesize{\euro{30}  today} & & \footnotesize{\euro{30}  today} & \footnotesize{\euro{1}  today} & \\
\footnotesize{2} & & \footnotesize{\euro{25}  today} & \footnotesize{\euro{25}  today} & & \footnotesize{\euro{30}  today} & \footnotesize{\euro{1}  today} & \\
\footnotesize{3} & & \footnotesize{\euro{20}  today} & \footnotesize{\euro{20}  today} & & \footnotesize{\euro{30} today} & \footnotesize{\euro{1}  today} & \\
\footnotesize{4} & & \footnotesize{\euro{17}  today} & \footnotesize{\euro{17}  today} & & \footnotesize{\euro{30}  today} & \footnotesize{\euro{1}  today} & \\
\footnotesize{5} & & \footnotesize{\euro{16}  today} & \footnotesize{\euro{16}  today} & & \footnotesize{\euro{30}  today} & \footnotesize{\euro{1}  today} & \\
\footnotesize{6} & & \footnotesize{\euro{15}  today} & \footnotesize{\euro{15}  today} & & \footnotesize{\euro{30}  today} & \footnotesize{\euro{1} today} & \\
\footnotesize{7} & &\footnotesize{\euro{12}  today} & \footnotesize{\euro{12}  today} & & \footnotesize{\euro{30}  today} & \footnotesize{\euro{1}  today} & \\
\footnotesize{8} & & \footnotesize{\euro{10}\  today} & \footnotesize{\euro{10}  today} & & \footnotesize{\euro{30}  today} & \footnotesize{\euro{1}  today} & \\
\footnotesize{9} & &\footnotesize{\euro{5}  today} & \footnotesize{\euro{5}  today} & & \footnotesize{\euro{30} today} & \footnotesize{\euro{1}  today} & \\
\footnotesize{10} & & \footnotesize{\euro{1}  today} & \footnotesize{\euro{1}  today} & & \footnotesize{\euro{30}  today} & \footnotesize{\euro{1}  today} &\\
\bottomrule
\end{tabularx}
\caption{Multiple price list of certain payments vs. risky gambles (MPL2)}
\label{tab:MPLII}
\end{table}

\begin{table}[htp]
\begin{tabularx}{\textwidth}{ 
c
@{\hspace{36pt}}
C{0.1cm} 
>{\centering\arraybackslash}X 
>{\centering\arraybackslash}X 
C{0.1cm} 
>{\centering\arraybackslash}X 
>{\centering\arraybackslash}X  
C{0.1cm}
}
\toprule
& & \multicolumn{2}{c}{\textbf{Option A}} & & \multicolumn{2}{c}{\textbf{Option B}} & \\
\cline{3-4} \cline{6-7}
\textbf{Choice}  & &  \makecell{\small{ Coin shows  } \\[-0.2cm] \small{ Heads }} & \makecell*{\small{ Coin shows  } \\[-0.2cm] \small{ Tails }} & & \makecell*{\small{ Coin shows  } \\[-0.2cm] \small{ Heads }} & \makecell*{\small{ Coin shows  } \\[-0.2cm] \small{ Tails }} & \\
\hline
\footnotesize{1} & & \footnotesize{\euro{14}  today} & \footnotesize{\euro{17}  today} & & \footnotesize{\euro{17}  today} & \footnotesize{\euro{1} today} & \\
\footnotesize{2} & & \footnotesize{\euro{14}  today} & \footnotesize{\euro{17}  today} & & \footnotesize{\euro{20}  today} & \footnotesize{\euro{1} today} & \\
\footnotesize{3} & & \footnotesize{\euro{14}  today} & \footnotesize{\euro{17}  today} & & \footnotesize{\euro{25}  today} & \footnotesize{\euro{1} today} & \\
\footnotesize{4} & & \footnotesize{\euro{14}  today} & \footnotesize{\euro{17} today} & & \footnotesize{\euro{28}  today} & \footnotesize{\euro{1} today} & \\
\footnotesize{5} & & \footnotesize{\euro{14}  today} & \footnotesize{\euro{17}  today} & & \footnotesize{\euro{29}  today} & \footnotesize{\euro{1} today} & \\
\footnotesize{6} & & \footnotesize{\euro{14}  today} & \footnotesize{\euro{17}  today} & & \footnotesize{\euro{30}  today} & \footnotesize{\euro{2}  today} & \\
\footnotesize{7} & &\footnotesize{\euro{14} today} & \footnotesize{\euro{17}  today} & & \footnotesize{\euro{30}  today} & \footnotesize{\euro{3}  today} & \\
\footnotesize{8} & & \footnotesize{\euro{14}  today} & \footnotesize{\euro{17}  today} & & \footnotesize{\euro{32}  today} & \footnotesize{\euro{8}  today} & \\
\footnotesize{9} & &\footnotesize{\euro{14}  today} & \footnotesize{\euro{17}  today} & & \footnotesize{\euro{32}  today} & \footnotesize{\euro{10}  today} & \\
\footnotesize{10} & & \footnotesize{\euro{14}  today} & \footnotesize{\euro{17}  today} & & \footnotesize{\euro{32}  today} & \footnotesize{\euro{14}  today} & \\
\bottomrule
\end{tabularx}
\caption{Multiple price list of less risky vs. more risky gambles (MPL3)}
\label{tab:MPLIII}
\end{table}

\begin{table}[htp]
\begin{tabularx}{\textwidth}{
    c
    @{\hspace{36pt}}
    C{0.1cm} 
  >{\centering\arraybackslash}X 
    C{0.1cm} 
  >{\centering\arraybackslash}X 
    C{0.1cm}
    }
\toprule
 & & \multicolumn{3}{c}{\textbf{Early saving contracts}} & \\
\cline{3-5}
\textbf{Choice} & & \makecell{\small{ Session 1 } \\[-0.2cm] \footnotesize{ (today) }} & & \makecell{\small{ Session 2 } \\[-0.2cm] \footnotesize{ (in 4 weeks) }} & \\

\hline
\footnotesize{1} & & \footnotesize{Pay an amount of \euro{15}} & & \footnotesize{Receive an amount of  \euro{45}}\\
\footnotesize{2} & & \footnotesize{Pay an amount of \euro{15}} & & \footnotesize{Receive an amount of  \euro{40}}\\
\footnotesize{3} & & \footnotesize{Pay an amount of \euro{15}} & & \footnotesize{Receive an amount of  \euro{36}}\\
\footnotesize{4} & & \footnotesize{Pay an amount of \euro{15}} & & \footnotesize{Receive an amount of  \euro{34}}\\
\footnotesize{5} & & \footnotesize{Pay an amount of \euro{15}} & & \footnotesize{Receive an amount of  \euro{32}}\\
\footnotesize{6} & & \footnotesize{Pay an amount of \euro{15}} & & \footnotesize{Receive an amount of  \euro{30}}\\
\footnotesize{7} & & \footnotesize{Pay an amount of \euro{15}} & & \footnotesize{Receive an amount of  \euro{28}}\\
\footnotesize{8} & & \footnotesize{Pay an amount of \euro{15}} & & \footnotesize{Receive an amount of  \euro{26}}\\
\footnotesize{9} & & \footnotesize{Pay an amount of \euro{15}} & & \footnotesize{Receive an amount of  \euro{24}}\\
\footnotesize{10} & & \footnotesize{Pay an amount of \euro{15}} & & \footnotesize{Receive an amount of  \euro{22}}\\
\footnotesize{11} & & \footnotesize{Pay an amount of \euro{15}} & & \footnotesize{Receive an amount of  \euro{20}}\\
\footnotesize{12} & & \footnotesize{Pay an amount of \euro{15}} & & \footnotesize{Receive an amount of  \euro{18}}\\
\footnotesize{13} & & \footnotesize{Pay an amount of \euro{15}} & & \footnotesize{Receive an amount of  \euro{16}}\\
\footnotesize{14} & & \footnotesize{Pay an amount of \euro{15}} & & \footnotesize{Receive an amount of  \euro{14}}\\
\footnotesize{15} & & \footnotesize{Pay an amount of \euro{15}} & & \footnotesize{Receive an amount of  \euro{12}}\\

\bottomrule
\end{tabularx}
\caption{Multiple price list of 4-week saving contracts starting at Session 1 (MPL4)}
\label{tab:MPLIV}
\end{table}

\begin{table}[htp]
\begin{tabularx}{\textwidth}{
    c
    @{\hspace{36pt}}
    C{0.1cm} 
  >{\centering\arraybackslash}X 
    C{0.1cm} 
  >{\centering\arraybackslash}X 
    C{0.1cm}
    }
\toprule
 & & \multicolumn{3}{c}{\textbf{Late saving contracts}} & \\
\cline{3-5}
\textbf{Choice} & & \makecell{\small{ Session 2 } \\[-0.2cm] \footnotesize{ (in 4 weeks) }} & & \makecell{\small{ Session 3 } \\[-0.2cm] \footnotesize{ (in 8 weeks) }} & \\
\hline
\footnotesize{1} & & \footnotesize{Pay an amount of \euro{15}} & & \footnotesize{Receive an amount of  \euro{40}}\\
\footnotesize{2} & & \footnotesize{Pay an amount of \euro{15}} & & \footnotesize{Receive an amount of  \euro{35}}\\
\footnotesize{3} & & \footnotesize{Pay an amount of \euro{15}} & & \footnotesize{Receive an amount of  \euro{31}}\\
\footnotesize{4} & & \footnotesize{Pay an amount of \euro{15}} & & \footnotesize{Receive an amount of  \euro{29}}\\
\footnotesize{5} & & \footnotesize{Pay an amount of \euro{15}} & & \footnotesize{Receive an amount of  \euro{27}}\\
\footnotesize{6} & & \footnotesize{Pay an amount of \euro{15}} & & \footnotesize{Receive an amount of  \euro{25}}\\
\footnotesize{7} & & \footnotesize{Pay an amount of \euro{15}} & & \footnotesize{Receive an amount of  \euro{23}}\\
\footnotesize{8} & & \footnotesize{Pay an amount of \euro{15}} & & \footnotesize{Receive an amount of  \euro{21}}\\
\footnotesize{9} & & \footnotesize{Pay an amount of \euro{15}} & & \footnotesize{Receive an amount of  \euro{19}}\\
\footnotesize{10} & & \footnotesize{Pay an amount of \euro{15}} & & \footnotesize{Receive an amount of  \euro{17}}\\
\footnotesize{11} & & \footnotesize{Pay an amount of \euro{15}} & & \footnotesize{Receive an amount of  \euro{15}}\\
\footnotesize{12} & & \footnotesize{Pay an amount of \euro{15}} & & \footnotesize{Receive an amount of  \euro{13}}\\
\footnotesize{13} & & \footnotesize{Pay an amount of \euro{15}} & & \footnotesize{Receive an amount of  \euro{11}}\\
\footnotesize{14} & & \footnotesize{Pay an amount of \euro{15}} & & \footnotesize{Receive an amount of  \euro{9}}\\
\footnotesize{15} & & \footnotesize{Pay an amount of \euro{15}} & & \footnotesize{Receive an amount of  \euro{7}}\\

\bottomrule
\end{tabularx}
\caption{Multiple price list of 4-week saving contracts starting at Session 2 (MPL5)}
\label{tab:MPLV}
\end{table}

\begin{table}[htp]
\begin{tabularx}{\textwidth}{
    c
    @{\hspace{36pt}}
    C{0.1cm} 
  >{\centering\arraybackslash}X 
    C{0.1cm} 
  >{\centering\arraybackslash}X 
    C{0.1cm}
    }
\toprule
 & & \multicolumn{3}{c}{\textbf{Early debt contracts}} & \\
\cline{3-5}
\textbf{Choice} & & \makecell{\small{ Session 1 } \\[-0.2cm] \footnotesize{ (today) }} & & \makecell{\small{ Session 2 } \\[-0.2cm] \footnotesize{ (in 4 weeks) }} & \\
\hline
\footnotesize{1} & & \footnotesize{Receive an amount of \euro{31}} & & \footnotesize{Pay an amount of  \euro{15}}\\
\footnotesize{2} & & \footnotesize{Receive an amount of \euro{27}} & & \footnotesize{Pay an amount of  \euro{15}}\\
\footnotesize{3} & & \footnotesize{Receive an amount of \euro{24}} & & \footnotesize{Pay an amount of  \euro{15}}\\
\footnotesize{4} & & \footnotesize{Receive an amount of \euro{21}} & & \footnotesize{Pay an amount of  \euro{15}}\\
\footnotesize{5} & & \footnotesize{Receive an amount of \euro{19}} & & \footnotesize{Pay an amount of  \euro{15}}\\
\footnotesize{6} & & \footnotesize{Receive an amount of \euro{17}} & & \footnotesize{Pay an amount of  \euro{15}}\\
\footnotesize{7} & & \footnotesize{Receive an amount of \euro{16}} & & \footnotesize{Pay an amount of  \euro{15}}\\
\footnotesize{8} & & \footnotesize{Receive an amount of \euro{15}} & & \footnotesize{Pay an amount of  \euro{15}}\\
\footnotesize{9} & & \footnotesize{Receive an amount of \euro{14}} & & \footnotesize{Pay an amount of  \euro{15}}\\
\footnotesize{10} & & \footnotesize{Receive an amount of \euro{13}} & & \footnotesize{Pay an amount of  \euro{15}}\\
\footnotesize{11} & & \footnotesize{Receive an amount of \euro{11}} & & \footnotesize{Pay an amount of  \euro{15}}\\
\footnotesize{12} & & \footnotesize{Receive an amount of \euro{9}} & & \footnotesize{Pay an amount of  \euro{15}}\\
\footnotesize{13} & & \footnotesize{Receive an amount of \euro{7}} & & \footnotesize{Pay an amount of  \euro{15}}\\
\footnotesize{14} & & \footnotesize{Receive an amount of \euro{5}} & & \footnotesize{Pay an amount of  \euro{15}}\\
\footnotesize{15} & & \footnotesize{Receive an amount of \euro{3}} & & \footnotesize{Pay an amount of  \euro{15}}\\

\bottomrule
\end{tabularx}
\caption{Multiple price list of 4-week debt contracts starting at Session 1 (MPL6)}
\label{tab:MPLVI}
\end{table}

\begin{table}[htp]
\begin{tabularx}{\textwidth}{
    c
    @{\hspace{36pt}}
    C{0.1cm} 
  >{\centering\arraybackslash}X 
    C{0.1cm} 
  >{\centering\arraybackslash}X 
    C{0.1cm}
    }
\toprule
 & & \multicolumn{3}{c}{\textbf{Late debt contracts}} & \\
\cline{3-5}
\textbf{Choice} & & \makecell{\small{ Session 2 } \\[-0.2cm] \footnotesize{ (in 4 weeks) }} & & \makecell{\small{ Session 3 } \\[-0.2cm] \footnotesize{ (in 8 weeks) }} & \\
\hline
\footnotesize{1} & & \footnotesize{Receive an amount of \euro{33}} & & \footnotesize{Pay an amount of  \euro{15}}\\
\footnotesize{2} & & \footnotesize{Receive an amount of \euro{30}} & & \footnotesize{Pay an amount of  \euro{15}}\\
\footnotesize{3} & & \footnotesize{Receive an amount of \euro{27}} & & \footnotesize{Pay an amount of  \euro{15}}\\
\footnotesize{4} & & \footnotesize{Receive an amount of \euro{24}} & & \footnotesize{Pay an amount of  \euro{15}}\\
\footnotesize{5} & & \footnotesize{Receive an amount of \euro{22}} & & \footnotesize{Pay an amount of  \euro{15}}\\
\footnotesize{6} & & \footnotesize{Receive an amount of \euro{20}} & & \footnotesize{Pay an amount of  \euro{15}}\\
\footnotesize{7} & & \footnotesize{Receive an amount of \euro{18}} & & \footnotesize{Pay an amount of  \euro{15}}\\
\footnotesize{8} & & \footnotesize{Receive an amount of \euro{16}} & & \footnotesize{Pay an amount of  \euro{15}}\\
\footnotesize{9} & & \footnotesize{Receive an amount of \euro{15}} & & \footnotesize{Pay an amount of  \euro{15}}\\
\footnotesize{10} & & \footnotesize{Receive an amount of \euro{14}} & & \footnotesize{Pay an amount of  \euro{15}}\\
\footnotesize{11} & & \footnotesize{Receive an amount of \euro{12}} & & \footnotesize{Pay an amount of  \euro{15}}\\
\footnotesize{12} & & \footnotesize{Receive an amount of \euro{10}} & & \footnotesize{Pay an amount of  \euro{15}}\\
\footnotesize{13} & & \footnotesize{Receive an amount of \euro{8}} & & \footnotesize{Pay an amount of  \euro{15}}\\
\footnotesize{14} & & \footnotesize{Receive an amount of \euro{6}} & & \footnotesize{Pay an amount of  \euro{15}}\\
\footnotesize{15} & & \footnotesize{Receive an amount of \euro{3}} & & \footnotesize{Pay an amount of  \euro{15}}\\

\bottomrule
\end{tabularx}
\caption{Multiple price list of 4-week debt contracts starting at Session 2 (MPL7)}
\label{tab:MPLVII}
\end{table}

\begin{table}[htp]
\begin{tabularx}{\textwidth}{
    c
    @{\hspace{36pt}}
    C{0.1cm} 
  >{\centering\arraybackslash}X 
    C{0.1cm} 
  >{\centering\arraybackslash}X 
    C{0.1cm}
    }
\toprule
 & & \multicolumn{3}{c}{\textbf{Long saving contracts}} & \\
\cline{3-5}
\textbf{Choice} & & \makecell{\small{ Session 1 } \\[-0.2cm] \footnotesize{ (today) }} & & \makecell{\small{ Session 3 } \\[-0.2cm] \footnotesize{ (in 8 weeks) }} & \\
\hline

\footnotesize{1} & & \footnotesize{Pay an amount of \euro{15}} & & \footnotesize{Receive an amount of  \euro{50}}\\
\footnotesize{2} & & \footnotesize{Pay an amount of \euro{15}} & & \footnotesize{Receive an amount of  \euro{45}}\\
\footnotesize{3} & & \footnotesize{Pay an amount of \euro{15}} & & \footnotesize{Receive an amount of  \euro{40}}\\
\footnotesize{4} & & \footnotesize{Pay an amount of \euro{15}} & & \footnotesize{Receive an amount of  \euro{36}}\\
\footnotesize{5} & & \footnotesize{Pay an amount of \euro{15}} & & \footnotesize{Receive an amount of  \euro{34}}\\
\footnotesize{6} & & \footnotesize{Pay an amount of \euro{15}} & & \footnotesize{Receive an amount of  \euro{32}}\\
\footnotesize{7} & & \footnotesize{Pay an amount of \euro{15}} & & \footnotesize{Receive an amount of  \euro{30}}\\
\footnotesize{8} & & \footnotesize{Pay an amount of \euro{15}} & & \footnotesize{Receive an amount of  \euro{28}}\\
\footnotesize{9} & & \footnotesize{Pay an amount of \euro{15}} & & \footnotesize{Receive an amount of  \euro{26}}\\
\footnotesize{10} & & \footnotesize{Pay an amount of \euro{15}} & & \footnotesize{Receive an amount of  \euro{24}}\\
\footnotesize{11} & & \footnotesize{Pay an amount of \euro{15}} & & \footnotesize{Receive an amount of  \euro{22}}\\
\footnotesize{12} & & \footnotesize{Pay an amount of \euro{15}} & & \footnotesize{Receive an amount of  \euro{20}}\\
\footnotesize{13} & & \footnotesize{Pay an amount of \euro{15}} & & \footnotesize{Receive an amount of  \euro{18}}\\
\footnotesize{14} & & \footnotesize{Pay an amount of \euro{15}} & & \footnotesize{Receive an amount of  \euro{16}}\\
\footnotesize{15} & & \footnotesize{Pay an amount of \euro{15}} & & \footnotesize{Receive an amount of  \euro{14}}\\

\bottomrule
\end{tabularx}
\caption{Multiple price list of 8-week saving contracts starting at Session 1 (MPL8)}
\label{tab:MPLIIX}
\end{table}

\begin{table}[htp]
\begin{tabularx}{\textwidth}{
    c
    @{\hspace{36pt}}
    C{0.1cm} 
  >{\centering\arraybackslash}X 
    C{0.1cm} 
  >{\centering\arraybackslash}X 
    C{0.1cm}
    }
\toprule
 & & \multicolumn{3}{c}{\textbf{Long debt contracts}} & \\
\cline{3-5}
\textbf{Choice} & & \makecell{\small{ Session 1 } \\[-0.2cm] \footnotesize{ (today) }} & & \makecell{\small{ Session 3 } \\[-0.2cm] \footnotesize{ (in 8 weeks) }} & \\
\hline
\footnotesize{1} & & \footnotesize{Receive an amount of \euro{39}} & & \footnotesize{Pay an amount of  \euro{15}}\\
\footnotesize{2} & & \footnotesize{Receive an amount of \euro{35}} & & \footnotesize{Pay an amount of  \euro{15}}\\
\footnotesize{3} & & \footnotesize{Receive an amount of \euro{31}} & & \footnotesize{Pay an amount of  \euro{15}}\\
\footnotesize{4} & & \footnotesize{Receive an amount of \euro{27}} & & \footnotesize{Pay an amount of  \euro{15}}\\
\footnotesize{5} & & \footnotesize{Receive an amount of \euro{24}} & & \footnotesize{Pay an amount of  \euro{15}}\\
\footnotesize{6} & & \footnotesize{Receive an amount of \euro{21}} & & \footnotesize{Pay an amount of  \euro{15}}\\
\footnotesize{7} & & \footnotesize{Receive an amount of \euro{19}} & & \footnotesize{Pay an amount of  \euro{15}}\\
\footnotesize{8} & & \footnotesize{Receive an amount of \euro{17}} & & \footnotesize{Pay an amount of  \euro{15}}\\
\footnotesize{9} & & \footnotesize{Receive an amount of \euro{16}} & & \footnotesize{Pay an amount of  \euro{15}}\\
\footnotesize{10} & & \footnotesize{Receive an amount of \euro{15}} & & \footnotesize{Pay an amount of  \euro{15}}\\
\footnotesize{11} & & \footnotesize{Receive an amount of \euro{14}} & & \footnotesize{Pay an amount of  \euro{15}}\\
\footnotesize{12} & & \footnotesize{Receive an amount of \euro{13}} & & \footnotesize{Pay an amount of  \euro{15}}\\
\footnotesize{13} & & \footnotesize{Receive an amount of \euro{11}} & & \footnotesize{Pay an amount of  \euro{15}}\\
\footnotesize{14} & & \footnotesize{Receive an amount of \euro{9}} & & \footnotesize{Pay an amount of  \euro{15}}\\
\footnotesize{15} & & \footnotesize{Receive an amount of \euro{7}} & & \footnotesize{Pay an amount of  \euro{15}}\\

\bottomrule
\end{tabularx}
\caption{Multiple price list of 8-week debt contracts starting at Session 1 (MPL9)}
\label{tab:MPLIX}
\end{table}
}
\FloatBarrier
\newpage

\section{Contract Form} \label{sec:contract}
\begin{center}
\includegraphics[width=0.95\textwidth]{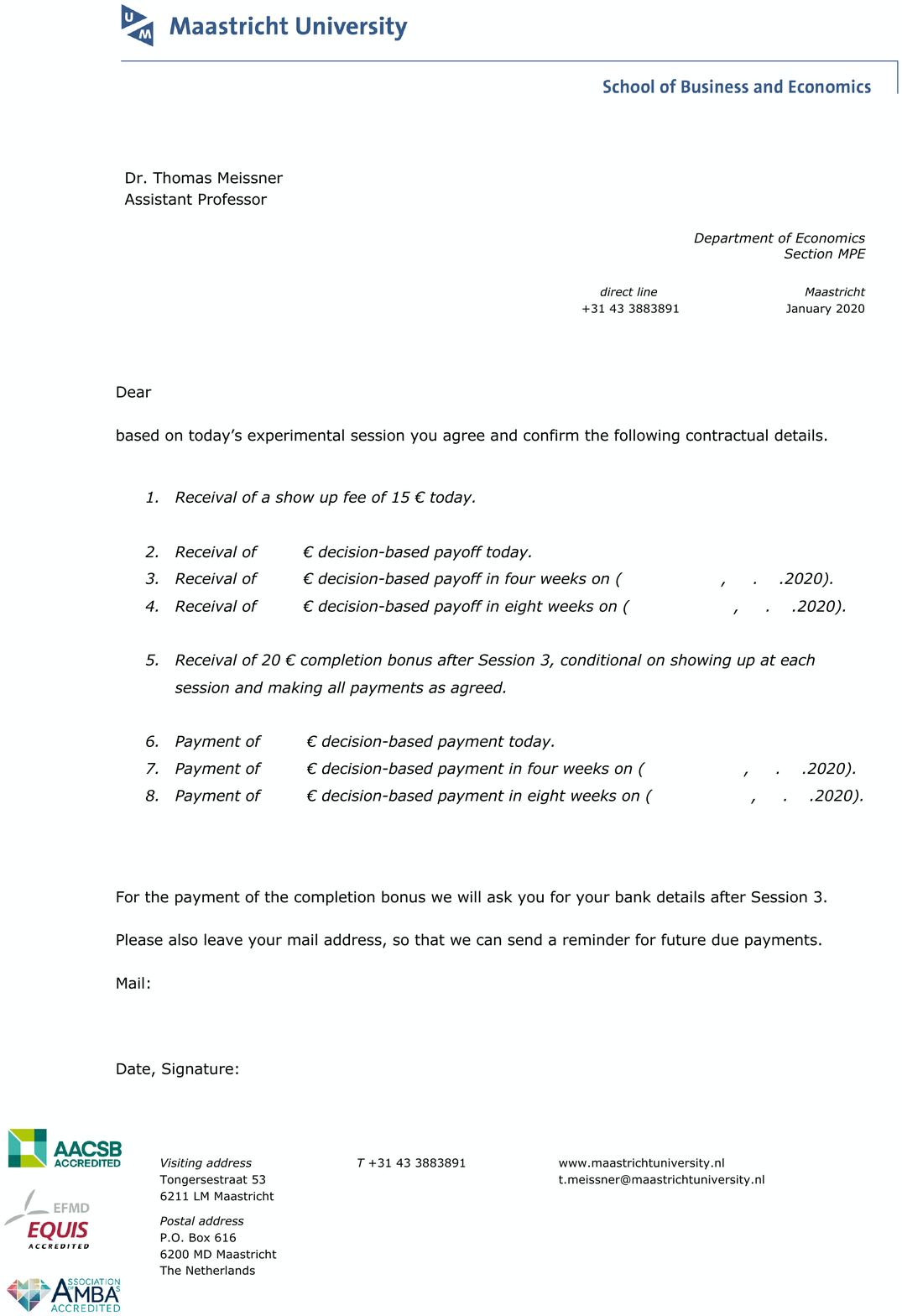}
\end{center}
\newpage

\section{Individual Characteristics} \label{sec:indiv_char_app}
 
\begin{table}[H]
\setlength{\tabcolsep}{14pt}
\begin{tabularx}{\textwidth}{
C{3.1cm}
>{\centering\arraybackslash}X
C{1.5cm}
C{3.7cm}
}

\toprule

\textbf{Individual Characteristic}
& \textbf{Measurement} 
& \textbf{No. of Items}
& \textbf{Source / Reference}\\
\midrule

\multirow{7}{*}{\footnotesize{Financial Literacy}} 
& \footnotesize{Global OECD/INFE FinLitSurvey}
& \footnotesize{5} 
& \footnotesize{\citep{atkinson2016}}\\

& \footnotesize{S\&P International FinLit Survey}
&  \footnotesize{3} 
& \footnotesize{\citep{klapper2015}}\\

& \footnotesize{Debt Literacy}
& \footnotesize{3} 
& \footnotesize{\citep{lusardi2015}}\\

& \footnotesize{FinLit Quiz}
&  \footnotesize{4} 
& \footnotesize{\citep{agnew2015}}\\

\hline

\multirow{2}{*}{\footnotesize{Numeracy}}
& \footnotesize{Abbreviated Numeracy Scale}
& \footnotesize{6} 
& \footnotesize{\citep{weller2013}}\\

\hline
\multirow{5}{*}{\footnotesize{Cognitive Reflection}}
& \footnotesize{CRT} 
& \footnotesize{3} 
& \footnotesize{\citep{frederick2005}}\\

& \footnotesize{CRT-2} 
& \footnotesize{4} 
& \footnotesize{\citep{thomson2016}}\\

& \footnotesize{CRT-long} 
& \footnotesize{3} 
& \footnotesize{\citep{primi2016}}\\

& \footnotesize{Extended CRT} 
& \footnotesize{3} 
& \footnotesize{\citep{toplak2013}}\\

\hline
\multirow{2}{*}{\footnotesize{Fluid Intelligence}}
& \footnotesize{Raven Progressive Matrices}
& \footnotesize{36} 
& \footnotesize{\citep{raven1960}}\\

\hline
\multirow{2}{*}{\footnotesize{Personality}} 
&\footnotesize{BFI-2-S} 
& \footnotesize{30} 
&\footnotesize{\citep{RN188}}\\

& \footnotesize{HEXACO-60 (Honesty)} 
& \footnotesize{8} 
&\footnotesize{\citep{ashton2009}}\\

\hline
\multirow{2}{*}{\footnotesize{Preferences}} 
& \footnotesize{Preference Survey Module}
& \footnotesize{12} 
& \footnotesize{\citep{falk2016}}\\

\hline
\multirow{2}{*}{\footnotesize{Planned Behavior}}
& \footnotesize{Purchases} 
& \footnotesize{7} 
& \footnotesize{--}\\
& \footnotesize{Financing} 
& \footnotesize{7} 
& \footnotesize{--}\\

\bottomrule
\end{tabularx}
\caption{Overview of Elicited Individual Characteristics}
\label{table:individual_char}
\end{table}
\FloatBarrier 
\newpage

\section{Actual Choices} \label{sec:choices_appendix}
\begin{figure}[ht]
\begin{subfigure}{1\textwidth}

\begin{minipage}{0.49\textwidth}
\includegraphics[width=0.85\textwidth]{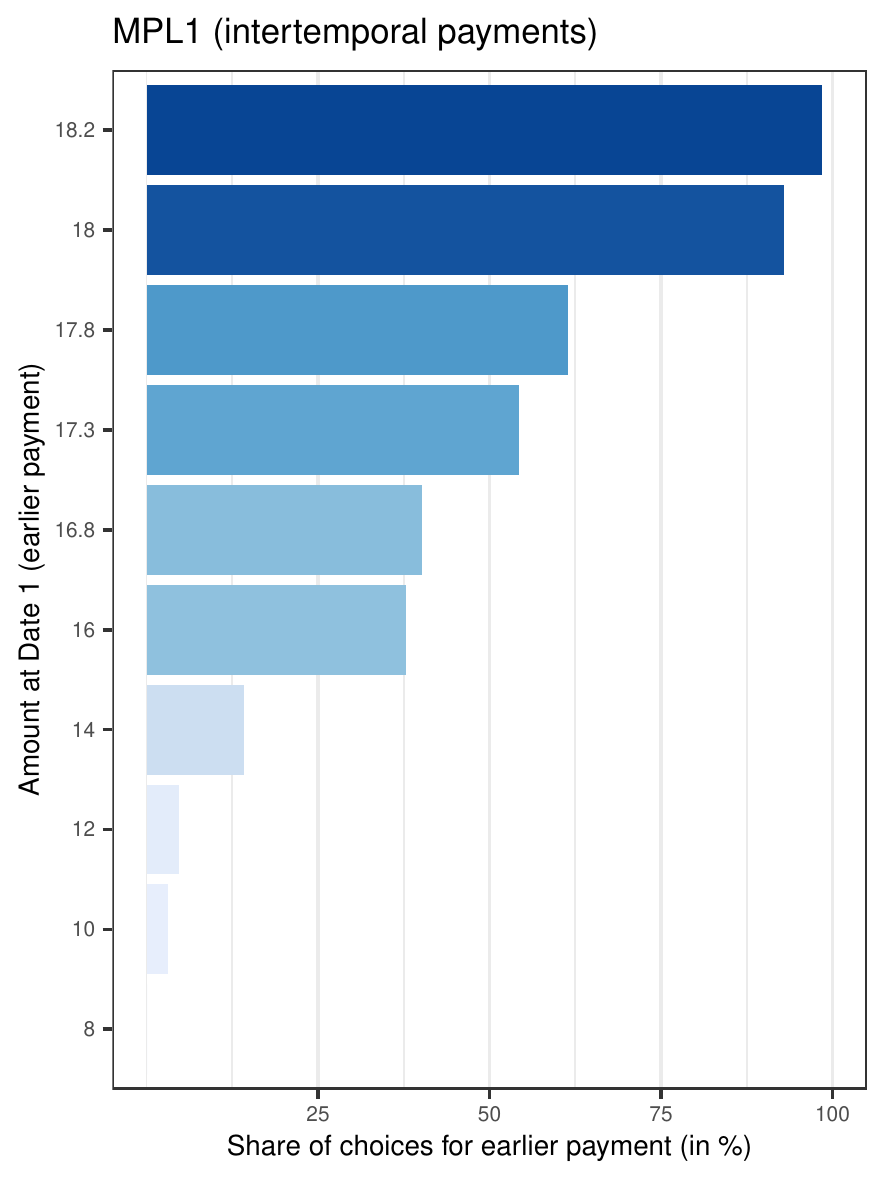}
\end{minipage}
\begin{minipage}{0.49\textwidth}
\end{minipage}
\end{subfigure}

\vspace{0.5cm}

\begin{subfigure}{1\textwidth}
\begin{minipage}{0.49\textwidth}
\includegraphics[width=0.85\textwidth]{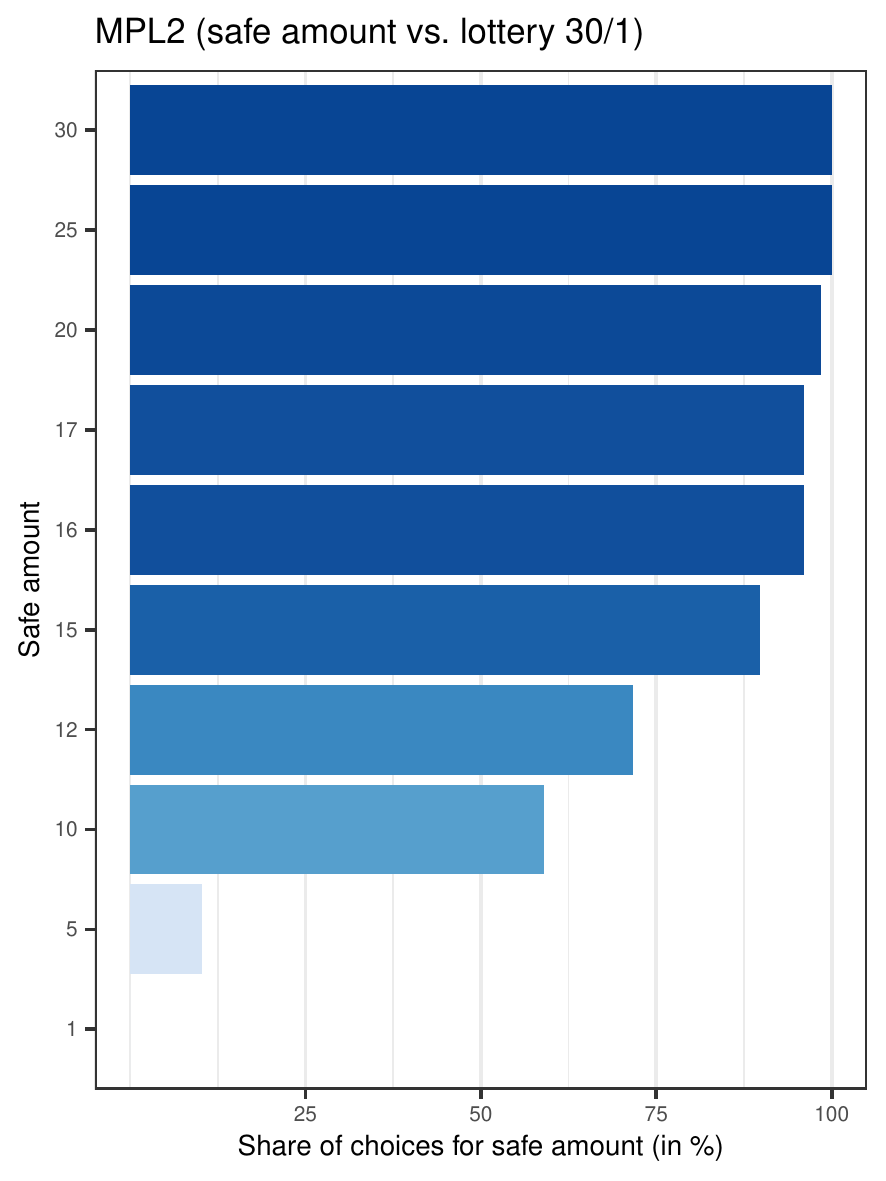}
\end{minipage}
\begin{minipage}{0.49\textwidth}
\includegraphics[width=0.85\textwidth]{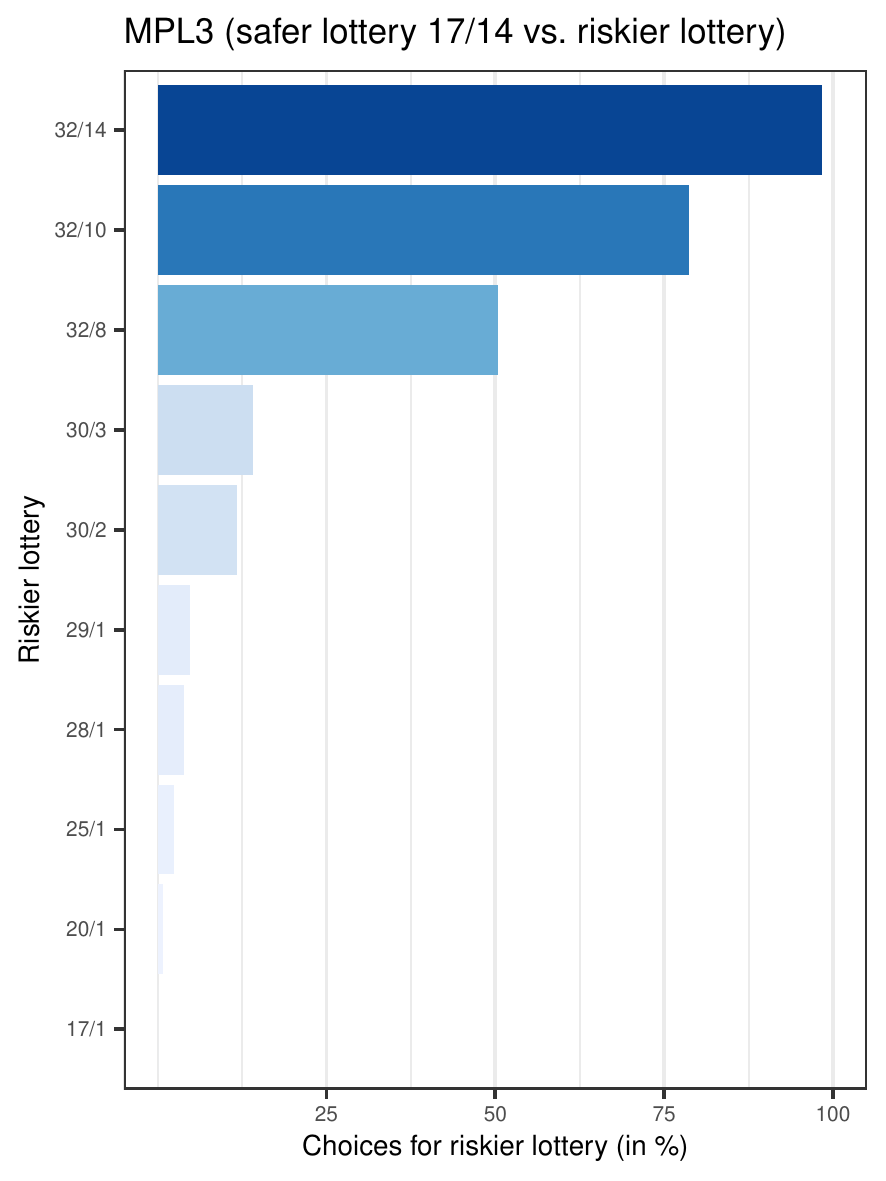}
\end{minipage}
\end{subfigure}
\end{figure}

\begin{figure}
\begin{subfigure}{1\textwidth}
\begin{minipage}{0.49\textwidth}
\includegraphics[width=\textwidth]{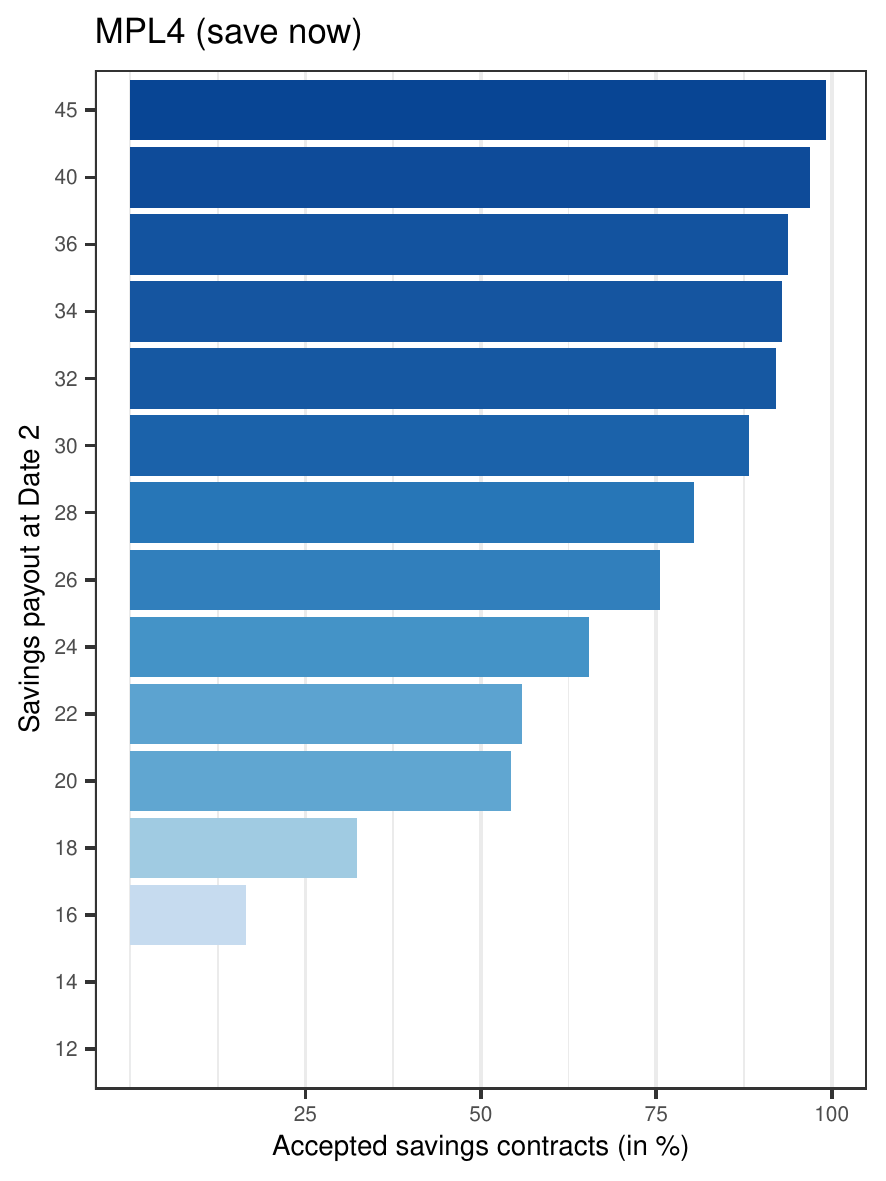}
\end{minipage}
\begin{minipage}{0.49\textwidth}
\includegraphics[width=\textwidth]{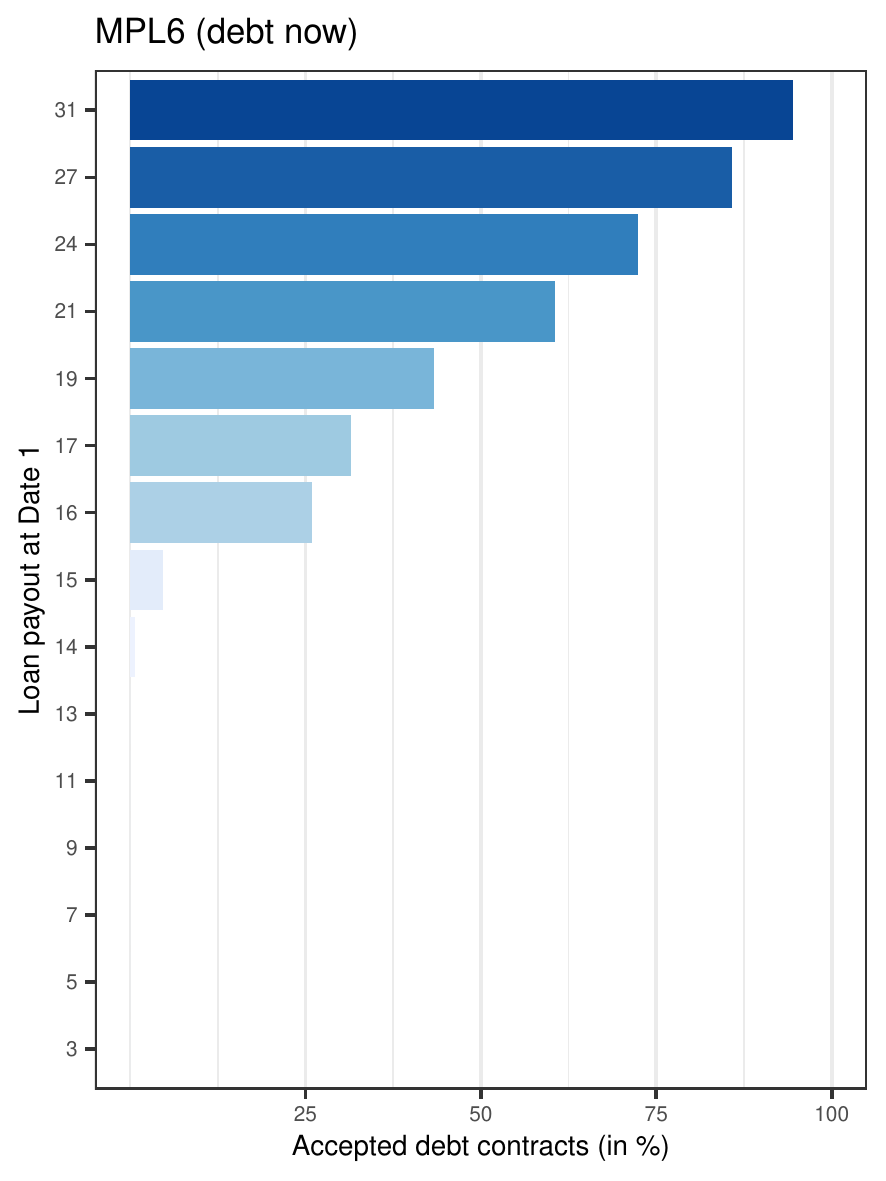}
\end{minipage}
\end{subfigure}

\vspace{0.5cm}

\begin{subfigure}{1\textwidth}
\begin{minipage}{0.49\textwidth}
\includegraphics[width=\textwidth]{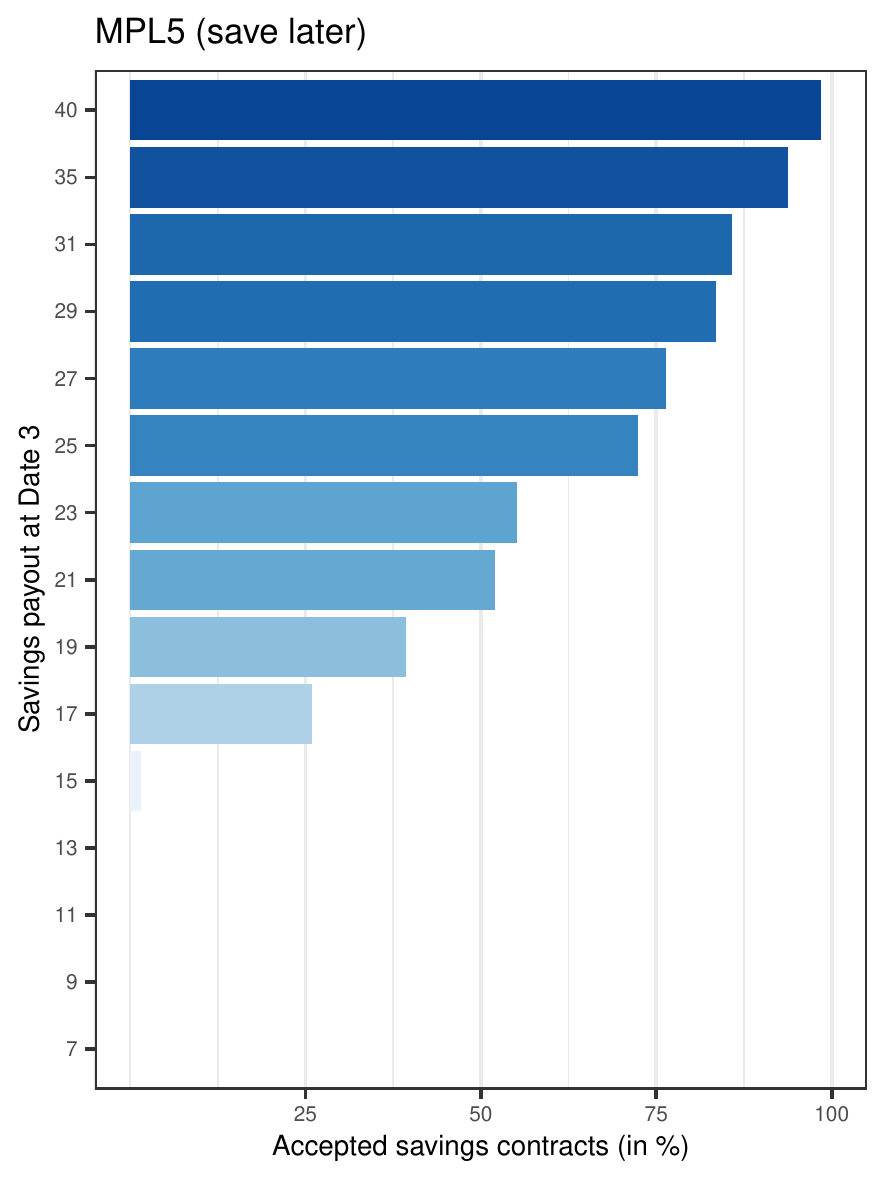}
\end{minipage}
\begin{minipage}{0.49\textwidth}
\includegraphics[width=\textwidth]{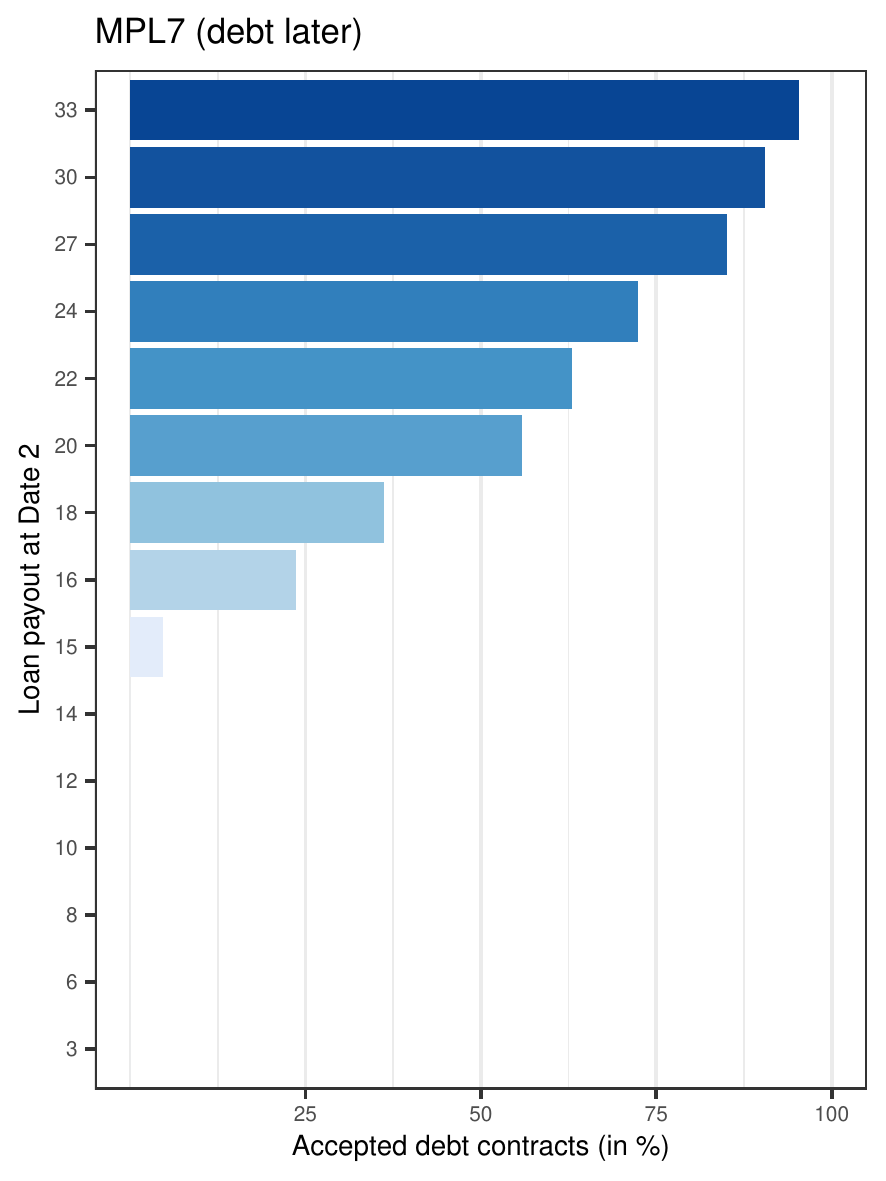}
\end{minipage}
\end{subfigure}
\end{figure}

\begin{figure}[ht]
\begin{subfigure}{1\textwidth}
\begin{minipage}{0.49\textwidth}
\includegraphics[width=\textwidth]{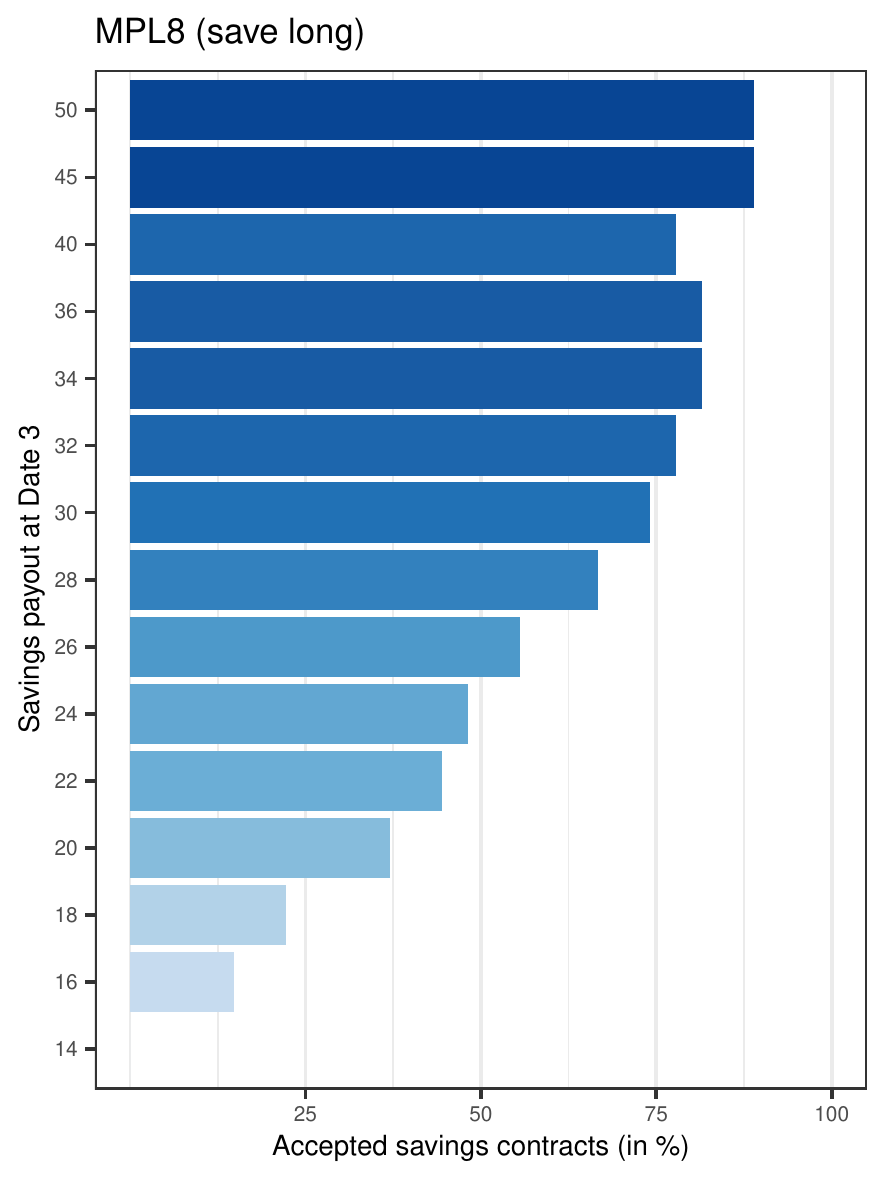}
\end{minipage}
\begin{minipage}{0.49\textwidth}
\includegraphics[width=\textwidth]{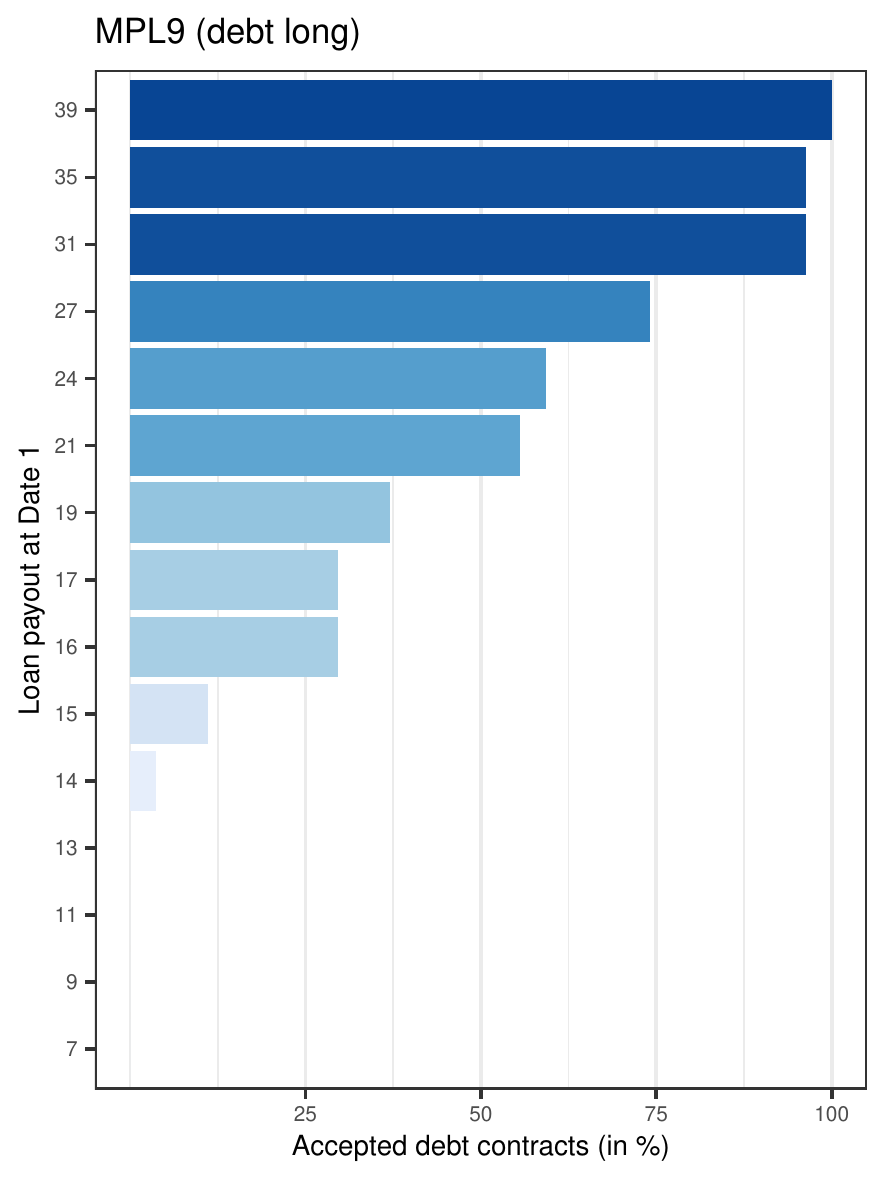}
\end{minipage}
\end{subfigure}

\caption{Distribution of actual choices made, displayed per multiple price list, excluding participants who dropped out before completing the entire experimental sequence (n=127 for MPL1 to MPL7 and n=27 for MPL8 and MPL9 in the extension.}
\label{fig:actual_choices}
\end{figure}

\FloatBarrier 
\newpage

\section[Robustness Checks]{Robustness Checks
\label{sec:robust_appendix}
}

In this section we will consider several different structural assumptions pertaining to the cost of being in debt, the utility and discount functions, as well as the decision error process. To keep the appendix size manageable, we will make use of the simple aggregate maximum likelihood specification, outlined in Section \ref{sec:agg_estimation} 

\subsection{Alternative Specifications of the Utility Cost of Borrowing}

{\bf Fixed cost of being indebted:} 
\begin{filecontents*}{RESULTS/rob_fixcost.csv}
"","result","alpha","delta","gamma","lambda","mu"
"1","parameter",0.6421,0.0371,0.4332,1.1064,0.449
"2","robust se",0.0344,0.0063,0.0911,0.0118,0.0402
"3","p-value",18.69085999915,5.84313037483579,4.75392471459189,94.0419596634451,11.1707010860354
"4","z",0.00000000000000000000000000000000000000000000000000000000000000000000000000000587581774143348,0.00000000512288920942833,0.000001995053536992,0,0.0000000000000000000000000000567313187493834
"5","lowerlim",0.57,0.02,0.25,1.08,0.37
"6","upperlim",0.71,0.05,0.61,1.13,0.53
"7",NA,NA,NA,NA,NA,NA
"8",NA,1.95996398454005,1.95996398454005,1.95996398454005,1.95996398454005,1.95996398454005
"9",NA,0,0,0,0,0
"10","log-likelihood",-4106.98,NA,NA,NA,NA
"11","n",12240,NA,NA,NA,NA
"12","AIC",8223.95,NA,NA,NA,NA
"13","BIC",8261.01,NA,NA,NA,NA
"14","cluster",127,NA,NA,NA,NA
\end{filecontents*}
\DTLloaddb{rob_fixcost}{RESULTS/rob_fixcost.csv} 

In our main specification, cost of debt depends on the timing and amount of the required repayment. Instead, debt aversion could also be modeled as fixed cost, which is incurred at the time of going into debt. Thus, we consider an alternative definition of $c(\boldsymbol{x})$ with $\gamma$ as fixed cost independent of the size of the loan:

\begin{align}
   c(\boldsymbol{x})=\gamma\phi(t).
\end{align}

For interpretation, $\gamma=0$ now corresponds to debt neutrality, i.e. no utility cost of borrowing, $\gamma>0$ corresponds to debt aversion and $\gamma<0$ to debt affinity. Estimation results are presented in Table~\ref{table:rob_fixcost}. Except for $\gamma$, parameter estimates remain virtually identical. The new debt aversion parameter is estimated at $\gamma=\DTLfetch{rob_fixcost}{result}{parameter}{gamma}$, implying that the average decision maker faces positive fixed utility costs when going into debt, i.e. they are debt averse.


\begin{table}[t!]
\caption{Aggregate parameter estimates with fixed cost of going into debt.}
\begin{tabularx}{\textwidth}{ 
    c
    @{\hspace{36pt}}
  >{\centering\arraybackslash}X 
  >{\centering\arraybackslash}X 
  >{\centering\arraybackslash}X
  >{\centering\arraybackslash}X
  >{\centering\arraybackslash}X
  }
\toprule
& $\boldsymbol{\alpha}$
& $\boldsymbol{\delta}$ 
&  $\boldsymbol{\gamma}$ 
& $\boldsymbol{\lambda}$ 
& $\boldsymbol{\mu}$\\[-6pt]

& \scriptsize{risk aversion} 
& \scriptsize{discounting} 
& \scriptsize{debt aversion} 
& \scriptsize{loss aversion} 
& \scriptsize{Fechner error} \\
 \midrule

\scriptsize{\textbf{point estimate}} 
& \DTLfetch{rob_fixcost}{result}{parameter}{alpha}
& \DTLfetch{rob_fixcost}{result}{parameter}{delta}
& \DTLfetch{rob_fixcost}{result}{parameter}{gamma}
& \DTLfetch{rob_fixcost}{result}{parameter}{lambda}
& \DTLfetch{rob_fixcost}{result}{parameter}{mu}\\[-6pt]

\scriptsize{95\% confidence interval} 
& \scriptsize{\DTLfetch{rob_fixcost}{result}{lowerlim}{alpha}\,/\,\DTLfetch{rob_fixcost}{result}{upperlim}{alpha}}
& \scriptsize{\DTLfetch{rob_fixcost}{result}{lowerlim}{delta}\,/\,\DTLfetch{rob_fixcost}{result}{upperlim}{delta}}
& \scriptsize{\DTLfetch{rob_fixcost}{result}{lowerlim}{gamma}\,/\,\DTLfetch{rob_fixcost}{result}{upperlim}{gamma}}
& \scriptsize{\DTLfetch{rob_fixcost}{result}{lowerlim}{lambda}\,/\,\DTLfetch{rob_fixcost}{result}{upperlim}{lambda}}
&\scriptsize{\DTLfetch{rob_fixcost}{result}{lowerlim}{mu}\,/\,\DTLfetch{rob_fixcost}{result}{upperlim}{mu}}\\[-9pt]

\scriptsize{robust standard error} 
& \scriptsize{\DTLfetch{rob_fixcost}{result}{robust se}{alpha}}
& \scriptsize{\DTLfetch{rob_fixcost}{result}{robust se}{delta}}
& \scriptsize{\DTLfetch{rob_fixcost}{result}{robust se}{gamma}}
& \scriptsize{\DTLfetch{rob_fixcost}{result}{robust se}{lambda}}
& \scriptsize{\DTLfetch{rob_fixcost}{result}{robust se}{mu}}\\
\midrule
 
\multicolumn{6}{l}{\scriptsize{estimation details: n = \DTLfetch{rob_fixcost}{result}{n}{alpha},  log-likelihood = \DTLfetch{rob_fixcost}{result}{log-likelihood}{alpha},
AIC = \DTLfetch{rob_fixcost}{result}{AIC}{alpha}, 
BIC = \DTLfetch{rob_fixcost}{result}{BIC}{alpha}, logit Fechner error}}\\

\bottomrule
\multicolumn{6}{l}{\scriptsize{Robust standard errors (SE) clustered at the individual level, \DTLfetch{rob_fixcost}{result}{cluster}{alpha} clusters}}
\end{tabularx}
\label{table:rob_fixcost}
\end{table}

\noindent{\bf Scaling utility of borrowed money:}
\begin{filecontents*}{RESULTS/rob_scaledloan.csv}
"","result","alpha","delta","gamma","lambda","mu"
"1","parameter",0.6421,0.034,0.9543,1.1111,0.4435
"2","robust se",0.0348,0.006,0.0108,0.0124,0.0399
"3","p-value",18.4688861237019,5.62887660754344,88.3420368293773,89.883981395036,11.1198562810935
"4","z",0.000000000000000000000000000000000000000000000000000000000000000000000000000367590879649198,0.0000000181387095289508,0,0,0.000000000000000000000000000100433647922126
"5","lowerlim",0.57,0.02,0.93,1.09,0.37
"6","upperlim",0.71,0.05,0.98,1.14,0.52
"7",NA,NA,NA,NA,NA,NA
"8",NA,1.95996398454005,1.95996398454005,1.95996398454005,1.95996398454005,1.95996398454005
"9",NA,0,0,0,0,0
"10","log-likelihood",-4112.31,NA,NA,NA,NA
"11","n",12240,NA,NA,NA,NA
"12","AIC",8234.62,NA,NA,NA,NA
"13","BIC",8271.68,NA,NA,NA,NA
"14","cluster",127,NA,NA,NA,NA
\end{filecontents*}
\DTLloaddb{rob_scaledloan}{RESULTS/rob_scaledloan.csv} 
As second alternative, we consider utility cost of borrowing that is dependent on timing and amount of loan receipt. Intuitively, one could think of the decision maker experiencing less pleasure from money that is actually borrowed compared to money from other sources. The cost of being in debt is thus defined as:

\begin{align}
   c(\boldsymbol{x})=(1-\gamma)\phi(t)v(x_t).
\end{align}

Interpretation of $\gamma$ is inverted as compared to the main specification, i.e. $\gamma<1$ corresponds to debt aversion and $\gamma>1$ to debt affinity. Estimation results are presented in Table~\ref{table:rob_scaledloan}. Again, except of $\gamma$, parameter estimates remain largely unchanged. The new debt aversion parameter is estimated at $\gamma=\DTLfetch{rob_scaledloan}{result}{parameter}{gamma}$, i.e. the average decision maker remains debt averse.

\begin{table}[t!]
\caption{Aggregate parameter estimates with gamma as scaling factor for utility from loan payments}
\begin{tabularx}{\textwidth}{ 
    c
    @{\hspace{36pt}}
  >{\centering\arraybackslash}X 
  >{\centering\arraybackslash}X 
  >{\centering\arraybackslash}X
  >{\centering\arraybackslash}X
  >{\centering\arraybackslash}X
  }
\toprule

& $\boldsymbol{\alpha}$
& $\boldsymbol{\delta}$ 
&  $\boldsymbol{\gamma}$ 
& $\boldsymbol{\lambda}$ 
& $\boldsymbol{\mu}$\\[-6pt]

& \scriptsize{risk aversion} 
& \scriptsize{discounting} 
& \scriptsize{debt aversion} 
& \scriptsize{loss aversion} 
& \scriptsize{Fechner error} \\
 \midrule

\scriptsize{\textbf{point estimate}} 
& \DTLfetch{rob_scaledloan}{result}{parameter}{alpha}
& \DTLfetch{rob_scaledloan}{result}{parameter}{delta}
& \DTLfetch{rob_scaledloan}{result}{parameter}{gamma}
& \DTLfetch{rob_scaledloan}{result}{parameter}{lambda}
& \DTLfetch{rob_scaledloan}{result}{parameter}{mu}\\[-6pt]

\scriptsize{95\% confidence interval} 
& \scriptsize{\DTLfetch{rob_scaledloan}{result}{lowerlim}{alpha}\,/\,\DTLfetch{rob_scaledloan}{result}{upperlim}{alpha}}
& \scriptsize{\DTLfetch{rob_scaledloan}{result}{lowerlim}{delta}\,/\,\DTLfetch{rob_scaledloan}{result}{upperlim}{delta}}
& \scriptsize{\DTLfetch{rob_scaledloan}{result}{lowerlim}{gamma}\,/\,\DTLfetch{rob_scaledloan}{result}{upperlim}{gamma}}
& \scriptsize{\DTLfetch{rob_scaledloan}{result}{lowerlim}{lambda}\,/\,\DTLfetch{rob_scaledloan}{result}{upperlim}{lambda}}
&\scriptsize{\DTLfetch{rob_scaledloan}{result}{lowerlim}{mu}\,/\,\DTLfetch{rob_scaledloan}{result}{upperlim}{mu}}\\[-9pt]

\scriptsize{robust standard error} 
& \scriptsize{\DTLfetch{rob_scaledloan}{result}{robust se}{alpha}}
& \scriptsize{\DTLfetch{rob_scaledloan}{result}{robust se}{delta}}
& \scriptsize{\DTLfetch{rob_scaledloan}{result}{robust se}{gamma}}
& \scriptsize{\DTLfetch{rob_scaledloan}{result}{robust se}{lambda}}
& \scriptsize{\DTLfetch{rob_scaledloan}{result}{robust se}{mu}}\\
\midrule
 
\multicolumn{6}{l}{\scriptsize{estimation details: n = \DTLfetch{rob_scaledloan}{result}{n}{alpha},  log-likelihood = \DTLfetch{rob_scaledloan}{result}{log-likelihood}{alpha},
AIC = \DTLfetch{rob_scaledloan}{result}{AIC}{alpha}, 
BIC = \DTLfetch{rob_scaledloan}{result}{BIC}{alpha}, logit Fechner error}}\\

\bottomrule
\multicolumn{6}{l}{\scriptsize{Robust standard errors (SE) clustered at the individual level, \DTLfetch{rob_scaledloan}{result}{cluster}{alpha} clusters}}
\end{tabularx}
\label{table:rob_scaledloan}
\end{table}

\noindent{\bf Assuming debt neutrality:} 
\begin{filecontents*}{RESULTS/rob_debtneutral.csv}
"","result","alpha","delta","gamma","lambda","mu"
"1","parameter",0.649,0.0165,1,1.1342,0.4496
"2","robust se",0.0346,0.0034,NA,0.0133,0.0403
"3","p-value",18.7608944905769,4.90015417573536,NA,85.1864608407867,11.1579750399314
"4","z",0.00000000000000000000000000000000000000000000000000000000000000000000000000000157722818880704,0.000000957614780054013,NA,0,0.0000000000000000000000000000654657159196463
"5","lowerlim",0.58,0.01,NA,1.11,0.37
"6","upperlim",0.72,0.02,NA,1.16,0.53
"7",NA,NA,NA,NA,NA,NA
"8",NA,1.95996398454005,1.95996398454005,1.95996398454005,1.95996398454005,1.95996398454005
"9",NA,0,0,0,0,0
"10","log-likelihood",-4128.24,NA,NA,NA,NA
"11","n",12240,NA,NA,NA,NA
"12","AIC",8264.48,NA,NA,NA,NA
"13","BIC",8294.13,NA,NA,NA,NA
"14","cluster",127,NA,NA,NA,NA
\end{filecontents*}
\DTLloaddb{rob_debtneutral}{RESULTS/rob_debtneutral.csv} 
Finally, we consider a model based on our main specification, but abstracting from debt aversion, i.e. assuming $\gamma=1$. Comparing our main specification to the debt neutral model, allows to scrutinize whether incorporating any form of cost of being indebted increases explanatory power.


\begin{table}[t!]
\caption{Aggregate parameter estimates assuming debt neutrality}
\begin{tabularx}{\textwidth}{ 
    c
    @{\hspace{36pt}}
  >{\centering\arraybackslash}X 
  >{\centering\arraybackslash}X 
  >{\centering\arraybackslash}X
  >{\centering\arraybackslash}X
  >{\centering\arraybackslash}X
  }
\toprule

& $\boldsymbol{\alpha}$ 
& $\boldsymbol{\delta}$ &  \textcolor{gray}{$\boldsymbol{\gamma}$} & $\boldsymbol{\lambda}$ & $\boldsymbol{\mu}$\\[-6pt]

& \scriptsize{risk aversion} 
& \scriptsize{discounting} 
& \textcolor{gray}{\scriptsize{debt aversion}} 
& \scriptsize{loss aversion} 
& \scriptsize{Fechner error} \\
 \midrule
 
\scriptsize{\textbf{point estimate}} 
& \DTLfetch{rob_debtneutral}{result}{parameter}{alpha}
& \DTLfetch{rob_debtneutral}{result}{parameter}{delta}
& 
& \DTLfetch{rob_debtneutral}{result}{parameter}{lambda}
& \DTLfetch{rob_debtneutral}{result}{parameter}{mu}\\[-6pt]

\scriptsize{95\% confidence interval} 
& \scriptsize{\DTLfetch{rob_debtneutral}{result}{lowerlim}{alpha}\,/\,\DTLfetch{rob_debtneutral}{result}{upperlim}{alpha}}
& \scriptsize{\DTLfetch{rob_debtneutral}{result}{lowerlim}{delta}\,/\,\DTLfetch{rob_debtneutral}{result}{upperlim}{delta}}
& 
& \scriptsize{\DTLfetch{rob_debtneutral}{result}{lowerlim}{lambda}\,/\,\DTLfetch{rob_debtneutral}{result}{upperlim}{lambda}}
&\scriptsize{\DTLfetch{rob_debtneutral}{result}{lowerlim}{mu}\,/\,\DTLfetch{rob_debtneutral}{result}{upperlim}{mu}}\\[-9pt]

\scriptsize{robust standard error} 
& \scriptsize{\DTLfetch{rob_debtneutral}{result}{robust se}{alpha}}
& \scriptsize{\DTLfetch{rob_debtneutral}{result}{robust se}{delta}}
& 
& \scriptsize{\DTLfetch{rob_debtneutral}{result}{robust se}{lambda}}
& \scriptsize{\DTLfetch{rob_debtneutral}{result}{robust se}{mu}}\\
\midrule
 
\multicolumn{6}{l}{\scriptsize{estimation details: n = \DTLfetch{rob_debtneutral}{result}{n}{alpha},  log-likelihood = \DTLfetch{rob_debtneutral}{result}{log-likelihood}{alpha},
AIC = \DTLfetch{rob_debtneutral}{result}{AIC}{alpha}, 
BIC = \DTLfetch{rob_debtneutral}{result}{BIC}{alpha}, logit Fechner error}}\\

\bottomrule
\multicolumn{6}{l}{\scriptsize{Robust standard errors (SE) clustered at the individual level, \DTLfetch{rob_debtneutral}{result}{cluster}{alpha} clusters}}
\end{tabularx}
\label{table:rob_debtneutral}
\end{table}

Estimation results are presented in Table~\ref{table:rob_debtneutral}. Comparing the information criteria AIC and BIC we observe that any model incorporating cost of debt is superior to the specification abstracting from debt attitudes, thus corroborating debt attitudes as distinct domain of individual preferences. This holds irrespective whether cost of debt is modelled as scaling disutility from repayments (main specification), fixed cost of being indebted or scaling utility from borrowed money.


\FloatBarrier 
\newpage
\subsection{Alternative Utility Specifications}
Besides cost of debt, we consider a wide range of alternatives to characteristics of our utility specification in general. First we allow certain preference parameters of our main specification to vary in the gain and loss domain.

\noindent{\bf Different utility curvature and time discounting in gains and losses:}
\begin{filecontents*}{RESULTS/rob_2alphas2deltas.csv}
"","result","alphagain","alphaloss","deltagain","deltaloss","gamma","lambda","mu"
"1","parameter",0.6425,0.7878,0.0369,0.0307,1.0485,1.0402,0.4484
"2","robust se",0.0346,0.0016,0.0061,0.0062,0.0109,0.0146,0.0402
"3","p-value",18.5777510131656,504.413416191409,6.03816347074287,4.92302471040346,95.8782814552748,71.2246859222853,11.1421368726643
"4","z",0.0000000000000000000000000000000000000000000000000000000000000000000000000000486469018124013,0,0.0000000015587811095504,0.000000852166872670848,0,0,0.0000000000000000000000000000782196980893615
"5","lowerlim",0.57,0.78,0.02,0.02,1.03,1.01,0.37
"6","upperlim",0.71,0.79,0.05,0.04,1.07,1.07,0.53
"7",NA,NA,NA,NA,NA,NA,NA,NA
"8",NA,1.95996398454005,1.95996398454005,1.95996398454005,1.95996398454005,1.95996398454005,1.95996398454005,1.95996398454005
"9",NA,0,0,0,0,0,0,0
"10","log-likelihood",-4106.69,NA,NA,NA,NA,NA,NA
"11","n",12240,NA,NA,NA,NA,NA,NA
"12","AIC",8227.39,NA,NA,NA,NA,NA,NA
"13","BIC",8279.27,NA,NA,NA,NA,NA,NA
"14","cluster",127,NA,NA,NA,NA,NA,NA
\end{filecontents*}
\DTLloaddb{rob_2alphas2deltas}{RESULTS/rob_2alphas2deltas.csv} 
Following the argumentation in
\cite{loewenstein1992} debt averse behavior may be explained by a combination of different utility curvatures as well as discount rates for gains and losses. In their model, the combination of a steeper value function for losses than for gains and a larger discount factor for gains than for losses will results in debt aversion. We test whether debt aversion is robust to this by adapting our main specification such that we allow distinct risk aversion and time discounting parameters for gains and losses: $\alpha^+,\alpha^-,\delta^+$ and $\delta^-$ respectively. Consequently, atemporal utility takes the form

\begin{align}
    u(x)=
    \begin{cases}
    \frac{(x+\varepsilon)^{1-\alpha^+}-\varepsilon^{1-\alpha^+}}{1-\alpha^+} & \mbox{if } x\geq 0\\
    \frac{(x+\varepsilon)^{1-\alpha^-}-\varepsilon^{1-\alpha^-}}{1-\alpha^-} & \mbox{if } x < 0,
    \end{cases}
\end{align}

\noindent and the discount function $\phi$ now also depends on the sign of $x$:

\begin{align}
    \phi(\tau, x) = 
\begin{cases}
\frac{1}{(1+\delta^+)^{\tau}}   & 
\mbox{ and } x\geq 0\\
\frac{1}{(1+\delta^-)^{\tau}}   & 
\mbox{ and } x<0.
\end{cases}
\end{align}

Estimation results are presented in Table~\ref{table:rob_2alphas2deltas}. In line with \cite{loewenstein1992}, we find $\alpha^+<\alpha^-$ and $\delta^+>\delta^-$, although only the difference in risk aversion is statistically significant ($P<0.01$). 
Importantly, the debt aversion parameter remains significantly larger than one and has a similar magnitude compared to the debt aversion parameter in our main specification, i.e. there is debt aversion that cannot be explained by differences in utility curvature and time discounting in the loss and gain domain.


\begin{table}[t!]
\caption{Aggregate parameter estimates with separate utility curvature  and time discounting in the gain and loss domain}
\begin{tabularx}{\textwidth}{ 
    c
    @{\hspace{36pt}}
  >{\centering\arraybackslash}X 
  >{\centering\arraybackslash}X 
  >{\centering\arraybackslash}X
  >{\centering\arraybackslash}X
  >{\centering\arraybackslash}X  >{\centering\arraybackslash}X  >{\centering\arraybackslash}X
  }
\toprule

& $\boldsymbol{\alpha^+}$ 
& $\boldsymbol{\alpha^-}$ 
& $\boldsymbol{\delta^+} $ 
&  $\boldsymbol{\delta^-} $ 
& $\boldsymbol{\gamma}$ 
& $\boldsymbol{\lambda}$ 
& $\boldsymbol{\mu}$ \\[-6pt]

& \scriptsize{RA gains} 
& \scriptsize{RA losses} 
& \scriptsize{TD gains} 
& \scriptsize{TD losses} 
& \makecell{\scriptsize{debt} \\[-12pt] \scriptsize{aversion}}
& \makecell{\scriptsize{loss} \\[-12pt] \scriptsize{aversion}}
& \makecell{\scriptsize{Fechner} \\[-12pt] \scriptsize{error}} \\
 \midrule
 
\makecell{\scriptsize{\textbf{point}} \\[-12pt] \scriptsize{\textbf{estimate}}} 
& \DTLfetch{rob_2alphas2deltas}{result}{parameter}{alphagain}
& \DTLfetch{rob_2alphas2deltas}{result}{parameter}{alphaloss}
& \DTLfetch{rob_2alphas2deltas}{result}{parameter}{deltagain}
& \DTLfetch{rob_2alphas2deltas}{result}{parameter}{deltaloss}
& \DTLfetch{rob_2alphas2deltas}{result}{parameter}{gamma}
& \DTLfetch{rob_2alphas2deltas}{result}{parameter}{lambda}
& \DTLfetch{rob_2alphas2deltas}{result}{parameter}{mu}\\[-9pt]

\scriptsize{95\% CI} 
& \scriptsize{\DTLfetch{rob_2alphas2deltas}{result}{lowerlim}{alphagain}\,/\,\DTLfetch{rob_2alphas2deltas}{result}{upperlim}{alphagain}}
& \scriptsize{\DTLfetch{rob_2alphas2deltas}{result}{lowerlim}{alphaloss}\,/\,\DTLfetch{rob_2alphas2deltas}{result}{upperlim}{alphaloss}}
& \scriptsize{\DTLfetch{rob_2alphas2deltas}{result}{lowerlim}{deltagain}\,/\,\DTLfetch{rob_2alphas2deltas}{result}{upperlim}{deltagain}}
& \scriptsize{\DTLfetch{rob_2alphas2deltas}{result}{lowerlim}{deltaloss}\,/\,\DTLfetch{rob_2alphas2deltas}{result}{upperlim}{deltaloss}}
& \scriptsize{\DTLfetch{rob_2alphas2deltas}{result}{lowerlim}{gamma}\,/\,\DTLfetch{rob_2alphas2deltas}{result}{upperlim}{gamma}}
& \scriptsize{\DTLfetch{rob_2alphas2deltas}{result}{lowerlim}{lambda}\,/\,\DTLfetch{rob_2alphas2deltas}{result}{upperlim}{lambda}}
&\scriptsize{\DTLfetch{rob_2alphas2deltas}{result}{lowerlim}{mu}\,/\,\DTLfetch{rob_2alphas2deltas}{result}{upperlim}{mu}}\\[-9pt]

\scriptsize{robust SE} 
& \scriptsize{\DTLfetch{rob_2alphas2deltas}{result}{robust se}{alphagain}}
& \scriptsize{\DTLfetch{rob_2alphas2deltas}{result}{robust se}{alphaloss}}
& \scriptsize{\DTLfetch{rob_2alphas2deltas}{result}{robust se}{deltagain}}
& \scriptsize{\DTLfetch{rob_2alphas2deltas}{result}{robust se}{deltaloss}}
& \scriptsize{\DTLfetch{rob_2alphas2deltas}{result}{robust se}{gamma}}
& \scriptsize{\DTLfetch{rob_2alphas2deltas}{result}{robust se}{lambda}}
& \scriptsize{\DTLfetch{rob_2alphas2deltas}{result}{robust se}{mu}}\\
\midrule
 
\multicolumn{8}{l}{\scriptsize{estimation details: n = \DTLfetch{rob_2alphas2deltas}{result}{n}{alphagain},  log-likelihood = \DTLfetch{rob_2alphas2deltas}{result}{log-likelihood}{alphagain},
AIC = \DTLfetch{rob_2alphas2deltas}{result}{AIC}{alphagain}, 
BIC = \DTLfetch{rob_2alphas2deltas}{result}{BIC}{alphagain}, logit Fechner error}}\\
\bottomrule
\multicolumn{8}{l}{\scriptsize{Robust standard errors (SE) clustered at the individual level, \DTLfetch{rob_2alphas2deltas}{result}{cluster}{alphagain} clusters}}
\end{tabularx}
\label{table:rob_2alphas2deltas}
\end{table}

\noindent{\bf Different present bias for gains and losses:} Besides utility curvature and time discounting, allowing for distinct present bias in gains and losses constitutes an appealing robustness check in the context of our experiment. In particular, present bias for payments in the gain domain, i.e. $\beta^+$ for $x>0$ also captures participants' trust in the experimenters' promises to affect payments in the future. Analogously, present bias in the loss domain, i.e. $\beta^-$ for $x<0$ captures participants' confidence that they will actually satisfy their own future payment obligations. If participants exhibit either mistrust or a lack of confidence in their own payment reliability, on average debt contracts would appear more and savings contracts less appealing to them. In this situation our estimate of debt aversion would actually be biased downwards. In this light, we consider different present bias in gains and losses through the discount function:

\begin{align}
    \phi(\tau, x) = 
\begin{cases}
1  & \mbox{if } \tau=0 \\
\frac{1}{1+\beta^+} \frac{1}{(1+\delta)^{\tau}}   & \mbox{if }\tau\neq0 \mbox{ and } x\geq 0\\
\frac{1}{1+\beta^-} \frac{1}{(1+\delta)^{\tau}}   & \mbox{if }\tau\neq0 \mbox{ and } x<0.
\end{cases}
\end{align}

Estimation results are presented in Table~\ref{table:rob_2betas}. We find neither evidence for  significant differences between $\beta^+$ and $\beta^-$, nor that any parameter of present bias is different from zero. Moreover, the debt aversion parameter remains larger than $1$, and has a similar magnitude as in our main specification. Summing up, debt aversion persists and is unlikely biased downwards due to mistrust in the experiment or participants' confidence in their payment morale.

\begin{filecontents*}{RESULTS/rob_2betas.csv}
"","result","alpha","betagain","betaloss","delta","gamma","lambda","mu"
"1","parameter",0.6431,-0.0008,-0.0018,0.0367,1.0537,1.1065,0.4484
"2","robust se",0.0345,0.0062,0.0055,0.0078,0.0144,0.0119,0.0402
"3","p-value",18.6382684188339,-0.124284325934108,-0.331270976285232,4.73351068191663,73.3643666771544,92.7416243224837,11.1538313574916
"4","z",0.0000000000000000000000000000000000000000000000000000000000000000000000000000157249078719799,0.90109015716884,0.740439812344519,0.0000022066934447977,0,0,0.0000000000000000000000000000685880825806119
"5","lowerlim",0.58,-0.01,-0.01,0.02,1.03,1.08,0.37
"6","upperlim",0.71,0.01,0.01,0.05,1.08,1.13,0.53
"7",NA,NA,NA,NA,NA,NA,NA,NA
"8",NA,1.95996398454005,1.95996398454005,1.95996398454005,1.95996398454005,1.95996398454005,1.95996398454005,1.95996398454005
"9",NA,0,0,0,0,0,0,0
"10","log-likelihood",-4107.84,NA,NA,NA,NA,NA,NA
"11","n",12240,NA,NA,NA,NA,NA,NA
"12","AIC",8229.69,NA,NA,NA,NA,NA,NA
"13","BIC",8281.57,NA,NA,NA,NA,NA,NA
"14","cluster",127,NA,NA,NA,NA,NA,NA
\end{filecontents*}
\DTLloaddb{rob_2betas}{RESULTS/rob_2betas.csv} 

\begin{table}[ht]
\caption{Aggregate parameter estimates allowing with separate present bias in the gain and loss domain}
\begin{tabularx}{\textwidth}{ 
    c
    @{\hspace{36pt}}
  >{\centering\arraybackslash}X 
  >{\centering\arraybackslash}X 
  >{\centering\arraybackslash}X
  >{\centering\arraybackslash}X
  >{\centering\arraybackslash}X  >{\centering\arraybackslash}X  >{\centering\arraybackslash}X
  }
\toprule
& $\boldsymbol{\alpha}$ & $\boldsymbol{\beta^+}$ & $\boldsymbol{\beta^-}$ & $\boldsymbol{\delta}$ &  $\boldsymbol{\gamma}$ & $\boldsymbol{\lambda}$ & $\boldsymbol{\mu}$ \\[-6pt]
& \makecell{\scriptsize{risk} \\[-12pt] \scriptsize{aversion}}
& \scriptsize{PB gains} 
& \scriptsize{PB losses} 
& \makecell{\scriptsize{time} \\[-12pt] \scriptsize{discounting}}
& \makecell{\scriptsize{debt} \\[-12pt] \scriptsize{aversion}} 
& \makecell{\scriptsize{loss} \\[-12pt] \scriptsize{aversion}}
& \makecell{\scriptsize{Fechner} \\[-12pt] \scriptsize{error}}\\
 \midrule
\makecell{\scriptsize{\textbf{point}} \\[-12pt] \scriptsize{\textbf{estimate}}} & \DTLfetch{rob_2betas}{result}{parameter}{alpha}
& \DTLfetch{rob_2betas}{result}{parameter}{betagain}
& \DTLfetch{rob_2betas}{result}{parameter}{betaloss}
& \DTLfetch{rob_2betas}{result}{parameter}{delta}
& \DTLfetch{rob_2betas}{result}{parameter}{gamma}
& \DTLfetch{rob_2betas}{result}{parameter}{lambda}
& \DTLfetch{rob_2betas}{result}{parameter}{mu}\\[-6pt]

\scriptsize{95\% CI} & \scriptsize{\DTLfetch{rob_2betas}{result}{lowerlim}{alpha}\,/\,\DTLfetch{rob_2betas}{result}{upperlim}{alpha}}
& \scriptsize{\DTLfetch{rob_2betas}{result}{lowerlim}{betagain}\,/\,\DTLfetch{rob_2betas}{result}{upperlim}{betagain}}
& \scriptsize{\DTLfetch{rob_2betas}{result}{lowerlim}{betaloss}\,/\,\DTLfetch{rob_2betas}{result}{upperlim}{betaloss}}
& \scriptsize{\DTLfetch{rob_2betas}{result}{lowerlim}{delta}\,/\,\DTLfetch{rob_2betas}{result}{upperlim}{delta}}
& \scriptsize{\DTLfetch{rob_2betas}{result}{lowerlim}{gamma}\,/\,\DTLfetch{rob_2betas}{result}{upperlim}{gamma}}
& \scriptsize{\DTLfetch{rob_2betas}{result}{lowerlim}{lambda}\,/\,\DTLfetch{rob_2betas}{result}{upperlim}{lambda}}
&\scriptsize{\DTLfetch{rob_2betas}{result}{lowerlim}{mu}\,/\,\DTLfetch{rob_2betas}{result}{upperlim}{mu}}\\[-9pt]

\scriptsize{robust SE} & \scriptsize{\DTLfetch{rob_2betas}{result}{robust se}{alpha}}
& \scriptsize{\DTLfetch{rob_2betas}{result}{robust se}{betagain}}
& \scriptsize{\DTLfetch{rob_2betas}{result}{robust se}{betaloss}}
& \scriptsize{\DTLfetch{rob_2betas}{result}{robust se}{delta}}
& \scriptsize{\DTLfetch{rob_2betas}{result}{robust se}{gamma}}
& \scriptsize{\DTLfetch{rob_2betas}{result}{robust se}{lambda}}
& \scriptsize{\DTLfetch{rob_2betas}{result}{robust se}{mu}}\\
\midrule
 
\multicolumn{8}{l}{\scriptsize{estimation details: n = \DTLfetch{rob_2betas}{result}{n}{alpha},  log-likelihood = \DTLfetch{rob_2betas}{result}{log-likelihood}{alpha},
AIC = \DTLfetch{rob_2betas}{result}{AIC}{alpha}, 
BIC = \DTLfetch{rob_2betas}{result}{BIC}{alpha}, logit Fechner error}}\\
\bottomrule
\multicolumn{8}{l}{\scriptsize{Robust standard errors (SE) clustered at the individual level, \DTLfetch{rob_2betas}{result}{cluster}{alpha} clusters}}
\end{tabularx}
\label{table:rob_2betas}
\end{table}

\noindent{\bf Risk neutrality:}
\begin{filecontents*}{RESULTS/rob_riskneutral.csv}
"","result","alpha","delta","gamma","lambda","mu"
"1","parameter","${\scriptstyle <}$.0001",0.1018,1.154,1.3533,3.0901
"2","robust se",NA,0.0175,0.033,0.0269,0.1468
"3","p-value",NA,5.8041292677698,34.93373911034,50.3431604034753,21.0428594435764
"4","z",NA,0.00000000647013471836593,0.00000000000000000000000000000000000000000000000000000000000000000000000000000000000000000000000000000000000000000000000000000000000000000000000000000000000000000000000000000000000000000000000000000000000000000000000000000000000000000000000000000000000000000000000000228666290699271,0,0.0000000000000000000000000000000000000000000000000000000000000000000000000000000000000000000000000265853165351598
"5","lowerlim",NA,0.07,1.09,1.3,2.8
"6","upperlim",NA,0.14,1.22,1.41,3.38
"7",NA,NA,NA,NA,NA,NA
"8",NA,"1.95996398454005",1.95996398454005,1.95996398454005,1.95996398454005,1.95996398454005
"9",NA,"0",0,0,0,0
"10","log-likelihood","-4813.36",NA,NA,NA,NA
"11","n","12240",NA,NA,NA,NA
"12","AIC","9634.73",NA,NA,NA,NA
"13","BIC","9664.38",NA,NA,NA,NA
"14","cluster","127",NA,NA,NA,NA
\end{filecontents*}
\DTLloaddb{rob_riskneutral}{RESULTS/rob_riskneutral.csv} 
As utility curvature, determined through $\alpha$, has a major effect on the size of the remaining preference parameters in our main specification we also consider an adaption abstracting from utility curvature. In particular, we consider the case of risk neutrality with $\alpha=0$.


\begin{table}[t!]
\caption{Aggregate parameter estimates assuming risk neutrality \label{tab:riskneutral}}
\begin{tabularx}{\textwidth}{ 
    c
    @{\hspace{36pt}}
  >{\centering\arraybackslash}X 
  >{\centering\arraybackslash}X 
  >{\centering\arraybackslash}X
  >{\centering\arraybackslash}X
  >{\centering\arraybackslash}X
  }
\toprule

& \textcolor{gray}{$\boldsymbol{\alpha}$} & $\boldsymbol{\delta}$ &  $\boldsymbol{\gamma}$ & $\boldsymbol{\lambda}$ & $\boldsymbol{\mu}$\\[-6pt]

& \textcolor{gray}{\scriptsize{risk aversion}} 
& \scriptsize{discounting} 
& \scriptsize{debt aversion} 
& \scriptsize{loss aversion} 
& \scriptsize{Fechner error} \\
 \midrule
 
\scriptsize{\textbf{point estimate}} 
&
& \DTLfetch{rob_riskneutral}{result}{parameter}{delta}
& \DTLfetch{rob_riskneutral}{result}{parameter}{gamma}
& \DTLfetch{rob_riskneutral}{result}{parameter}{lambda}
& \DTLfetch{rob_riskneutral}{result}{parameter}{mu}\\[-6pt]

\scriptsize{95\% confidence interval} 
& 
& \scriptsize{\DTLfetch{rob_riskneutral}{result}{lowerlim}{delta}\,/\,\DTLfetch{rob_riskneutral}{result}{upperlim}{delta}}
& \scriptsize{\DTLfetch{rob_riskneutral}{result}{lowerlim}{gamma}\,/\,\DTLfetch{rob_riskneutral}{result}{upperlim}{gamma}}
& \scriptsize{\DTLfetch{rob_riskneutral}{result}{lowerlim}{lambda}\,/\,\DTLfetch{rob_riskneutral}{result}{upperlim}{lambda}}
&\scriptsize{\DTLfetch{rob_riskneutral}{result}{lowerlim}{mu}\,/\,\DTLfetch{rob_riskneutral}{result}{upperlim}{mu}}\\[-9pt]

\scriptsize{robust standard error} 
& 
& \scriptsize{\DTLfetch{rob_riskneutral}{result}{robust se}{delta}}
& \scriptsize{\DTLfetch{rob_riskneutral}{result}{robust se}{gamma}}
& \scriptsize{\DTLfetch{rob_riskneutral}{result}{robust se}{lambda}}
& \scriptsize{\DTLfetch{rob_riskneutral}{result}{robust se}{mu}}\\
\midrule
 
\multicolumn{6}{l}{\scriptsize{estimation details: n = \DTLfetch{rob_riskneutral}{result}{n}{alpha},  log-likelihood = \DTLfetch{rob_riskneutral}{result}{log-likelihood}{alpha},
AIC = \DTLfetch{rob_riskneutral}{result}{AIC}{alpha}, 
BIC = \DTLfetch{rob_riskneutral}{result}{BIC}{alpha}, logit Fechner error}}\\
\bottomrule
\multicolumn{6}{l}{\scriptsize{Robust standard errors (SE) clustered at the individual level, \DTLfetch{rob_riskneutral}{result}{cluster}{alpha} clusters}}
\end{tabularx}
\label{table:rob_riskneutral}
\end{table}

Estimation results are presented in Table~\ref{table:rob_riskneutral}. Debt aversion persists, and appears higher than in our main specification allowing for risk aversion. As expected, also the remaining parameter estimates change substantially, which makes intuitive sense due to the different shape of the atemporal utility function $u(x)$ when assuming $\alpha=0$.

\noindent{\bf CARA utility:} Our main specification assumes atemporal utility to be in line with constant relative risk aversion (CRRA). We examine the robustness of our findings, towards alternative forms of atemporal utility. In particular, we consider exponential utility characterized by constant absolute risk aversion (CARA):

\begin{align}
    u(x)=
    \frac{1-e^{- \varphi x}}{\varphi}
\end{align}

Here, $\varphi$ is the parameter of absolute risk aversion. Estimation results are presented in Table~\ref{table:rob_cara}. Again as in the previous robustness check, this adaption leads to a substantial change in the shape of the atemporal utility function $u(x)$. As a consequence, also parameter estimates beyond $\varphi$ change considerably. Debt aversion, however, persists.

\begin{filecontents*}{RESULTS/rob_cara.csv}
"","result","alpha","delta","gamma","lambda","mu"
"1","parameter","0.0112",0.0898,1.1358,1.3055,2.4346
"2","robust se","0.0042",0.0141,0.027,0.0252,0.1805
"3","p-value","2.68888684767646",6.36186680543448,42.0580928333374,51.791639404059,13.4901732366525
"4","z","0.00716907154112477",0.000000000199316251619629,0,0,0.0000000000000000000000000000000000000000178680216534737
"5","lowerlim","${\scriptscriptstyle <}$.01",0.06,1.08,1.26,2.08
"6","upperlim","0.02",0.12,1.19,1.35,2.79
"7",NA,NA,NA,NA,NA,NA
"8",NA,"1.95996398454005",1.95996398454005,1.95996398454005,1.95996398454005,1.95996398454005
"9",NA,"0",0,0,0,0
"10","log-likelihood","-4182.05",NA,NA,NA,NA
"11","n","12240",NA,NA,NA,NA
"12","AIC","8374.1",NA,NA,NA,NA
"13","BIC","8411.16",NA,NA,NA,NA
"14","cluster","127",NA,NA,NA,NA
\end{filecontents*}
\DTLloaddb{rob_cara}{RESULTS/rob_cara.csv} 

\begin{table}[ht]
\caption{Aggregate parameter estimates with CARA utility.}
\begin{tabularx}{\textwidth}{ 
    c
    @{\hspace{36pt}}
  >{\centering\arraybackslash}X 
  >{\centering\arraybackslash}X 
  >{\centering\arraybackslash}X
  >{\centering\arraybackslash}X
  >{\centering\arraybackslash}X
  }
\toprule
& $\boldsymbol{\varphi}$ 
& $\boldsymbol{\delta}$ 
&  $\boldsymbol{\gamma}$ 
& $\boldsymbol{\lambda}$ 
& $\boldsymbol{\mu}$\\[-6pt]

& \scriptsize{risk aversion} 
& \scriptsize{discounting} 
& \scriptsize{debt aversion} 
& \scriptsize{loss aversion} 
& \scriptsize{Fechner error} \\
 \midrule
 
\scriptsize{\textbf{point estimate}} 
& \DTLfetch{rob_cara}{result}{parameter}{alpha}
& \DTLfetch{rob_cara}{result}{parameter}{delta}
& \DTLfetch{rob_cara}{result}{parameter}{gamma}
& \DTLfetch{rob_cara}{result}{parameter}{lambda}
& \DTLfetch{rob_cara}{result}{parameter}{mu}\\[-6pt]

\scriptsize{95\% confidence interval} 
& \scriptsize{\DTLfetch{rob_cara}{result}{lowerlim}{alpha}\,/\,\DTLfetch{rob_cara}{result}{upperlim}{alpha}}
& \scriptsize{\DTLfetch{rob_cara}{result}{lowerlim}{delta}\,/\,\DTLfetch{rob_cara}{result}{upperlim}{delta}}
& \scriptsize{\DTLfetch{rob_cara}{result}{lowerlim}{gamma}\,/\,\DTLfetch{rob_cara}{result}{upperlim}{gamma}}
& \scriptsize{\DTLfetch{rob_cara}{result}{lowerlim}{lambda}\,/\,\DTLfetch{rob_cara}{result}{upperlim}{lambda}}
&\scriptsize{\DTLfetch{rob_cara}{result}{lowerlim}{mu}\,/\,\DTLfetch{rob_cara}{result}{upperlim}{mu}}\\[-9pt]

\scriptsize{robust standard error} 
& \scriptsize{\DTLfetch{rob_cara}{result}{robust se}{alpha}}
& \scriptsize{\DTLfetch{rob_cara}{result}{robust se}{delta}}
& \scriptsize{\DTLfetch{rob_cara}{result}{robust se}{gamma}}
& \scriptsize{\DTLfetch{rob_cara}{result}{robust se}{lambda}}
& \scriptsize{\DTLfetch{rob_cara}{result}{robust se}{mu}}\\
\midrule
 
\multicolumn{6}{l}{\scriptsize{estimation details:  n = \DTLfetch{rob_cara}{result}{n}{alpha},  log-likelihood = \DTLfetch{rob_cara}{result}{log-likelihood}{alpha},
AIC = \DTLfetch{rob_cara}{result}{AIC}{alpha}, 
BIC = \DTLfetch{rob_cara}{result}{BIC}{alpha}, logit Fechner error}}\\
\bottomrule
\multicolumn{6}{l}{\scriptsize{Robust standard errors (SE) clustered at the individual level, \DTLfetch{rob_cara}{result}{cluster}{alpha} clusters}}
\end{tabularx}
\label{table:rob_cara}
\end{table}

\noindent{\bf No $\varepsilon$-transformation:} Lastly, in our main specification we consider CRRA utility including an $\varepsilon$-transformation, because of its beneficial properties for estimation. As a final alteration of the utility specification we consider a robustness check without the $\varepsilon$-transformation, i.e. atemporal utility takes the form:

\begin{align}
    u(x)=\frac{x^{1-\alpha}}{1-\alpha}
\end{align}

Estimation results are presented in Table~\ref{table:rob_noepsilon}. Parameter estimates remain largely unchanged compared to the main specification. Debt aversion remains robust.

\begin{filecontents*}{RESULTS/rob_noepsilon.csv}
"","result","alpha","delta","gamma","lambda","mu"
"1","parameter",0.6429,0.0355,1.0528,1.106,0.4487
"2","robust se",0.0344,0.006,0.0111,0.0121,0.0402
"3","p-value",18.6692889964921,5.92065933380389,94.4903656229843,91.4300640409961,11.1693716839664
"4","z",0.00000000000000000000000000000000000000000000000000000000000000000000000000000880166942991735,0.00000000320653461082401,0,0,0.0000000000000000000000000000575867822252939
"5","lowerlim",0.58,0.02,1.03,1.08,0.37
"6","upperlim",0.71,0.05,1.07,1.13,0.53
"7",NA,NA,NA,NA,NA,NA
"8",NA,1.95996398454005,1.95996398454005,1.95996398454005,1.95996398454005,1.95996398454005
"9",NA,0,0,0,0,0
"10","log-likelihood",-4108,NA,NA,NA,NA
"11","n",12240,NA,NA,NA,NA
"12","AIC",8226,NA,NA,NA,NA
"13","BIC",8263.06,NA,NA,NA,NA
"14","cluster",127,NA,NA,NA,NA
\end{filecontents*}
\DTLloaddb{rob_noepsilon}{RESULTS/rob_noepsilon.csv} 

\begin{table}[t!]
\caption{Aggregate parameter estimates estimated without $\varepsilon$-transformation.}
\begin{tabularx}{\textwidth}{ 
    c
    @{\hspace{36pt}}
  >{\centering\arraybackslash}X 
  >{\centering\arraybackslash}X 
  >{\centering\arraybackslash}X
  >{\centering\arraybackslash}X
  >{\centering\arraybackslash}X
  }
\toprule

& $\boldsymbol{\alpha}$ 
& $\boldsymbol{\delta}$ 
&  $\boldsymbol{\gamma}$ 
& $\boldsymbol{\lambda}$ 
& $\boldsymbol{\mu}$\\[-6pt]

& \scriptsize{risk aversion} 
& \scriptsize{discounting} 
& \scriptsize{debt aversion} 
& \scriptsize{loss aversion} 
& \scriptsize{Fechner error} \\
 \midrule
 
\scriptsize{\textbf{point estimate}} 
& \DTLfetch{rob_noepsilon}{result}{parameter}{alpha}
& \DTLfetch{rob_noepsilon}{result}{parameter}{delta}
& \DTLfetch{rob_noepsilon}{result}{parameter}{gamma}
& \DTLfetch{rob_noepsilon}{result}{parameter}{lambda}
& \DTLfetch{rob_noepsilon}{result}{parameter}{mu}\\[-6pt]

\scriptsize{95\% confidence interval} 
& \scriptsize{\DTLfetch{rob_noepsilon}{result}{lowerlim}{alpha}\,/\,\DTLfetch{rob_noepsilon}{result}{upperlim}{alpha}}
& \scriptsize{\DTLfetch{rob_noepsilon}{result}{lowerlim}{delta}\,/\,\DTLfetch{rob_noepsilon}{result}{upperlim}{delta}}
& \scriptsize{\DTLfetch{rob_noepsilon}{result}{lowerlim}{gamma}\,/\,\DTLfetch{rob_noepsilon}{result}{upperlim}{gamma}}
& \scriptsize{\DTLfetch{rob_noepsilon}{result}{lowerlim}{lambda}\,/\,\DTLfetch{rob_noepsilon}{result}{upperlim}{lambda}}
&\scriptsize{\DTLfetch{rob_noepsilon}{result}{lowerlim}{mu}\,/\,\DTLfetch{rob_noepsilon}{result}{upperlim}{mu}}\\[-9pt]

\scriptsize{robust standard error} 
& \scriptsize{\DTLfetch{rob_noepsilon}{result}{robust se}{alpha}}
& \scriptsize{\DTLfetch{rob_noepsilon}{result}{robust se}{delta}}
& \scriptsize{\DTLfetch{rob_noepsilon}{result}{robust se}{gamma}}
& \scriptsize{\DTLfetch{rob_noepsilon}{result}{robust se}{lambda}}
& \scriptsize{\DTLfetch{rob_noepsilon}{result}{robust se}{mu}}\\
\midrule
 
\multicolumn{6}{l}{\scriptsize{estimation details: n = \DTLfetch{rob_noepsilon}{result}{n}{alpha},  log-likelihood = \DTLfetch{rob_noepsilon}{result}{log-likelihood}{alpha},
AIC = \DTLfetch{rob_noepsilon}{result}{AIC}{alpha}, 
BIC = \DTLfetch{rob_noepsilon}{result}{BIC}{alpha}, logit Fechner error}}\\
\bottomrule
\multicolumn{6}{l}{\scriptsize{Robust standard errors (SE) clustered at the individual level, \DTLfetch{rob_noepsilon}{result}{cluster}{alpha} clusters}}
\end{tabularx}
\label{table:rob_noepsilon}
\end{table}

\subsection{Alternative Error Structures}
\noindent In line with \cite{drichoutis2014}, we acknowledge that parameter estimates may depend on assumptions about the decision error process. Therefore, we employ three alternative error structures as robustness checks.

\noindent{\bf Probit-link Fechner error:} First, we consider a Fechner error with probit link instead of logit link as in our main specification. Technically, $F(\xi)$ is no longer a standard logistic distribution function but the standard normal distribution function, i.e $F(\xi)=\Phi(\xi)$, where $\Phi$ represents the standard normal CDF. Estimation results are presented in Table~\ref{table:rob_probit}. Parameter estimates, except for the Fechner error term, remain largely unchanged compared to the main specification. Debt aversion is robust.

\begin{filecontents*}{RESULTS/rob_probit.csv}
"","result","alpha","delta","gamma","lambda","mu"
"1","parameter",0.6411,0.0373,1.0559,1.1099,0.7963
"2","robust se",0.0327,0.0063,0.0118,0.0117,0.0671
"3","p-value",19.6273730766445,5.96648822275726,89.5540898835895,94.9726186743865,11.8589712168529
"4","z",0.0000000000000000000000000000000000000000000000000000000000000000000000000000000000000902623493769632,0.00000000242414226924827,0,0,0.0000000000000000000000000000000193335995772879
"5","lowerlim",0.58,0.03,1.03,1.09,0.66
"6","upperlim",0.71,0.05,1.08,1.13,0.93
"7",NA,NA,NA,NA,NA,NA
"8",NA,1.95996398454005,1.95996398454005,1.95996398454005,1.95996398454005,1.95996398454005
"9",NA,0,0,0,0,0
"10","log-likelihood",-4109.29,NA,NA,NA,NA
"11","n",12240,NA,NA,NA,NA
"12","AIC",8228.58,NA,NA,NA,NA
"13","BIC",8265.65,NA,NA,NA,NA
"14","cluster",127,NA,NA,NA,NA
\end{filecontents*}
\DTLloaddb{rob_probit}{RESULTS/rob_probit.csv} 

\begin{table}[t!]
\caption{Aggregate parameter estimates with probit instead Fechner error.}
\begin{tabularx}{\textwidth}{ 
    c
    @{\hspace{36pt}}
  >{\centering\arraybackslash}X 
  >{\centering\arraybackslash}X 
  >{\centering\arraybackslash}X
  >{\centering\arraybackslash}X
  >{\centering\arraybackslash}X
  }
\toprule

& $\boldsymbol{\alpha}$ 
& $\boldsymbol{\delta}$ 
&  $\boldsymbol{\gamma}$ 
& $\boldsymbol{\lambda}$ 
& $\boldsymbol{\mu}$\\[-6pt]

& \scriptsize{risk aversion} 
& \scriptsize{discounting} 
& \scriptsize{debt aversion} 
& \scriptsize{loss aversion} 
& \scriptsize{Fechner error} \\

\midrule
\scriptsize{\textbf{point estimate}} 
& \DTLfetch{rob_probit}{result}{parameter}{alpha}
& \DTLfetch{rob_probit}{result}{parameter}{delta}
& \DTLfetch{rob_probit}{result}{parameter}{gamma}
& \DTLfetch{rob_probit}{result}{parameter}{lambda}
& \DTLfetch{rob_probit}{result}{parameter}{mu}\\[-6pt]

\scriptsize{95\% confidence interval} 
& \scriptsize{\DTLfetch{rob_probit}{result}{lowerlim}{alpha}\,/\,\DTLfetch{rob_probit}{result}{upperlim}{alpha}}
& \scriptsize{\DTLfetch{rob_probit}{result}{lowerlim}{delta}\,/\,\DTLfetch{rob_probit}{result}{upperlim}{delta}}
& \scriptsize{\DTLfetch{rob_probit}{result}{lowerlim}{gamma}\,/\,\DTLfetch{rob_probit}{result}{upperlim}{gamma}}
& \scriptsize{\DTLfetch{rob_probit}{result}{lowerlim}{lambda}\,/\,\DTLfetch{rob_probit}{result}{upperlim}{lambda}}
&\scriptsize{\DTLfetch{rob_probit}{result}{lowerlim}{mu}\,/\,\DTLfetch{rob_probit}{result}{upperlim}{mu}}\\[-9pt]

\scriptsize{robust standard error} 
& \scriptsize{\DTLfetch{rob_probit}{result}{robust se}{alpha}}
& \scriptsize{\DTLfetch{rob_probit}{result}{robust se}{delta}}
& \scriptsize{\DTLfetch{rob_probit}{result}{robust se}{gamma}}
& \scriptsize{\DTLfetch{rob_probit}{result}{robust se}{lambda}}
& \scriptsize{\DTLfetch{rob_probit}{result}{robust se}{mu}}\\
\midrule
 
\multicolumn{6}{l}{\scriptsize{estimation details: n = \DTLfetch{rob_probit}{result}{n}{alpha},  log-likelihood = \DTLfetch{rob_probit}{result}{log-likelihood}{alpha},
AIC = \DTLfetch{rob_probit}{result}{AIC}{alpha}, 
BIC = \DTLfetch{rob_probit}{result}{BIC}{alpha}, \textbf{probit Fechner error}}}\\

\bottomrule
\multicolumn{6}{l}{\scriptsize{Robust standard errors (SE) clustered at the individual level, \DTLfetch{rob_probit}{result}{cluster}{alpha} clusters}}
\end{tabularx}
\label{table:rob_probit}
\end{table}

\noindent{\bf Additional Tremble error:} Second, we introduce a second type of error aside the logit Fechner error of our main specification. In particular, we consider that decision makers may make a tremble error, i.e. randomize choice between both options with some probability $|\kappa|$  as e.g. in \cite{Andersson2020}. Consequently, the probability of observing choice B is given by:

\begin{align}
\label{eq:cond_prob_tremble}
P^B(\theta') & = (1-|\kappa|)F\left(\frac{U(X^B,p)-U(X^A,p)}{\mu}\right)+\frac{|\kappa|}{2} = (1-|\kappa|)F(\Delta U(\theta))+\frac{|\kappa|}{2},
\end{align}

where $\theta'=(\alpha,\delta,\gamma,\lambda,\mu,\kappa)$. The corresponding log-likelihood function writes as:


\begin{align}
\label{eq:logL1_tremble}
ln \ L(\theta')= \sum_i\sum_j\left[ln\left(P^B(\theta')\right)c_{ij}+ln\left(1-P^B(\theta')\right)(1-c_{ij})\right]
\end{align}

Intuitively, the error parameter can be interpreted as follows: for $|\kappa| \rightarrow 0$ the tremble error has no effect on choices, while for $|\kappa| \rightarrow 1$ choices approach uniform randomization.

Estimation results are presented in Table~\ref{table:rob_logit2errors}. The estimated tremble error probability $|\kappa|$ is statistically indistinguishable from $0$, and the remaining parameter estimates are virtually unchanged compared to the main specification. Debt aversion also persists when considering an additional tremble error.

\begin{filecontents*}{RESULTS/rob_logit2errors.csv}
"","result","alpha","delta","gamma","lambda","mu","kappa"
"1","parameter",0.643,0.036,1.0535,1.1074,0.4484,"\ensuremath{-0.000\cdot}"
"2","robust se",0.0345,0.006,0.0112,0.0118,0.0402,"\ensuremath{0.000\cdot}"
"3","p-value",18.6449626839952,6.00345541538156,93.9752974765787,93.7220132004413,11.1450069093351,"-0.0805842796093929"
"4","z",0.000000000000000000000000000000000000000000000000000000000000000000000000000013875115177259,0.00000000193161825410223,0,0,0.0000000000000000000000000000757384145150522,"0.935772568595555"
"5","lowerlim",0.58,0.02,1.03,1.08,0.37,"\ensuremath{-0.0\cdot}"
"6","upperlim",0.71,0.05,1.08,1.13,0.53," \ensuremath{0.0\cdot}"
"7",NA,NA,NA,NA,NA,NA,NA
"8",NA,1.95996398454005,1.95996398454005,1.95996398454005,1.95996398454005,1.95996398454005,"1.95996398454005"
"9",NA,0,0,0,0,0,"0"
"10","log-likelihood",-4107.91,NA,NA,NA,NA,NA
"11","n",12240,NA,NA,NA,NA,NA
"12","AIC",8225.81,NA,NA,NA,NA,NA
"13","BIC",8262.88,NA,NA,NA,NA,NA
"14","cluster",127,NA,NA,NA,NA,NA
\end{filecontents*}
\DTLloaddb{rob_logit2errors}{RESULTS/rob_logit2errors.csv} 

\begin{table}[t!]
\caption{Aggregate parameter estimates with tremble and logit Fechner error.}
\begin{tabularx}{\textwidth}{ 
    c
    @{\hspace{36pt}}
  >{\centering\arraybackslash}X 
  >{\centering\arraybackslash}X 
  >{\centering\arraybackslash}X
  >{\centering\arraybackslash}X
  >{\centering\arraybackslash}X  >{\centering\arraybackslash}X
  }
\toprule

& $\boldsymbol{\alpha}$ 
& $\boldsymbol{\delta}$ 
&  $\boldsymbol{\gamma}$ 
& $\boldsymbol{\lambda}$ 
& $\boldsymbol{\mu}$ 
& $\boldsymbol{\kappa}$ \\[-6pt]

& \scriptsize{risk aversion} 
& \scriptsize{discounting} 
& \scriptsize{debt aversion} 
& \scriptsize{loss aversion} 
& \scriptsize{Fechner error} 
& \scriptsize{tremble error} \\
 \midrule
 
\makecell{\scriptsize{\textbf{point}} \\[-12pt] \scriptsize{\textbf{estimate}}} 
& \DTLfetch{rob_logit2errors}{result}{parameter}{alpha}
& \DTLfetch{rob_logit2errors}{result}{parameter}{delta}
& \DTLfetch{rob_logit2errors}{result}{parameter}{gamma}
& \DTLfetch{rob_logit2errors}{result}{parameter}{lambda}
& \DTLfetch{rob_logit2errors}{result}{parameter}{mu} 
& \DTLfetch{rob_logit2errors}{result}{parameter}{kappa} \\[-6pt]

\scriptsize{95\% CI} 
& \scriptsize{\DTLfetch{rob_logit2errors}{result}{lowerlim}{alpha}\,/\,\DTLfetch{rob_logit2errors}{result}{upperlim}{alpha}}
& \scriptsize{\DTLfetch{rob_logit2errors}{result}{lowerlim}{delta}\,/\,\DTLfetch{rob_logit2errors}{result}{upperlim}{delta}}
& \scriptsize{\DTLfetch{rob_logit2errors}{result}{lowerlim}{gamma}\,/\,\DTLfetch{rob_logit2errors}{result}{upperlim}{gamma}}
& \scriptsize{\DTLfetch{rob_logit2errors}{result}{lowerlim}{lambda}\,/\,\DTLfetch{rob_logit2errors}{result}{upperlim}{lambda}}
&\scriptsize{\DTLfetch{rob_logit2errors}{result}{lowerlim}{mu}\,/\,\DTLfetch{rob_logit2errors}{result}{upperlim}{mu}} 
&\scriptsize{\DTLfetch{rob_logit2errors}{result}{lowerlim}{kappa}\,/\,\DTLfetch{rob_logit2errors}{result}{upperlim}{kappa}}\\[-9pt]

\scriptsize{robust SE} 
& \scriptsize{\DTLfetch{rob_logit2errors}{result}{robust se}{alpha}}
& \scriptsize{\DTLfetch{rob_logit2errors}{result}{robust se}{delta}}
& \scriptsize{\DTLfetch{rob_logit2errors}{result}{robust se}{gamma}}
& \scriptsize{\DTLfetch{rob_logit2errors}{result}{robust se}{lambda}}
& \scriptsize{\DTLfetch{rob_logit2errors}{result}{robust se}{mu}}
& \scriptsize{\DTLfetch{rob_logit2errors}{result}{robust se}{kappa}}\\
\midrule
 
\multicolumn{7}{l}{\scriptsize{estimation details: n = \DTLfetch{rob_logit2errors}{result}{n}{alpha},  log-likelihood = \DTLfetch{rob_logit2errors}{result}{log-likelihood}{alpha},
AIC = \DTLfetch{rob_logit2errors}{result}{AIC}{alpha}, 
BIC = \DTLfetch{rob_logit2errors}{result}{BIC}{alpha},  \textbf{logit FE + tremble error}}}\\[-9pt]

\multicolumn{7}{l}{\scriptsize{$-0.0 \cdot$ (resp. $0.0 \cdot$) is between $0$ and $-0.01$ resp. ($0.01$); \  $-0.000 \cdot$ (resp. $0.000 \cdot$) is between $0$ and $-0.0001$ resp. ($0.0001$)}}\\

\bottomrule
\multicolumn{7}{l}{\scriptsize{Robust standard errors (SE) clustered at the individual level,  \DTLfetch{rob_logit2errors}{result}{cluster}{alpha} clusters}}

\end{tabularx}
\label{table:rob_logit2errors}
\end{table}

\noindent{\bf Multiple Fechner errors:} Third, we consider a specification with distinct Fechner error terms for all domains, as represented through our set of seven different MPLs. Accordingly, we end up with seven Fechner error terms $\mu_1$, ..., $\mu_7$ for MPL1 to MPL7, respectively. Estimation results are presented in Appendix~Table~\ref{table:rob_7errors}. While we observe significant variation in the Fechner errors across some MPLs, the remaining parameter estimates are virtually unchanged compared to the main specification. Debt aversion persists.

\begin{filecontents*}{RESULTS/rob_7errors.csv}
"","result","alpha","delta","gamma","lambda","mu1","mu2","mu3","mu4","mu5","mu6","mu7"
"1","parameter",0.6553,0.0326,1.0442,1.1029,0.2169,0.4142,0.5186,0.3886,0.3905,0.48,0.4404
"2","robust se",0.0379,0.0054,0.0097,0.014,0.0339,0.0433,0.0478,0.0485,0.0485,0.065,0.0564
"3","p-value",17.2815787454528,6.00700269909636,107.234052210496,78.6474028183151,6.39100522116363,9.55906514585298,10.8441223385214,8.00480259436283,8.05408704281826,7.38445518983422,7.80715203133904
"4","z",0.000000000000000000000000000000000000000000000000000000000000000000647511259938652,0.00000000188984371621127,0,0,0.0000000001647987544635,0.00000000000000000000118825163642618,0.00000000000000000000000000212667167539225,0.0000000000000011965846950478,0.000000000000000800742219764035,0.000000000000153078898400869,0.00000000000000584946943698787
"5","lowerlim",0.58,0.02,1.03,1.08,0.15,0.33,0.42,0.29,0.3,0.35,0.33
"6","upperlim",0.73,0.04,1.06,1.13,0.28,0.5,0.61,0.48,0.49,0.61,0.55
"7",NA,NA,NA,NA,NA,NA,NA,NA,NA,NA,NA,NA
"8",NA,1.95996398454005,1.95996398454005,1.95996398454005,1.95996398454005,1.95996398454005,1.95996398454005,1.95996398454005,1.95996398454005,1.95996398454005,1.95996398454005,1.95996398454005
"9",NA,0,0,0,0,0,0,0,0,0,0,0
"10","log-likelihood",-3676.81,NA,NA,NA,NA,NA,NA,NA,NA,NA,NA
"11","n",11430,NA,NA,NA,NA,NA,NA,NA,NA,NA,NA
"12","AIC",7375.62,NA,NA,NA,NA,NA,NA,NA,NA,NA,NA
"13","BIC",7456.41,NA,NA,NA,NA,NA,NA,NA,NA,NA,NA
"14","cluster",127,NA,NA,NA,NA,NA,NA,NA,NA,NA,NA
\end{filecontents*}
\DTLloaddb{rob_7errors}{RESULTS/rob_7errors.csv} 

\begin{table}[t!]
\caption{Aggregate parameter estimates with multiple Fechner errors}
\begin{tabularx}{\textwidth}{ 
    c
    @{\hspace{36pt}}
  >{\centering\arraybackslash}X 
  >{\centering\arraybackslash}X 
  >{\centering\arraybackslash}X
  >{\centering\arraybackslash}X
  >{\centering\arraybackslash}X  >{\centering\arraybackslash}X
  }
\toprule
& $\boldsymbol{\alpha}$ 
& $\boldsymbol{\delta}$ 
& $\boldsymbol{\gamma}$ 
& $\boldsymbol{\lambda}$ 
&
& $\boldsymbol{\mu_1}$ \\[-6pt]

& \scriptsize{risk aversion}
& \scriptsize{discounting} 
& \scriptsize{debt aversion} 
& \scriptsize{loss aversion} 
&
& \scriptsize{FE MPL1} \\
 \midrule
 
\makecell{\scriptsize{\textbf{point}} \\[-12pt] \scriptsize{\textbf{estimate}}} 
& \DTLfetch{rob_7errors}{result}{parameter}{alpha}
& \DTLfetch{rob_7errors}{result}{parameter}{delta}
& \DTLfetch{rob_7errors}{result}{parameter}{gamma}
& \DTLfetch{rob_7errors}{result}{parameter}{lambda}
&
& \DTLfetch{rob_7errors}{result}{parameter}{mu1} \\[-6pt]

\scriptsize{95\% CI} 
& \scriptsize{\DTLfetch{rob_7errors}{result}{lowerlim}{alpha}\,/\,\DTLfetch{rob_7errors}{result}{upperlim}{alpha}}
& \scriptsize{\DTLfetch{rob_7errors}{result}{lowerlim}{delta}\,/\,\DTLfetch{rob_7errors}{result}{upperlim}{delta}}
& \scriptsize{\DTLfetch{rob_7errors}{result}{lowerlim}{gamma}\,/\,\DTLfetch{rob_7errors}{result}{upperlim}{gamma}}
& \scriptsize{\DTLfetch{rob_7errors}{result}{lowerlim}{lambda}\,/\,\DTLfetch{rob_7errors}{result}{upperlim}{lambda}}
&
&\scriptsize{\DTLfetch{rob_7errors}{result}{lowerlim}{mu1}\,/\,\DTLfetch{rob_7errors}{result}{upperlim}{mu1}}\\[-9pt]

\scriptsize{robust SE}  & \scriptsize{\DTLfetch{rob_7errors}{result}{robust se}{alpha}}
& \scriptsize{\DTLfetch{rob_7errors}{result}{robust se}{delta}}
& \scriptsize{\DTLfetch{rob_7errors}{result}{robust se}{gamma}}
& \scriptsize{\DTLfetch{rob_7errors}{result}{robust se}{lambda}}
&
& \scriptsize{\DTLfetch{rob_7errors}{result}{robust se}{mu1}}\\
 \midrule
 
 & $\boldsymbol{\mu_2}$ & $\boldsymbol{\mu_3}$ & $\boldsymbol{\mu_4}$ &  $\boldsymbol{\mu_5}$ & $\boldsymbol{\mu_6}$ & $\boldsymbol{\mu_7}$ \\[-6pt]
& \scriptsize{FE MPL2} & \scriptsize{FE MPL3} & \scriptsize{FE MPL4} & \scriptsize{FE MPL5} & \scriptsize{FE MPL6} & \scriptsize{FE MPL7} \\
 \midrule
\makecell{\scriptsize{\textbf{point}} \\[-12pt] \scriptsize{\textbf{estimate}}}  
& \DTLfetch{rob_7errors}{result}{parameter}{mu2}
& \DTLfetch{rob_7errors}{result}{parameter}{mu3}
& \DTLfetch{rob_7errors}{result}{parameter}{mu4}
& \DTLfetch{rob_7errors}{result}{parameter}{mu5}
& \DTLfetch{rob_7errors}{result}{parameter}{mu6}
& \DTLfetch{rob_7errors}{result}{parameter}{mu7} \\[-6pt]

\scriptsize{95\% CI} & \scriptsize{\DTLfetch{rob_7errors}{result}{lowerlim}{mu2}\,/\,\DTLfetch{rob_7errors}{result}{upperlim}{mu2}}
& \scriptsize{\DTLfetch{rob_7errors}{result}{lowerlim}{mu3}\,/\,\DTLfetch{rob_7errors}{result}{upperlim}{mu3}}
& \scriptsize{\DTLfetch{rob_7errors}{result}{lowerlim}{mu4}\,/\,\DTLfetch{rob_7errors}{result}{upperlim}{mu4}}
& \scriptsize{\DTLfetch{rob_7errors}{result}{lowerlim}{mu5}\,/\,\DTLfetch{rob_7errors}{result}{upperlim}{mu5}}
& \scriptsize{\DTLfetch{rob_7errors}{result}{lowerlim}{mu6}\,/\,\DTLfetch{rob_7errors}{result}{upperlim}{mu6}}&
\scriptsize{\DTLfetch{rob_7errors}{result}{lowerlim}{mu7}\,/\,\DTLfetch{rob_7errors}{result}{upperlim}{mu7}}\\[-9pt]

\scriptsize{robust SE}  & \scriptsize{\DTLfetch{rob_7errors}{result}{robust se}{mu2}}
& \scriptsize{\DTLfetch{rob_7errors}{result}{robust se}{mu3}}
& \scriptsize{\DTLfetch{rob_7errors}{result}{robust se}{mu4}}
& \scriptsize{\DTLfetch{rob_7errors}{result}{robust se}{mu5}}
& \scriptsize{\DTLfetch{rob_7errors}{result}{robust se}{mu6}}
& \scriptsize{\DTLfetch{rob_7errors}{result}{robust se}{mu7}}\\
 \midrule
 
\multicolumn{7}{l}{\scriptsize{estimation details: n = \DTLfetch{rob_7errors}{result}{n}{alpha},  ln-likelihood = \DTLfetch{rob_7errors}{result}{log-likelihood}{alpha},
AIC = \DTLfetch{rob_7errors}{result}{AIC}{alpha}, 
BIC = \DTLfetch{rob_7errors}{result}{BIC}{alpha}, \textbf{distinct errors per MPL}}}\\
\bottomrule
\multicolumn{7}{l}{\scriptsize{Robust standard errors (SE) clustered at the individual level, \DTLfetch{rob_7errors}{result}{cluster}{alpha} clusters}}
\end{tabularx}
\label{table:rob_7errors}
\end{table}

\subsection{Sample Variations}
Finally, we check whether the composition of the sample used to estimate preference parameters does have an effect on the estimation. To this end, we scrutinize two variations.

\noindent{\bf Drop-outs:} First, we take into account, that participants who completed the entire experimental sequence of three sessions might be systematically different from those who dropped out along the way. Estimation results presented in Table~\ref{table:rob_all}, are based on all observations including participants who dropped out prematurely. Parameter estimates do change to some degree compared to estimated parameters of the main modelling specification based on the more restrictive sample of people who completed the entire experimental sequence. However, debt aversion also characterizes this extended population.

\begin{filecontents*}{RESULTS/rob_all.csv}
"","result","alpha","delta","gamma","lambda","mu"
"1","parameter",0.6624,0.0311,1.0499,1.1056,0.4356
"2","robust se",0.0311,0.0054,0.0099,0.0109,0.0357
"3","p-value",21.3226628079721,5.77031881082639,106.154051374219,101.155693612857,12.2102410269653
"4","z",0.0000000000000000000000000000000000000000000000000000000000000000000000000000000000000000000000000000699622305180803,0.00000000791216978510054,0,0,0.000000000000000000000000000000000274076032424712
"5","lowerlim",0.6,0.02,1.03,1.08,0.37
"6","upperlim",0.72,0.04,1.07,1.13,0.51
"7",NA,NA,NA,NA,NA,NA
"8",NA,1.95996398454005,1.95996398454005,1.95996398454005,1.95996398454005,1.95996398454005
"9",NA,0,0,0,0,0
"10","log-likelihood",-4848.86,NA,NA,NA,NA
"11","n",14310,NA,NA,NA,NA
"12","AIC",9707.72,NA,NA,NA,NA
"13","BIC",9745.56,NA,NA,NA,NA
"14","cluster",148,NA,NA,NA,NA
\end{filecontents*}
\DTLloaddb{rob_all}{RESULTS/rob_all.csv} 

\begin{table}[ht]
\caption{Aggregate parameter estimates including drop-outs}
\begin{tabularx}{\textwidth}{ 
    c
    @{\hspace{36pt}}
  >{\centering\arraybackslash}X 
  >{\centering\arraybackslash}X 
  >{\centering\arraybackslash}X
  >{\centering\arraybackslash}X
  >{\centering\arraybackslash}X
  }
\toprule

& $\boldsymbol{\alpha}$ 
& $\boldsymbol{\delta}$ 
&  $\boldsymbol{\gamma}$ 
& $\boldsymbol{\lambda}$ 
& $\boldsymbol{\mu}$\\[-6pt]

& \scriptsize{risk aversion} 
& \scriptsize{discounting} 
& \scriptsize{debt aversion} 
& \scriptsize{loss aversion} 
& \scriptsize{Fechner error} \\
 \midrule
 
\scriptsize{\textbf{point estimate}} 
& \DTLfetch{rob_all}{result}{parameter}{alpha}
& \DTLfetch{rob_all}{result}{parameter}{delta}
& \DTLfetch{rob_all}{result}{parameter}{gamma}
& \DTLfetch{rob_all}{result}{parameter}{lambda}
& \DTLfetch{rob_all}{result}{parameter}{mu}\\[-6pt]

\scriptsize{95\% confidence interval} 
& \scriptsize{\DTLfetch{rob_all}{result}{lowerlim}{alpha}\,/\,\DTLfetch{rob_all}{result}{upperlim}{alpha}}
& \scriptsize{\DTLfetch{rob_all}{result}{lowerlim}{delta}\,/\,\DTLfetch{rob_all}{result}{upperlim}{delta}}
& \scriptsize{\DTLfetch{rob_all}{result}{lowerlim}{gamma}\,/\,\DTLfetch{rob_all}{result}{upperlim}{gamma}}
& \scriptsize{\DTLfetch{rob_all}{result}{lowerlim}{lambda}\,/\,\DTLfetch{rob_all}{result}{upperlim}{lambda}}
&\scriptsize{\DTLfetch{rob_all}{result}{lowerlim}{mu}\,/\,\DTLfetch{rob_all}{result}{upperlim}{mu}}\\[-9pt]

\scriptsize{robust standard error} 
& \scriptsize{\DTLfetch{rob_all}{result}{robust se}{alpha}}
& \scriptsize{\DTLfetch{rob_all}{result}{robust se}{delta}}
& \scriptsize{\DTLfetch{rob_all}{result}{robust se}{gamma}}
& \scriptsize{\DTLfetch{rob_all}{result}{robust se}{lambda}}
& \scriptsize{\DTLfetch{rob_all}{result}{robust se}{mu}}\\
\midrule
 
\multicolumn{6}{l}{\scriptsize{estimation details: n = \DTLfetch{rob_all}{result}{n}{alpha},  log-likelihood = \DTLfetch{rob_all}{result}{log-likelihood}{alpha},
AIC = \DTLfetch{rob_all}{result}{AIC}{alpha}, 
BIC = \DTLfetch{rob_all}{result}{BIC}{alpha}, logit Fechner error}}\\
\bottomrule
\multicolumn{6}{l}{\scriptsize{Robust standard errors (SE) clustered at the individual level, \DTLfetch{rob_all}{result}{cluster}{alpha} clusters}}
\end{tabularx}
\label{table:rob_all}
\end{table}

\noindent{\bf Trust and confidence:} Second, we consider a sample variation along the dimensions of trust and confidence of participants. As outlined earlier, if participants exhibit either mistrust in the payment reliability of the experimenter, or a lack of confidence in their own payment reliability, on average debt contracts would appear more and savings contracts less appealing to them. In this situation our estimate of debt aversion would be biased downwards. To investigate this possibility, we make use of participants' self reported ratings on two questions presented at the very end of the experimental sequence at Session~3: \textit{``Back then} [in the first session when you made all financial decisions]\textit{, 1. how sure have you been that the experimenters will make the promised payments in the future in case such a decision has been chosen as the decision that counts? 2. how sure have you been that you will make the promised payments in the future in case such a decision has been chosen as the decision that counts?"} To derive parameter estimates unperturbed by suboptimal trust or confidence, we consider a sample excluding all people who did not answer in the most positive way \textit{``very sure"}.

Estimation results are presented in Table~\ref{table:rob_mistrust}. Albeit our very strict exclusion criteria affects around 50\% of participants, parameter estimates remain similar to those of the main specification estimated on the entire sample. Debt aversion persists.

\begin{filecontents*}{RESULTS/rob_mistrust.csv}
"","result","alpha","delta","gamma","lambda","mu"
"1","parameter",0.6248,0.0379,1.0575,1.0943,0.444
"2","robust se",0.0436,0.0101,0.0178,0.015,0.0544
"3","p-value",14.3253710762092,3.7444954666329,59.2451604330212,73.0497633819222,8.16202912567006
"4","z",0.000000000000000000000000000000000000000000000151914488139044,0.00018075664051838,0,0,0.0000000000000003294427324459
"5","lowerlim",0.54,0.02,1.02,1.06,0.34
"6","upperlim",0.71,0.06,1.09,1.12,0.55
"7",NA,NA,NA,NA,NA,NA
"8",NA,1.95996398454005,1.95996398454005,1.95996398454005,1.95996398454005,1.95996398454005
"9",NA,0,0,0,0,0
"10","log-likelihood",-1837.52,NA,NA,NA,NA
"11","n",5700,NA,NA,NA,NA
"12","AIC",3685.05,NA,NA,NA,NA
"13","BIC",3718.29,NA,NA,NA,NA
"14","cluster",60,NA,NA,NA,NA
\end{filecontents*}
\DTLloaddb{rob_mistrust}{RESULTS/rob_mistrust.csv} 

\begin{table}[t!]
\caption{Aggregate parameter estimates including only participants who perfectly trust the experiment}
\begin{tabularx}{\textwidth}{ 
    c
    @{\hspace{36pt}}
  >{\centering\arraybackslash}X 
  >{\centering\arraybackslash}X 
  >{\centering\arraybackslash}X
  >{\centering\arraybackslash}X
  >{\centering\arraybackslash}X
  }
\toprule

& $\boldsymbol{\alpha}$ 
& $\boldsymbol{\delta}$ 
&  $\boldsymbol{\gamma}$ 
& $\boldsymbol{\lambda}$ 
& $\boldsymbol{\mu}$\\[-6pt]

& \scriptsize{risk aversion} 
& \scriptsize{discounting} 
& \scriptsize{debt aversion} 
& \scriptsize{loss aversion} 
& \scriptsize{Fechner error} \\
 \midrule
 
\scriptsize{\textbf{point estimate}} 
& \DTLfetch{rob_mistrust}{result}{parameter}{alpha}
& \DTLfetch{rob_mistrust}{result}{parameter}{delta}
& \DTLfetch{rob_mistrust}{result}{parameter}{gamma}
& \DTLfetch{rob_mistrust}{result}{parameter}{lambda}
& \DTLfetch{rob_mistrust}{result}{parameter}{mu}\\[-6pt]

\scriptsize{95\% confidence interval} 
& \scriptsize{\DTLfetch{rob_mistrust}{result}{lowerlim}{alpha}\,/\,\DTLfetch{rob_mistrust}{result}{upperlim}{alpha}}
& \scriptsize{\DTLfetch{rob_mistrust}{result}{lowerlim}{delta}\,/\,\DTLfetch{rob_mistrust}{result}{upperlim}{delta}}
& \scriptsize{\DTLfetch{rob_mistrust}{result}{lowerlim}{gamma}\,/\,\DTLfetch{rob_mistrust}{result}{upperlim}{gamma}}
& \scriptsize{\DTLfetch{rob_mistrust}{result}{lowerlim}{lambda}\,/\,\DTLfetch{rob_mistrust}{result}{upperlim}{lambda}}
&\scriptsize{\DTLfetch{rob_mistrust}{result}{lowerlim}{mu}\,/\,\DTLfetch{rob_mistrust}{result}{upperlim}{mu}}\\[-9pt]

\scriptsize{robust standard error} 
& \scriptsize{\DTLfetch{rob_mistrust}{result}{robust se}{alpha}}
& \scriptsize{\DTLfetch{rob_mistrust}{result}{robust se}{delta}}
& \scriptsize{\DTLfetch{rob_mistrust}{result}{robust se}{gamma}}
& \scriptsize{\DTLfetch{rob_mistrust}{result}{robust se}{lambda}}
& \scriptsize{\DTLfetch{rob_mistrust}{result}{robust se}{mu}}\\
\midrule
 
\multicolumn{6}{l}{\scriptsize{estimation details: n = \DTLfetch{rob_mistrust}{result}{n}{alpha},  log-likelihood = \DTLfetch{rob_mistrust}{result}{log-likelihood}{alpha},
AIC = \DTLfetch{rob_mistrust}{result}{AIC}{alpha}, 
BIC = \DTLfetch{rob_mistrust}{result}{BIC}{alpha},  logit Fechner error}}\\
\bottomrule
\multicolumn{6}{l}{\scriptsize{Robust standard errors (SE) clustered at the individual level, \DTLfetch{rob_mistrust}{result}{cluster}{alpha} clusters}}
\end{tabularx}
\label{table:rob_mistrust}
\end{table}

\section{Experimental Instructions} \label{sec:inst}

Upon first arrival at the lab detailed instructions regarding the experiment as a whole and Session~1 in particular were given as printed handout. Identical instructions were displayed on screen through the course of the experiment. Task-specific instructions where displayed on screen sequentially before the respective tasks  in all sessions.  After reading instructions, participants completed the tasks, and then received instructions for the following task. Participants were given time to carefully read the instructions and ask questions.

The study design as delineated in the main paper section \hyperref[sec:experiment]{Experiment} and in the instructions was approved by the Ethics Review Committee of Maastricht University (Reference Number: ERCIC\_138\_07\_05\_2019).

\subsection{Instructions at the beginning of Session 1 (on screen + printout to reread)}

\subsubsection{Overview}
As announced in the invitation email, this is a three-part experiment. Today is the first part of the experiment (Session~1). The second part (Session~2) will take place in exactly four weeks from now (\textit{Day, Date}, at the same starting time as today). The third part (Session~3) will take place in exactly eight weeks from now (\textit{Day, Date}, at the same starting time as today). The experiment today will last 90 minutes, Session~2 and Session~3 will last 30 minutes respectively. To participate in today's experiment, you have to be able to participate in all sessions. If you cannot participate at one of these dates, please raise your hand now.\bigbreak

\noindent The following will happen during the three sessions:

\subsubsection{Session 1 (today)}

Today you will make a total of 90 \textit{(120 in extension)} financial decisions, involving real money. The choices are simple and not meant to test you - the only correct answers are the ones you really think are best for you.\bigbreak

\noindent In the financial decisions you either have the choice between two options (Option A and Option B), or you have the choice of accepting or not accepting a savings or debt contract.\bigbreak

\noindent Generally, the financial decisions specify amounts of money that you will receive at different dates with different probabilities, or that you have to pay to the experimenter at different dates. The timing of the payments corresponds to the timing of the sessions. For instance, a financial decision may look as follows:

\begin{table}[H]
\setlength{\tabcolsep}{9pt}
\begin{tabularx}{\textwidth}{C{4cm} C{1pt} C{4cm} C{1pt} c}

\small{\textbf{Option A}} && \small{\textbf{Option B}} && \small{\textbf{Your choice}}\\ 
\cline{1-1} \cline{3-3} \cline{5-5}
\small{Receive \euro{18} today} && \small{Receive \euro{20} in 4 weeks} && \small{\ding{114} Option A \hspace{0.5cm} \ding{114} Option B}\\

\end{tabularx}
\end{table}

\noindent In this case, you have the choice between receiving \euro{18} today (i.e. at the end of today's session) or in four weeks, at the end of Session~2. A financial decision may also only involve future dates: 

\begin{table}[H]
\setlength{\tabcolsep}{9pt}
\begin{tabularx}{\textwidth}{C{4cm} C{1pt} C{4cm} C{1pt} c}

\small{\textbf{Option A}} && \small{\textbf{Option B}} && \small{\textbf{Your choice}}\\ 
\cline{1-1} \cline{3-3} \cline{5-5}
\small{Receive \euro{18} in 4 weeks} && \small{Receive \euro{20} in 8 weeks} && \small{\ding{114} Option A \hspace{0.5cm} \ding{114} Option B}\\

\end{tabularx}
\end{table}

\noindent In this case you choose between monetary amounts to be paid either in four weeks, at the end of Session~2 (Option~A), or in eight weeks, at the end of Session~3 (Option~B).\bigbreak

\noindent At the end of today's session, you will be asked to fill a short questionnaire. Afterwards, one of the 90 \textit{(120 in extension)} decision situations will be drawn randomly as the `decision that counts'. Your choice in that decision situation will then actually be implemented, and you will receive or pay the specified monetary amounts depending on your actual decision. Each decision situation has the same chance to be selected as the `decision that counts'. It is therefore in your interest to consider all decision situations with equal care. 

\subsubsection{Session 2 (In four weeks)}
In four weeks, we will ask you to complete a questionnaire. Additionally, all monetary payments that are due at Session~2 will be implemented. Please note that because of the questionnaire you will have to show up at this date, even if you will not receive or pay any monetary amounts at this date.

\subsubsection{Session 3 (In eight weeks)}
In eight weeks, we will ask you to complete a questionnaire. Additionally, all monetary payments that are due at Session~3 will be implemented. Please note that because of the questionnaire you will have to show up at this date, even if you will not receive or pay any monetary amounts at this date. 

\subsubsection{Your Payment}
The selection of the `decision that counts' will be made randomly and individually for each participant at the end of today's session. This selection will be made with the help of a bingo cage with 90 \textit{(120 in extension)} numbered balls. All decision situations are numbered, and the number drawn by the bingo cage will be the `decision that counts'. This decision will then actually be implemented, and you will receive or pay monetary amounts, as specified in the `decision that counts'.\bigbreak

\noindent At the beginning of today's session, you already received a show up fee of \euro{15} for all three sessions in cash. On top of that money you will receive the money earned from your decisions. Additionally, you will receive a completion bonus of \euro{20} after Session~3, provided you have shown up on time at each session, and have made all payments as agreed (more on this later). This completion bonus will be transferred to your bank account around one week after Session~3. For this payment, we will ask you for your bank details at the end of Session~3.\bigbreak

\noindent Please note, should you, arising through your own fault, fail to attend all sessions or fail to make any payments agreed upon you will be excluded from the remaining experiment and all payments associated with it. You will also be removed from the BEElab participant pool and thus not receive any further invitations for economic experiments.

\subsection{Instructions throughout Session~1 (on screen before respective task + printout to reread)}

\subsubsection{Let's go}
The 90 \textit{(120 in extension)} decision situations are separated into four parts. You will now receive the specific instructions for part 1. 

\subsubsection{Part1}
In this part, you will make a total of 10 decisions. In each decision you can choose between receiving monetary amounts today, or in one month, at Session~2. For instance, one of these decisions could look like this:

\begin{table}[H]
\setlength{\tabcolsep}{9pt}
\begin{tabularx}{\textwidth}{C{4cm} C{1pt} C{4cm} C{1pt} c}

\small{\textbf{Option A}} && \small{\textbf{Option B}} && \small{\textbf{Your choice}}\\ 
\cline{1-1} \cline{3-3} \cline{5-5}
\small{Receive \euro{18} today} && \small{Receive \euro{20} in 4 weeks} && \small{\ding{114} Option A \hspace{0.5cm} \ding{114} Option B}\\

\end{tabularx}
\end{table}

\noindent If you choose Option~A in this decision situation, you will receive \euro{18} today. If you choose Option~B, you will receive \euro{20} in four weeks, at Session~2.\bigbreak

\noindent If you prefer to receive \euro{18} today and nothing in four weeks, choose Option~A.\bigbreak

\noindent If you prefer to receive \euro{20} in four weeks and nothing today, choose Option~B.\bigbreak

\noindent Please note that we guarantee the later payment, even if you cannot participate on the due date for any unforeseen reason. In that case, we will transfer the money to your bank account, or you can pick it up at the secretarial office of the department of economics (MPE) at the School of Business and Economics. At the end of today's session, you will receive a receipt containing the email address of the principal investigator, who you can contact should there be any issues with the payment process.\bigbreak

\begin{center}
\textit{Subsequently, participants made the 10 decisions of MPL1 (Appendix~Table~\ref{tab:MPLI}).}
\end{center}

\subsubsection{Part 2}

In the following part you will make a total of 20 choices. All payments occur today but depend on the outcome of a coin flip. If a decision situation from this part has been randomly selected as the `decision that counts', you will make this coin flip yourself after the experiment today. The coin is fair. There is an equal chance of observing HEADS or TAILS.
\bigbreak

\noindent For example, you might be asked to choose between the following options:

\begin{table}[H]
\begin{tabularx}{\textwidth}{
C{1.75cm} 
>{\centering\arraybackslash}X 
C{1.75cm} 
>{\centering\arraybackslash}X 
C{1.75cm} 
>{\centering\arraybackslash}X 
C{1.75cm} 
>{\centering\arraybackslash}X 
c}

\multicolumn{3}{c}{\textbf{Option A}} && \multicolumn{3}{c}{\textbf{Option B}} && \multirow{2}{*}{\textbf{Your choice}}\\ 
\cline{1-3} \cline{5-7}

\footnotesize{Coin shows HEADS} && \footnotesize{Coin shows TAILS} && \footnotesize{Coin shows HEADS} && \footnotesize{Coin shows TAILS} &&\\
\cline{1-1} \cline{3-3} \cline{5-5} \cline{7-7} \cline{9-9}
\small{\euro{5}} && \small{\euro{4}} && \small{\euro{10}} && \small{\euro{1}} && \small{\ding{114} Option A \hspace{0.5cm} \ding{114} Option B}\\
\end{tabularx}
\end{table}

\noindent In this decision situation, if you choose Option~A and the coin shows HEADS, you win \euro{5}; if the coin shows TAILS, you win \euro{4}. 
If you choose Option~B and the coin shows HEADS, you win \euro{10}; if the coin shows TAILS, you win \euro{1}.
\bigbreak

\noindent In some decision situations, one option will be a safe amount and in the other option the amount depends on a coin flip. For instance, such a decision situation may look as follows: 

\begin{table}[H]
\begin{tabularx}{\textwidth}{
C{4.25cm}   
>{\centering\arraybackslash}X  
C{1.75cm} 
>{\centering\arraybackslash}X 
C{1.75cm}   
>{\centering\arraybackslash}X  
c}

\textbf{Option A} && \multicolumn{3}{c}{\textbf{Option B}} && \multirow{2}{*}{\textbf{Your choice}}\\ 
\cline{1-1} \cline{3-5}

\footnotesize{safe} && \footnotesize{Coin shows HEADS} && \footnotesize{Coin shows TAILS} &&\\
\cline{1-1} \cline{3-3} \cline{5-5} \cline{7-7}
\small{\euro{5}} && \small{\euro{10}} && \small{\euro{1}} && \small{\ding{114} Option A \hspace{0.5cm} \ding{114} Option B}\\
\end{tabularx}
\end{table}

\noindent In this decision situation, if you choose Option~A you receive \euro{5} for sure. \bigbreak

\noindent If you choose Option~B and the coin shows HEADS, you win \euro{10}; if the coin shows TAILS, you win \euro{1}. \bigbreak

\begin{center}
\textit{Subsequently, participants made the 20 decisions of MPL2 and~III (Appendix~Tables~\ref{tab:MPLII}~and~\ref{tab:MPLIII}).}
\end{center}

\subsubsection{Part 3}

In the following part you will make a total of 30 \textit{(45 in extension)} choices. This time, you will be offered a series of real savings contracts, that you can either accept or not accept. Savings contracts involve the payment of some monetary amount by you to the experimenter at an earlier date, and the repayment of a monetary amount to you at a later date. \bigbreak

\noindent For example, consider the following contract:

\begin{table}[H]
\setlength{\tabcolsep}{9pt}
\begin{tabularx}{\textwidth}{
C{3.5cm} 
>{\centering\arraybackslash}X 
C{4cm} 
>{\centering\arraybackslash}X 
c
}

\multicolumn{3}{c}{\textbf{Savings Contract}} && \small{\textbf{Your choice}}\\ 
\cline{1-3} \cline{5-5}
\small{Pay \euro{10} today} && \small{Receive \euro{12} in 4 weeks} && \small{\ding{114} Accept \hspace{1cm} \ding{114} Not Accept}\\

\end{tabularx}
\end{table}

\noindent Under such a contract, you pay the experimenter \euro{10} today, and receive \euro{12} in four weeks, at Session~2. Note that if you have accepted one of these contracts and in case it has been selected as the `decision that counts', you may use your show up fee of \euro{15}, to pay this amount today. \bigbreak

\noindent Please note that we guarantee the later payment, even if you cannot participate on the due date for any unforeseen reason. In that case, we will transfer the money to your bank account, or you can pick it up at the secretarial office of the department of economics (MPE) at the School of Business and Economics. \bigbreak

\noindent Some savings contracts are defined over dates in the future. This is an example of such a savings contract:

\begin{table}[H]
\setlength{\tabcolsep}{9pt}
\begin{tabularx}{\textwidth}{
C{3.5cm} 
>{\centering\arraybackslash}X 
C{4cm} 
>{\centering\arraybackslash}X 
c
}

\multicolumn{3}{c}{\textbf{Savings Contract}} && \small{\textbf{Your choice}}\\ 
\cline{1-3} \cline{5-5}
\small{Pay \euro{10} in 4 weeks} && \small{Receive \euro{15} in 8 weeks} && \small{\ding{114} Accept \hspace{1cm} \ding{114} Not Accept}\\

\end{tabularx}
\end{table}

\noindent Under such a contract, you pay the experimenter \euro{10} in four weeks, at Session~2; and receive \euro{12} in eight weeks, at Session~3. Note that if you have accepted one of these contracts and in case it has been selected as the `decision that counts', you need to bring the specified amount in cash at Session~2.  In any case, you will receive a receipt today, specifying what payments you agreed to make at what session. Additionally, we will send a reminder email before the session at which your payment is due, specifying the amount you need to bring to the session. \bigbreak

\noindent The receipt you get also contains the email address of the principal investigator, who you can contact should there be any issues with the payment process. \bigbreak

\noindent Should you fail to pay the specified amount at the specified Session, you will be excluded from the experiment, and will not receive any further payments, including the completion bonus payment of \euro{20}. \bigbreak

\noindent Note that you always have the choice to not accept a savings contract! If you do not accept, you won't pay any money at the earlier date, and won't receive any money at the later date. \bigbreak

\begin{center}
\textit{Subsequently, participants made the 30 (45 in extension) decisions of MPL4 and MPL5 (and MPL8 in extension) (Appendix~Tables~\ref{tab:MPLIV}, \ref{tab:MPLV}~and~\ref{tab:MPLIIX}).}
\end{center}

\subsubsection{Part 4}

In the following task you will make a total of 30 \textit{(45 in extension)} choices. This time, you will be offered a series of real debt contracts, that you can either accept or not accept. Debt contracts involve the payment of some monetary amount by the experimenter to you at an earlier date, and the repayment of a monetary amount by you to the experimenter at a later date. \bigbreak

\noindent For example, consider this following contract:

\begin{table}[H]
\setlength{\tabcolsep}{9pt}
\begin{tabularx}{\textwidth}{
C{4cm} 
>{\centering\arraybackslash}X 
C{3.5cm} 
>{\centering\arraybackslash}X 
c
}

\multicolumn{3}{c}{\textbf{Debt Contract}} && \small{\textbf{Your choice}}\\ 
\cline{1-3} \cline{5-5}
\small{Receive \euro{10} today} && \small{Pay \euro{12} in 4 weeks} && \small{\ding{114} Accept \hspace{1cm} \ding{114} Not Accept}\\

\end{tabularx}
\end{table}

\noindent Under such a contract, the experimenter pays you \euro{10} today, and you have to pay back \euro{12} to the experimenter in four weeks, at Session~2. 
Note that similar to the Saving Contracts, some debt contracts are defined over dates in the future. Here is an example of such a debt contract:

\begin{table}[H]
\setlength{\tabcolsep}{9pt}
\begin{tabularx}{\textwidth}{
C{4cm} 
>{\centering\arraybackslash}X 
C{3.5cm} 
>{\centering\arraybackslash}X 
c
}

\multicolumn{3}{c}{\textbf{Debt Contract}} && \small{\textbf{Your choice}}\\ 
\cline{1-3} \cline{5-5}
\small{Receive \euro{10} in 4 weeks} && \small{Pay \euro{12} in 8 weeks} && \small{\ding{114} Accept \hspace{1cm} \ding{114} Not Accept}\\

\end{tabularx}
\end{table}

\noindent Under such a contract, the experimenter pays you \euro{10} in four weeks, at Session~2; and you have to pay back \euro{12} to the experimenter in eight weeks, at Session~3.

\noindent Please note that should you accept a debt contract, we expect you to repay your debt in full, even if you cannot participate on the due date for any unforeseen reason. If you have accepted one of these contracts and in case it has been selected as the `decision that counts', you need to bring the specified amount in cash at the respective session.\footnote{Alternatively, you may also pay the specified amount via PayPal to the experimenter at the respective session.}  In any case, you will receive a receipt today, specifying what payments you agreed to make at what session. Additionally, we will send a reminder email before the session at which your payment is due, specifying the amount you need to bring to the session. \bigbreak

\noindent Should you fail to pay the specified amount at the specified session, you will be excluded from the experiment, and will not receive any further payments, including the completion bonus payment of \euro{20}. \bigbreak

\noindent Note that you always have the choice to not accept a debt contract! If you do not accept, you won't receive any money at the earlier date, and won't have to pay back any money at the later date. \bigbreak

\begin{center}
\textit{Subsequently, participants made the 30 (45 in extension) decisions of MPL6 and MPL7 (and MPL9 in extension) (Appendix~Tables~\ref{tab:MPLVI}, \ref{tab:MPLVII}~and~\ref{tab:MPLIX}).}
\end{center}

\subsubsection{Check-out questionnaire}
You completed all decisions. Now you still need to fill a questionnaire and then you are done with today's session. \bigbreak

\noindent In this part we ask you to answer some questions and rate some statements about yourself. Some of them you need to classify according to how much they resemble yourself, others need to be ranked according to how much you agree with them or you think society agrees with them, accordingly. \bigbreak

\begin{center}
\textit{Subsequently, participants provided basic sociodemographic information and answered the collection of 54 items on debt behavior and attitudes as described in detail in \cite{albrecht2022the}.}
\end{center}

\subsubsection{Random Draw}
Please give a sign to the experimenter. The experimenter will then come to you in order to draw the decision that counts from the bingo cage and implement your choice. Afterwards today's payments will be completed and you can leave the lab.

\subsection{Instructions at Session 2 (on screen before respective task)}

\subsubsection{General Intro}
Today we ask you to solve some logical tasks and to answer a set of questions. Additionally, at the end of the session all monetary payments that are due today will be implemented.

\subsubsection{CRT and Numeracy}
You will start by solving 19 logical tasks, afterwards there will be a questionnaire. Before you start with the logical tasks you will see an example on the next screen. \bigbreak

\noindent In each task there will be a short text explaining an issue. Underneath you will find a box where you can type your answer. The answer may be in form of a number or text depending on the task. When appropriate the answer's unit of measurement is already given. \bigbreak

\noindent Note, once you typed your answer and hit the continue button you will proceed to the next task and not be able to return. \bigbreak

\begin{center}
\textit{Subsequently, participants completed tasks on numeracy and cognitive reflection (see~\ref{sec:indiv_char_app}).}
\end{center}

\subsubsection{BFI}
You finished all logical tasks. In the next section we ask you to answer some questions about yourself. \bigbreak

\noindent All questions have the same structure: ``I am someone who ..." followed by something like``is outgoing and sociable." and need to be rated according to your level of agreement. \bigbreak

\begin{center}
\textit{Subsequently, participants completed 30 items of the Big Five Inventory-2-S (see Appendix~\ref{sec:indiv_char_app}).}
\end{center}

\subsubsection{Preference Module}
There is one more section with questions about yourself to go. \bigbreak

\begin{center}
\textit{Subsequently, participants completed 12 items of the Preference Survey Module (see Appendix~\ref{sec:indiv_char_app}).}
\end{center}

\subsection{Instructions at Session 3 (on screen before respective task)}

\subsubsection{General Intro}
Today we ask you to solve some logical tasks and to answer a set of questions. Additionally, at the end of the session all monetary payments that are due today will be implemented. 

\subsubsection{Raven}
You will start by solving 36 logical tasks, afterwards there will be a questionnaire. Before you start with the logical tasks you will see an example on the next screen. \bigbreak

\noindent In each task there will be a picture on the left side of the screen. In the upper half of the picture you may see a puzzle with different pieces. Most pieces are shown while the space for the last piece is left blank. You need to chose from the suggestions in the lower half, which piece fits the blank in the puzzle best. \bigbreak

\noindent Note, once you typed your answer and hit the continue button you will proceed to the next task and not be able to return. \bigbreak

\begin{center}
\textit{Subsequently, participants completed 36 Raven matrices (see Appendix~\ref{sec:indiv_char_app}).}
\end{center}

\subsubsection{Planned Behavior}
In the next section we ask you to answer some questions about the likelihood that you will make certain purchases in the future and how you will finance them. \bigbreak

\begin{center}
\textit{Subsequently, participants completed each seven items on their likelihood to purchase certain thins within the one year and the likelihood to loan-finance these purchases (see Appendix~\ref{sec:indiv_char_app}).}
\end{center}

\subsubsection{Financial Literacy}
In the next section we ask you to answer 16 financial questions. \bigbreak

\noindent Please note, for your convenience you may use the Windows built-in calculator. To start the calculator use the calculator-button in the bottom right-hand corner of the screen. If you want to, you can try that now. Please note, you can also set the calculator to scientific mode in case you want to do calculations involving exponents or the like. \bigbreak

\begin{center}
\textit{Subsequently, participants completed 36 financial literacy items (see Appendix~\ref{sec:indiv_char_app}).}
\end{center}

\subsubsection{Hypothetical Debt Contracts}
There is one more section to go. You will be asked how you would behave in a series of four different hypothetical situations. \bigbreak

\noindent Imagine your bank offered you a debt contract. Under this contract you receive \euro{100} from your bank today and have to pay back some amount in 6 months. \bigbreak

\noindent Please assume that in all these choices you must pay the full amount you owe to the bank on time. \bigbreak

\begin{center}
\textit{Subsequently, participants completed the four item staircase measure (see \cite{albrecht2022the} for details).}
\end{center}

\subsubsection{Additional Questions on Honesty and Trustworthiness of Experimental Environment}
In the next section we ask you to answer some questions about yourself and how you think about certain things.

\begin{center}
\textit{Subsequently, answered eight items from the HEXACO-60 Inventory in the honesty domain (see Appendix~\ref{sec:indiv_char_app}) and four further questions on the trustworthiness of the experimental environment.}
\end{center}

\end{document}